%% file: ms.tex
\documentclass[iop]{./emulateapj}
\usepackage{xcolor}
\usepackage{amsmath}

\newcommand\txred[1]{{\color{black}#1}}

\newcommand\Tstrut{\rule{0pt}{2.6ex}}       
\newcommand\Bstrut{\rule[-1.1ex]{0pt}{0pt}} 

\shortauthors{Hargis and Rhode}
\shorttitle{The Spatial Distributions of GC Systems}

\received{}
\accepted{}

\begin{document}

\title{Globular Cluster Systems and their Host Galaxies: Comparison of Spatial Distributions and Colors}

\author{Jonathan R. Hargis\altaffilmark{1,2} and Katherine L. Rhode\altaffilmark{2}} 

\altaffiltext{1}{Department of Astronomy, Haverford College, 370 Lancaster Avenue, Haverford, PA 19041, USA; jhargis@haverford.edu}

\altaffiltext{2}{Department of Astronomy, Indiana University, 727 East 3rd Street, Swain West 319, Bloomington, IN 47405, USA.}

\begin{abstract}
We present a study of the spatial and color distributions of four early-type galaxies and their globular cluster (GC) systems observed as part of our ongoing wide-field imaging survey.  We use $BVR$ KPNO-4m+MOSAIC imaging data to characterize the galaxies' GC populations, perform surface photometry of the galaxies, and compare the projected two-dimensional shape of the host galaxy light to that of the GC population.  The GC systems of the ellipticals NGC 4406 and NGC 5813 both show an elliptical  distribution consistent with that of the host galaxy light.  Our analysis suggests a similar result for the giant elliptical NGC 4472, but a smaller GC candidate sample precludes a definite conclusion. For the S0 galaxy NGC 4594, the GCs have a circular projected distribution, in contrast to the host galaxy light which is flattened in the inner regions.  For NGC 4406 and NGC 5813 we also examine the projected shapes of the metal-poor and metal-rich GC subpopulations and find that both subpopulations have elliptical shapes that are consistent with those of the host galaxy light.   Lastly, we use integrated colors and color profiles to compare the stellar populations of the galaxies to their GC systems.  \txred{For each galaxy, we explore the possibility of color gradients in the individual metal-rich and metal-poor GC subpopulations.  We find statistically significant color gradients in both GC subpopulations of NGC~4594 over the inner $\sim 5$ effective radii ($\sim 20$ kpc).}  We compare our results to scenarios for the formation and evolution of giant galaxies and their GC systems.
\\
\end{abstract}


\section{Introduction}
\label{sec:chap4_intro}

Globular cluster (GC) populations are powerful probes of the assembly histories of galaxies (e.g., Ashman \& Zepf 1998; Brodie \& Strader 2006).  GCs are one of the few stellar populations in galaxies that survive for a Hubble time (Milky Way GCs have ages of ${\sim}11-13$ Gyr; \citealt{do10}) and therefore they contain an observational record of the chemical and dynamical conditions present in their host galaxy at the earliest stages of the galaxy's formation.  In the context of hierarchical galaxy formation theory, GC populations have been used to probe a number of important physical processes.  For example, observations of the spatial distribution, stellar populations, and kinematics/proper motions of Milky Way GCs have been used to study the role of accretion and \textit{in situ} formation in building the Galaxy (e.g., \citealt{ke12,ma04,fo10,di99,se78}). 

Studies of GC populations outside the Local Group have extended this type of work to galaxies with a wider range of masses, morphologies, and environments \citep{br06,fe12,br14}.  We have been carrying out a wide-field optical imaging survey of giant spiral, elliptical, and lenticular galaxies at distances from ${\sim} 10 - 30$ Mpc in order to characterize the global properties of giant galaxy GC systems.  The survey design is presented in Rhode \& Zepf (2001; hereafter RZ01) and previous results are given in Rhode \& Zepf (2001, 2003, 2004; hereafter RZ01, RZ03, RZ04), Rhode et al. (2005, 2007, 2010), Hargis et al. (2011; hereafter H11), Hargis \& Rhode (2012; hereafter HR12) and Young, Dowell, \& Rhode (2012). 

The goal of the survey is to provide robust constraints on galaxy formation scenarios by quantifying the fundamental ensemble properties of galaxy GC populations (e.g, total number and specific frequency of GCs; fractions of metal-rich and metal-poor clusters).  Accurate measurements of these properties allow for key tests of the phenomenological GC/galaxy formation models of the previous two decades -- galaxy mergers \citep{as92}, in-situ multiphase dissipational collapse \citep{fo97}, and collapse plus accretion \citep{co98} -- and provide constraints on the latest hierarchical galaxy formation simulations that have investigated GC system properties in a $\Lambda {\rm CDM}$ context (e.g, \citealt{mo06,pr08,mu10,gr10}).  \txred{GC populations are also being used to study the formation and evolution of early-type galaxies in the context of a ``two-phase'' (or ``inside-out'') assembly scenario (see \citealt{br14} and references therein), where giant galaxies experience an early, rapid, dissipational, \textit{in situ} phase of growth followed by an ongoing dissipationless, accretion phase \citep{na09, va10,os10}.}

In this paper we study the spatial and color distributions of a subset of four giant E/S0 galaxies in our survey.  We perform surface photometry of the host galaxies and compare the radial and azimuthal structure of the galaxy light to the GC system spatial distribution.  We also compare the stellar populations of the host galaxies to their GC systems using radial color profiles (e.g., $B-R$ versus projected radius).  Because both the GC systems and host galaxies are studied using the identical imaging data (see $\S\ref{sec:chap4_data}$), we are able to minimize systematic errors in our comparisons.  Our primary goal is to quantify the shape of the projected two-dimensional (2D) distribution of the GC systems and compare the results to the shape of the host galaxy light distribution.  Whenever possible in the analysis, we consider both the total GC population as well as the metal-rich and metal-poor GC subpopulations.  

The majority of the work on the spatial distributions of galaxy GC populations has focused on the azimuthally-averaged {\it radial} distribution of clusters.  Consequently, the connection between the properties of GC system azimuthal distributions and their host galaxies remains poorly understood \citep{as98,br06}.  The largest systematic study of the azimuthal distribution of GC systems was by \citet{wa13}, who performed a statistical study of the 2D shape of the 94 early-type galaxies in the ACS Virgo Cluster Survey (hereafter ACSVCS; \citealt{co04}).  \citet{wa13} found that in modestly flattened galaxies (ellipticity $\epsilon > 0.2$) both the metal-rich and metal-poor GC subpopulations tend to be aligned to the host galaxy.  Detailed results on the GC system azimuthal distributions of 23 ACSVCS galaxies were recently presented by \citet{pa13}.  They confirmed the position angle alignment of GCs with the host galaxies discussed by \citet{wa13} and also found that the ellipticity of the metal-rich GC subpopulations shows a better correlation with the ellipticity of the bulk starlight of the host galaxies than the metal-poor GC subpopulations.  However, in both the \citet{wa13} and \citet{pa13} studies, the field of view of the ACS+{\it HST} configuration limits the analysis to only the central ${\sim} 2.4\arcmin$ of the galaxies  (${\sim} 12$ kpc at the the distance of Virgo). The radial distributions of giant galaxy GC systems typically show a significantly larger extent (see Rhode et al. 2010), and the ACSVCS imaging only provides $50\%$ spatial coverage for moderate-luminosity galaxies ($M_V \gtrsim -21$; Rhode 2012). 

Studies of individual galaxies that do cover a larger spatial area have reached differing conclusions regarding the alignment of galaxy GC systems. \citet{go04} studied the azimuthal distribution of the giant elliptical NGC~4374 and found that the metal-rich and metal-poor subpopulations are similar in shape (both to each other and to the host galaxy light).  However, their study of NGC~1316, the central elliptical galaxy in Fornax, found that only the metal-rich subpopulation has a similar 2D shape to the galaxy while the metal-poor population has a nearly spherical (circular) spatial distribution \citep{go01}.  A follow-up study of NGC~1316 by \citet{ri12} found the opposite result for the metal-poor population, namely that the blue GCs {\it do} show an elliptical spatial distribution.  \citet{ri12} suggests that the differing results are likely due to minimized contamination and a more precise sampling of the GC subpopulations through the inclusion of the Washington $C$ filter in their imaging (compared to the $BVI$ selection of GC candidates in the \citet{go01} work).  The study of M87's GC system by \citet{st11} also found that the metal-poor GC subpopulation is non-circular and has an azimuthal distribution consistent with the host galaxy light.  \txred{However, a compilation of published and new wide-field imaging results on the ellipticities of the GC systems of six galaxies by \citep{ka13} may suggest a better correlation between the azimuthal distributions of the metal-rich GC subpopulations and the host galaxy light.}

The general conclusion reached by \citet{br06} in their \textit{Annual Review of Astronomy and Astrophysics} article on GCs and galaxy formation was that the azimuthal distributions of galaxy GC systems follow that of the host galaxy. At present, however, the small number of galaxies with detailed GC azimuthal distribution subpopulation studies {\it over a large radial extent} makes general conclusions about the degree of agreement or disagreement between {\it subpopulations} less secure.  Our imaging has a radial coverage of ${\sim} 25\arcmin$ from the galaxy center (${\sim} 120$ kpc at Virgo distances), allowing us to study the azimuthal distribution out to several effective radii.

Additional motivation for azimuthal distribution studies comes from the connection between the projected 2D shapes of GC systems, GC system kinematics, and the signature of galaxy formation processes. The spatial distribution of point-particles in a large gravitational potential (i.e., GCs in a galaxy+dark matter potential) should be closely tied to the particles' kinematic properties (e.g., rotational flattening, pressure support; Binney \& Tremaine 2008), and therefore the azimuthal distribution of a galaxy's GC system should be linked to the dynamics of the system.  The differing spatial and kinematic properties of the metal-rich and metal-poor subpopulations of Milky Way GCs provide a local example of this connection.  The Milky Way metal-rich GC subpopulation is not only more centrally concentrated (radially) than the metal-poor GC subpopulation,  it is also spatially flattened and shows a modest rotation compared to the spherical distribution and low-rotation of the metal-poor GCs \citep{zi85}.  In addition, because the dynamical timescales of GCs in galaxy halos (in particular) are long, the kinematics -- and hence the spatial distribution -- may retain the imprint of the formation processes that created the GC subsystems (e.g., dissipation in the metal-rich population; accretion in the metal-poor population).   Put another way, the GC system shape may reflect the long-timescale kinematics (particularly in GCs at large radii) and therefore give insights into the various physical processes at work in the formation of galaxies. 

This paper is organized as follows.  In $\S\ref{sec:chap4_data}$ we give an overview of the data sets used in this study and discuss the galaxy sample selection.  In $\S\ref{sec:chap4_analysis}$ we give a detailed description of the data analysis methods (galaxy surface photometry and the azimuthal distribution analysis of the GC system).  In $\S\ref{sec:chap4_results}$ we present the results of the surface photometry and GC system analysis for the individual galaxies analyzed in this work.  In $\S\ref{sec:chap4_comparisons}$ we discuss the results and present our conclusions.


\section{Data Sets and Sample Selection}
\label{sec:chap4_data}

For this study, we selected a sub-sample of galaxies from our survey with the best overall GC system statistics so that we could explore the host galaxy-GC connection by examining the metal-rich (red) and metal-poor (blue) GC subpopulations.  We selected galaxies that (1) show clear statistical evidence of GC system color bimodality and (2) have sufficient numbers of GC candidates in each of the subpopulations.  Our previous work studying the azimuthal distribution of the GC system of NGC 7457 (H11) showed that at least 50 GC candidates per subpopulation are necessary for statistically meaningful results using our adopted methodology.
 
Two further constraints reduce the number of galaxies in the sub-sample constructed for this study.  First, for observations of a particular galaxy, the differing magnitude incompleteness in the individual $BVR$ filters (due to different image exposure times, filter/CCD sensitivity, GC color, etc.) means that we only consider the number of GCs in the {\it color-complete} GC candidate lists (i.e. the ``$90\%$ sample'', which is $90\%$ complete in each filter).  That is, the $90\%$ color-complete sample accounts for the fact that either blue or red GC subpopulations could be preferentially detected.  Using these samples, rather than the full GC candidate lists, is therefore important for minimizing observational biases.  Second, in order to minimize contamination, we impose a radial cut at the observed radial extent of the GC system (the radius at which the surface density of GCs is consistent with the background within the errors, as defined by our survey methods).  We therefore require a galaxy to have $N>50$ GC candidates (a) in both the red and blue subpopulations, (b) in the $90\%$ color-complete sample, and (c) within the radial extent of the GC system.  Four of the survey galaxies pass these criteria: NGC 4472, NGC 4406, NGC 4594, and NGC 5813.  The basic data on these galaxies and their GC systems are shown in Table~\ref{tab:chap4_table1}.  

\begin{deluxetable*}{lclcccccccc}
\tablecolumns{11}
\tablewidth{0pc}
\tabletypesize{\scriptsize}
\tablecaption{Basic Properties of Sample Galaxies and their GC Systems\label{tab:chap4_table1}}
\tablehead{
\colhead{Name} & \colhead{Type} & \colhead{$M_V^T$} & \colhead{$m-M$} & \colhead{Distance} & \colhead{$N_{\rm GC}$} & \colhead{$S_N$} & \colhead{Radial Extent}          & \colhead{$f_{\rm blue}$}                             & \colhead{Environment}  & \colhead{Reference\tablenotemark{1}} \\
                          &                          &                               &                              &   \colhead{(Mpc)}    &                                    &                            &     \colhead{(arcmin and kpc)}             &                                                                 &                                      &
}
\startdata
NGC~4472 (M49)& E2                    &   $-23.1$               & $31.12$                 & $16.7$                    &  $5900 \pm700$        & $3.6 \pm0.6$    &  $21\arcmin (100~{\rm kpc})$ & $0.60$                                                  &  Virgo Cluster               & $1$  \\
NGC~4406 (M86)& E3                    &   $-22.3$               & $31.12$                 & $16.7$                    &  $2900\pm400$            &  $3.5\pm0.5$ &  $17\arcmin (80~{\rm kpc})$ & $0.62$                                                     & Virgo Cluster               & $2$ \\
NGC~5813         &  E1                    &   $-22.3$               & $32.54$                  & $32.2$                   &  $2900\pm400$        & $3.6\pm0.8$      & $13\arcmin (120~{\rm kpc})$ &  $0.68$                                                   & NGC~5846 Group        &  $3$\\
NGC~4594 (M104)& S0                  &   $-22.4$               & $29.95$                 & $9.8$                      &  $1900\pm200$          & $2.1\pm0.3$    &  $19\arcmin(50~{\rm kpc})$   &  $0.63$                                                    & Field                           & $2$ 
\enddata
\tablecomments{$M_V^T$ are computed by combining $m-M$ with $V_T^0$ from the RC3.  $V_T^0$ are from \citet{RC3}, except for NGC~5813 where $V_T^0$ was calculated using $V_T$ from \citet{RC3} and the Galactic extinction $A_V$ from \citet{sfd98}. $m-M$ are from \citet{wh95} for NGC~4472 and NGC~4406 (distance to Virgo Cluster from {\it HST} GCLF observations of M87); others \txred{are SBF distances} from \citet{to01}.  Total number of GCs $N_{\rm GC}$ from references. $V$-band normalized specific frequency $S_N$ from references.  Radial extent of GC systems is defined as the radius at which the corrected surface density profile becomes consistent with the background within the uncertainties.  Fraction of metal-poor GCs $f_{\rm blue}$ from references.
}
\tablenotetext{1}{References: (1) \citealt{rh01}; (2) \citealt{rh04}; (3) Hargis \& Rhode 2012}
\end{deluxetable*}

The GC candidate lists and deep optical imaging for these galaxies are taken from the results of our ongoing survey: RZ01 (NGC 4472), RZ04 (NGC 4406, NGC 4594\footnote{For NGC~4594, the GC candidate list utilized in this work is slightly different from the original list derived in RZ04.  Subsequent work examining GC candidates close to the galaxy center resulted in a reduction in the size of the central mask, so we supplemented the original GC candidate list from RZ04 with 120 additional GC candidates located at small galactocentric radii.}), and HR12 (NGC 5813).    Here we give a brief overview of the observational and data analysis methods used in the survey (see RZ01, RZ04 for additional details).  In order to characterize the GC systems of galaxies outside the Local Group, we require deep, wide-field, high spatial resolution imaging in three filters.  This provides (a) sufficient depth of imaging to cover a significant fraction of the GC luminosity function (GCLF) at the distances of the galaxies, (b) large spatial coverage to trace galaxy GC systems over their full radial extent, (c) excellent image quality, allowing us to resolve many contaminating background galaxies, and (d) multi-color photometry, which further reduces contaminating objects in our GC candidate lists.  Our survey has been conducted using $BVR$ imaging with the KPNO Mayall 4m telescope with the Mosaic imager ($36\arcmin \times 36\arcmin$ field of view; $0.26\arcsec$ pixels) and the WIYN 3.5m telescope with the Minimosaic imager ($9.6\arcmin \times 9.6\arcmin$ field of view; $0.14\arcsec$ pixels).  All four galaxies in this study were imaged using the Mayall 4m+Mosaic configuration.    

The resulting deep imaging data are used for both galaxy surface photometry and the GC system analyses in our survey.  We derive GC candidate lists using the techniques described in RZ01 and RZ04. Contamination is significantly reduced by eliminating extended objects (resolved background galaxies) and selecting GC candidates in the $V-R$ vs $B-V$ color-color plane.  The total number of GC candidates in the $90\%$ color-complete samples are 366 for NGC 4472, 1031 for NGC 4406, 1286 for NGC 4594, and 809 for NGC 5813.


\section{Data Analysis and Methodology}
\label{sec:chap4_analysis}

\subsection{Galaxy Surface Photometry}
\label{sec:chap4_sfc_phot_analysis}

We used the ELLIPSE routine \citep{je87} in IRAF to fit a series of ellipses to the galaxy light after accounting for the sky background level and masking resolved and unresolved sources. Although the galaxy light will dominate the sky background in the inner regions of the galaxy, in the outskirts the galaxy flux will become comparable to the noise in the sky background.  Therefore the accurate determination of the sky background (and its associated uncertainty) is critical for proper photometry.  We mark a series of rectangular regions across the image that are free of bright stars and sufficiently far from the galaxy.  We adopt the average of the median sky values in these regions as our (constant) sky value and subtract it from the image.  For the associated statistical uncertainty in the sky background $\sigma_{\rm sky}$, we adopt the standard deviation of the median values.  We next mask the light from resolved and unresolved objects on the frame to avoid the contribution of flux from foreground stars and background galaxies.  Regions of the image containing saturated stars and bleed trails were also masked.  The ELLIPSE task was run on the masked, background subtracted $V$ band image of each galaxy, resulting in a series of best-fit isophotes.  Because our images are deep, the centers of the host galaxies (typically the inner ${\sim} 10\arcsec$) are saturated in our images.  We therefore hold the $X,Y$ centers of the ellipses fixed during the isophotal fitting process. These fixed positions were determined from an initial run of ELLIPSE performed {\it without} holding the center constant and adopting the $X,Y$ center of the innermost isophotes as calculated by ELLIPSE.  In general we saw good agreement between the $X,Y$ centers of the innermost isophotes (e.g. very little drift in the centering) and the adopted center of the galaxy used for the calculation of the GC radial surface density profiles.  

We fitted ellipses at semi-major axes ranging from just outside the central, saturated region of the galaxies to the radius at which the mean isophotal intensity became consistent with the overall variation in the sky background across the frame (see $\S\ref{sec:chap4_sfc_phot_errors}$).  After running ELLIPSE on the $V$ band image, we used ELLIPSE to measure the mean intensity in the $B$ and $R$ images along the resulting $V$ band isophotes.  Because this procedure measures the $BVR$ flux at identical semi-major axes, we can use the results to construct radial color profiles for the galaxy light.  We use observations of Landolt (1992) standard stars to determine color coefficients and magnitude zero points to convert the instrumental magnitudes and colors to the standard system. We directly calibrate the $V$ magnitudes and $B-V$ and $V-R$ colors and provide the corresponding $B-R$ colors as calculated from the $B-V$ and $V-R$ results.   The results of the surface photometry are discussed with the individual galaxy results in $\S\ref{sec:chap4_results}$.  The surface photometry data are presented in Tables~\ref{tab:chap4_n4406_sfc_phot}-\ref{tab:chap4_n4594_sfc_phot} of Appendix A and have been corrected for Galactic extinction using the reddening maps of \citet{sfd98}.

\subsubsection{Surface Photometry Uncertainties}
\label{sec:chap4_sfc_phot_errors}

The primary results of interest from the ELLIPSE fitting are the mean isophotal intensity, mean color, ellipticity, and position angle at each semi-major axis, as well as their associated uncertainties.  For the ellipticity and position angle, the uncertainties are calculated by ELLIPSE based on the internal fitting errors.  In general we consider three sources of uncertainty in the surface brightness: (1) the error on the mean isophotal intensity $\sigma_{\rm INT}$, (2) the statistical uncertainty in the sky background $\sigma_{\rm sky}$, and (3) the systematic uncertainty in the determination of the sky background $\sigma_{\rm sys}$.  The error on the mean isophotal intensity $\sigma_{\rm INT}$  is calculated by ELLIPSE from the RMS scatter in the intensity about the isophote and the number of data points used in the fitting.  As discussed above, we adopt the standard deviation of the median sky values (as measured in the box regions) as the statistical uncertainty in the sky level $\sigma_{\rm sky}$.  Because we perform the ELLIPSE fitting on an image which has had a constant sky level removed, the total uncertainty in the mean intensity $\sigma_{\rm tot}$ is simply the quadrature sum of the RMS uncertainty and the error in the sky level.  For the innermost isophotes near the galaxy center, the total error is often very small ($ < 0.01$).  So, where $\sigma_{\rm tot}$ is less than the uncertainty in the photometric zero point, we adopt the uncertainty in the photometric zero point (typically ${\sim} 0.01 - 0.03$) as the error on the surface brightness (i.e., the lower limit to the surface brightness errors).  

Lastly we consider the systematic uncertainty in the sky background determination.  We found that due to the presence of bright stars on or just off some of the images, scattered light artifacts sometimes contributed to a variable background level in ways that were not easily modeled using high-order polynomial surfaces.  In general the sky background variation across a frame was typically of order $1\%$ of the mean background level. Although this variation is small when compared to the overall background, such a variation will in general {\it not} be small when compared to the faint flux in the outskirts of galaxies.  The lack of a perfectly uniform sky background therefore sets a practical limit on the faintness level reached by our surface photometry.  For each $BVR$ image for our target galaxies, we adopt the overall variation in the sky background as a systematic uncertainty $\sigma_{\rm sys}$ in our choice of background level.  That is, we use the $\sigma_{\rm sys}$ term to account for the fact that our mean background level could be higher or lower by $\pm \sigma_{\rm sys}$.  Although the standard deviation in the sky $\sigma_{\rm sky}$ will reflect this variation as well, adopting a ``range'' for the sky background determination more closely reflects the practical difficulties in setting an absolute sky level.  We incorporate $\sigma_{\rm sys}$ into our analysis by increasing or decreasing the mean isophotal intensity $I$ at each radius by $\pm \sigma_{\rm sys}$ counts and recalculating the surface brightness.  This produces the ``envelopes'' about the fiducial surface brightness shown in panels (a) of Figures~\ref{fig:chap4_n4406_sfc_phot}, \ref{fig:chap4_n4472_sfc_phot}, \ref{fig:chap4_n5813_sfc_phot}, and \ref{fig:chap4_n4594_sfc_phot}.  In cases where there was little scattered light on the images, we find $\sigma_{\rm sys} < \sigma_{\rm sky}$ and therefore the statistical uncertainty in the sky dominates the error budget.  Alternatively, in cases where scattered light was problematic, we found  $\sigma_{\rm sys} > \sigma_{\rm sky}$.  In practice we fitted ellipses to the galaxy light out to the radius at which the intensity becomes comparable to the larger of $\sigma_{\rm sys}$ or $\sigma_{\rm sky}$.

\subsubsection{Galaxy Surface Brightness and GC System Radial Profile Comparisons}
\label{sec:chap4_profile_fitting}

The $V$-band surface brightness profiles of the target galaxies were fitted with de Vaucouleurs law ($r^{1/4}$) \txred{and S\'ersic ($r^{1/n}$) profiles} over the full radial extent of the data as a function of the geometric mean radius $r \equiv \sqrt{ab}$ of the isophote (where $a,b$ are the semi-major and semi-minor isophote axes, respectively).  The effective radius $r_e$ of the de Vaucouleurs profile was determined from the slope of a linear fit.  \txred{For the S\'ersic model, the effective radius was determined from the S\'ersic index $n$ (where $n=4$ for the $r^{1/4}$ law) and scale radius coefficient from a non-linear least squares fit \citep{ca93,gr05}.  }

To compare the radial distribution of the GC system to the host galaxy surface brightness profiles, we fitted both $r^{1/4}$ and S\'ersic profiles to the GC radial surface density profiles.  We use the results of the de Vaucouleurs law fits from HR12 for NGC~5813 and from RZ01, RZ04 for NGC~4472, NGC~4406, and NGC~4594, respectively.  The corrected radial surface density profiles were fit with a linear function of the form

\begin{equation}
\log{\sigma_{\rm GC}}=a_0 + a_1 r^{1/4}, \label{sigma_gc}
\end{equation}

\noindent where the resulting slope $a_1$ determines the effective radius of the GC system.  \txred{Our S\'ersics fits are of the form}

\begin{equation}
\log{\sigma_{\rm GC}}=a_0 + a_1 r^{1/n}, \label{sigma_gc_sersic}
\end{equation}

\noindent \txred{where the values of both $n$ and $a_1$ determine the effective radius.}

We emphasize that the effective radius as derived from the de Vaucouleurs law or S\'ersic profile is the radius that contains half the total luminosity (or number of GCs) {\it integrated over all radii}.  That is, the profiles explicitly assume an infinite radius for the system (either galaxy or GC system).  This is problematic when the model does not provide a particularly good fit out to infinity or, for example, a GC system has a shallow $r^{1/4}$ law slope.  In these cases, the effective radius will be large and implies a significant number of GCs at large radii.  We therefore make a distinction between the ``theoretical effective radius'' (determined from the model fits) and an ``empirical effective radius''.  Because we have an estimate of the radial extent of the GC system from the GC surface density profile and an estimate of the total number of GCs, we can use this to directly measure $r_e$.  That is, we determine an ``empirical $r_e$'' by integrating the best-fit $r^{1/4}$ profile to the radius that includes half the the total number of GCs in the system as derived from our data (i.e., the half-total number radius).  \txred{We adopt the $r^{1/4}$ profile rather than the S\'ersic profile for consistency with our total number of GC calculations in previous survey papers (RZ01, RZ04, HR12).}  While a first-order comparison of effective radii for the galaxy and GC system can be done directly using the slopes of the best-fit profiles, the empirical $r_e$ value for the GC system is a  more realistic estimate of the half-total number radius. 

\subsubsection{Comparing GC System and Galaxy Color Profiles}
\label{sec:chap4_gc_gal_grad}

We compare the mean colors and color gradients of the GC systems and host galaxies using the following procedure. The ELLIPSE fitting procedure returns the mean host galaxy color as measured along each isophote, with the isophotes spaced equidistant in log-space (i.e. scaled by a constant mutiplicative factor).  To obtain a mean GC system color measured in a similar fashion, we use the galaxy isophotes as determined by ELLIPSE to define elliptical bins.  We extend the elliptical bins beyond the last measured isophote for the galaxy using a fixed position angle and ellipticity set by the shape of the outermost measured isophote.  

The elliptical isophotes for the galaxy are narrowly spaced relative to the GC system, so we use every fourth isophote to define the edges of the elliptical bins in order to obtain a larger number of GC candidates per bin.  We calculate the mean $B-R$ color in each bin and adopt the standard error on the mean as the uncertainty in the color.  We adopt the geometric mean radius $r \equiv \sqrt{\bar{a}\bar{b}}$ as the radius of the bin, consistent with what we have adopted for the galaxy isophote fitting.  We use the inner and outer edges of the bin to calculate the average semi-major $\bar{a}$ and semi-minor axes $\bar{b}$.

For the host galaxy, we calculate an integrated $B-R$ color using the total magnitudes derived from de Vaucouleurs law fits to the $B$ and $R$ surface brightness profiles. For the GC system, we calculated the mean using the elliptical bins (and weighted by the inverse square of the uncertainties), and as a check, the calculation of the mean based on the individual GC candidate colors gave identical results within the uncertainties.  We also used a radial cut for the GC system calculation, using only GC candidates within the measured radial extent of the system in order to minimize contamination. The error in the weighted mean was chosen as the uncertainty in the average color of the GC system.

We also compare the mean colors of the blue and red GC subpopulations to the integrated host galaxy color.  We adopt the homoscedastic results (assumption of same dispersion for both subpopulations) of the KMM analysis from HR12 (NGC~5813) and RZ01, RZ04 (NGC~4472, NGC~4406, NGC~4594) for the mean GC subpopulation colors.  The uncertainties in the blue and red color distribution peaks were determined using a Monte Carlo bootstrapping algorithm.  

The presence of galaxy and GC system color gradients was explored by fitting a weighted linear function to the data in log-space due to the non-linear spacing of the elliptical bins.  \txred{The ``global'' GC system color gradients -- which consider both the metal-rich and metal-poor GC subpopulations together -- have been investigated previously for our target galaxies (HR12 for NGC~5813; RZ01 for NGC~4472, and RZ04 for NGC~4406 and NGC~4594).  Numerous studies have shown that global GC system color gradients arise from the changing mix of metal-rich and metal-poor GCs with projected radius from the galaxy center \citep{ge96, ha98, le98, rh01, rh04}.  In this study we explore the possibility of color gradients within the \textit{individual} GC subpopulations (see also Harris et al. 2009, Forbes et al. 2011) and compare the results to the host galaxy color gradients.}  In general the color profile data were fitted with a linear function of the form

\begin{equation}
(B-R)= a_0 + a_1 \log{(r)}, \label{eq_color_grad}
\end{equation}

\noindent where the slope $a_1$ is the color gradient in units of ${\rm mag ~dex}^{-1}$.  The inverse squares of the uncertainties were chosen as the weights for the least-squares fitting.  This weighting is critical for measuring a galaxy color gradient in particular.  The uncertain sky background (in {\it both} filter bandpasses) causes both significant statistical and systematic uncertainties in $B-R$ with increasing radius, and the weighting minimizes the contribution to the results.  

The results of the host galaxy and GC color profiles are presented in $\S\ref{sec:chap4_results}$.  The colors of galaxies and GC systems reflect the ages and metallicities of their underlying stellar populations, and ideally one would like to make direct comparisons to these parameters rather than color.  Unfortunately, age and metallicity are degenerate when using optical colors, i.e., redder photometric colors can be caused by both older stellar populations and by more metal-rich populations \citep{wo94}.  However, because GCs are many Gyrs old, broadband colors are sensitive to metallicity, with blue colors being more metal-poor than red colors \citep{wo94,br03}.  \txred{For consistency with our previous GC survey papers, we adopt the same GC color to metallicity conversion used in RZ01, which is}

\txred{
\small
\begin{equation}
{\rm [Fe/H]} = -3.692+ 1.069{(B-R)} + 0.7901{(B-R)}^{2}.
\end{equation}
\normalsize
}

Unlike GCs, galaxies contain stellar populations with a much wider range of ages and metallicities which complicates the interpretation of a galaxy's optical colors. 
Numerous studies have used both spectroscopic and photometric techniques to distinguish between age or metallicity in the colors of the stellar populations of galaxies (e.g., \citealt{tr00,ta00,ku10,ro11a,ro11b}). The results have shown that the mean colors and color gradients in early-type galaxies primarily reflect variations in metallicity rather than age.  Because an accurate decomposition of age and metallicity would require either a larger photometric baseline (e.g. a UV-optical or optical-IR color; \citealt{ro11a,ro11b}) or spectroscopic data \citep{tr00}, we simply perform comparisons to the host galaxy by $B-R$ color.


\subsection{Azimuthal Distribution Analysis}
\label{sec:chap4_ellip_analysis}

In order to investigate the two-dimensional (projected) shapes of the GC systems, we applied an iterative method of moments algorithm \citep{ca80} to measure the ellipticity $\epsilon$ and position angle $\theta$ of the galaxy GC systems in our sample.  This method has been used by others to quantify the shape of galaxy GC systems (e.g., \citealt{mc94,fo01}) and we used this method to study the azimuthal distribution of the GC system of the field S0 galaxy NGC~7457 (H11).  

We adopt several additional procedures and constraints when implementing the method of moments.  First, we hold the centroid of the ellipse fixed at the galaxy center due to the lack of GC candidates (because of image saturation and high background) towards the center of the galaxy.  Second, because the method of moments requires that any measurement annuli lie entirely on the frame, we are limited in the range over which we can measure the $\epsilon$ and $\theta$ by the frame edges and any large masked regions.  Thirdly, we use an iterative method to determine $\epsilon$ and $\theta$ which accounts for the contaminating background.  For wide-field imaging studies which have covered the full radial extent of a galaxy's GC system, the surface density of GC candidates far from the galaxy center will approach a limiting (asymptotic) value.  This asymptotic value represents the surface density of contaminating objects, which we assume to be uniformly distributed across the frames.  The presence of this uniform background will, in general, yield a more circular (smaller $\epsilon$) measurement of an intrinsically elliptical distribution of points.  One can minimize the influence of this background by calculating $\epsilon$ and $\theta$ within a measurement annulus as determined from a {\it previous}  estimate of $\epsilon$ and $\theta$ (i.e. in an iterative fashion).  

The method of moments does not allow for a straightforward calculation of the statistical uncertainty in the ellipticity and position angle and so we used a Monte Carlo bootstrapping technique to estimate the uncertainties. We used 10,000 bootstrap samples to determine the distributions of $\epsilon$ and $\theta$ and adopt the standard deviations of these distributions as the uncertainty in $\epsilon$ and $\theta$. In general we find that the bootstrap distributions of ellipticities and position angles are quite symmetric and have mean values that are in excellent agreement with the measured values (i.e., no measurable bias is evident in the bootstrap).  We also simulate intrinsically circular (azimuthally uniform) distributions to characterize the statistical significance with which we detect a non-uniform spatial distribution.   For each galaxy, we generate 10,000 circular GC spatial distributions and examine the probability that one would obtain an ellipticity as larger or larger than the measured $\epsilon$  by chance.  We match the total number of GCs and their radial surface density distribution by using the (circular) radial profile bins that were adopted to generate the GC surface density profiles.  The results of these significance tests are described along with the individual galaxy results below. 

We explored various means of making systematic comparisons of the method of moments results to the galaxy isophote fitting.  Although the isophote fitting and method of moments yield measurements of $\epsilon$ and $\theta$, the ellipse-fitting measures the galaxy shape at discrete semi-major axes while the method of moments yields a single cumulative (or integrated) result for the GC system.  We attempted to measure the GC system $\epsilon$ and $\theta$ in discrete, circular annuli by applying the method of moments at various radii, adopting the NGC~4406 GC system as our test case.  Using a series of 10,000 simulated GC systems (matching the measured $\epsilon,\theta$ of NGC~4406), we found that although the algorithm is quite robust using circular annuli, the lower number of objects per bin degrades the statistical significance of the results.  Put another way, our method does allow for the GC system shape to be measured over smaller annular regions (similar to the more discrete ellipse fitting), but the lower number of objects increases the uncertainties.  Thus using an ensemble measure for the $\epsilon, \theta$ of the GC system gives both a global picture of the GC shape and leverages the larger number of data points to decrease statistical uncertainties.

For the galaxy light, obtaining a ``global'' measure of $\epsilon, \theta$ is more problematic due to the changing ellipticities and position angles (particularly in the central regions) as well as the radially-decreasing surface brightness.  The issue has been addressed in a number of ways in the literature, including using a surface brightness weighted mean \citep{lam92} or adopting the ellipticity at a particular isophote (see, for example, \citealt{fa91,ch10}).  Because we are interested in the physical properties of the outer regions of galaxies, we adopted a simplistic approach and adopt the mean $\epsilon,\theta$ of the outer regions (where the ellipticity ``asymptotes") as our ``global'' value of the galaxy ellipticity and position angle.  Furthermore, we are primarily interested in using GCs to probe the physical properties of the halos of galaxies, and therefore the central regions of galaxies are of less significance in this study.


\section{Results for Individual Galaxies }
\label{sec:chap4_results}

The results of the surface photometry and azimuthal distribution analysis for the four target galaxies are discussed below in $\S\ref{sec:chap4_n4406_results}-\S\ref{sec:chap4_n4594_results}$.

\subsection{NGC~4406}
\label{sec:chap4_n4406_results}

\subsubsection{Galaxy Surface Photometry}
\label{sec:chap4_n4406_sfcphot}

\begin{figure}
\epsscale{1.2}
\plotone{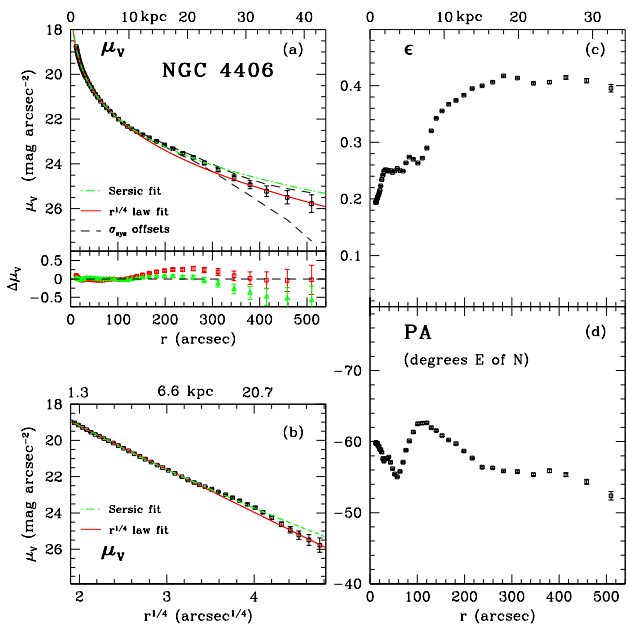}
\caption[Surface Photometry of NGC 4406]{Results of the surface photometry for NGC 4406.  Panels (a) and (b) show the surface brightness as a function of projected radius in (a) linear space and (b) $r^{1/4}$ space. The best-fit de Vaucouleurs law and S\'ersic profiles are shown as the red solid line and dash-dot green line, respectively.  The fit residuals ($\Delta \mu_{V}$) are shown below panel (a) as red squares for the de Vaucouleurs law fit and green triangles for the S\'ersic profile fit. The effects of the systematic uncertainty in the sky background ($\sigma_{\rm sys}$; see $\S \ref{sec:chap4_sfc_phot_errors}$) are shown as dashed lines.  The ellipticity ($\epsilon$) and position angle (PA) are shown in panels (c) and (d), respectively.  The surface photometry data are listed in Table~\ref{tab:chap4_n4406_sfc_phot}.
  \label{fig:chap4_n4406_sfc_phot}}
\end{figure}

We fitted isophotal ellipses to the galaxy light on the masked image from a semi-major axis of ${\sim} 13\arcsec$ 
to $650\arcsec$. 
The surface brightness, ellipticity, and position angle are shown as a function of geometric mean radius in Figure~\ref{fig:chap4_n4406_sfc_phot}.  The data are listed in Table~\ref{tab:chap4_n4406_sfc_phot}.  Comparing our surface brightness profile to the $V$-band surface photometry of NGC~4406 by \citet{ja10}, we see excellent agreement ($< 0.1$ mag difference) in the inner $r < 300\arcsec$.  At larger radii we measure a surface brightness which is approximately $0.2-0.3$ magnitudes fainter than \citet{ja10} but consistent with their results within our systematic uncertainties.   Although NGC~4374 is relatively nearby in projection, \citet{ja10} find that the contribution of diffuse light by this galaxy  to the NGC~4406 surface brightness profile is likely only important at $\mu_V \gtrsim 27~ {\rm mag~arcsec}^{-2}$, so we expect no significant contribution of light from NGC~4374.

Our $V$-band surface brightness profile is well-fit by an $r^{1/4}$ (de Vaucouleurs law) fit in the inner $r {\sim} 120\arcsec$ and outer $r > 350\arcsec$ (see Figure~\ref{fig:chap4_n4406_sfc_phot}).  The $r^{1/4}$ law fit yields an effective radius of $r_e = 151 \pm 1\arcsec$.  The profile shows a slight increase or bump in the surface brightness from ${\sim} 120\arcsec < R < 320\arcsec$, a feature that has been seen in other surface photometry of this galaxy \citep{ja10, ko09}.  This feature significantly degrades the quality of the $r^{1/4}$ fit, resulting in a reduced chi-square $\chi^2/\nu = 14.1$.  \txred{A S\'ersic fit yields a significantly larger effective radius of $r_e = 350 \pm 25\arcsec$ with $n=6.5 \pm 0.2$ and a reduced chi-square $\chi^2/\nu = 2.4$.  Although the S\'ersic model provides a statistically better fit, the outermost isophotes are better described by the $r^{1/4}$ profile (see Figure~\ref{fig:chap4_n4406_sfc_phot}).}

Our estimates of the effective radius fall between several literature estimates.  The RC3 \citep{RC3} gives an effective radius of $104\arcsec$, while \citet{mi94} and \citet{ca93} find $r_e = 158\arcsec$ ($r^{1/4}$ fit) and $r_e=167\arcsec$  (S\'ersic fit), respectively.  \txred{Our results are in good agreement with the S\'ersic fit from \citet{ja10} who find $r_e = 372 \pm 17\arcsec$ but a smaller $n=5.2\pm 0.2$.}  We find a changing ellipticity and position angle in the inner $r {\sim} 100\arcsec$, which is consistent with other results in the literature.  The ellipticity increases smoothly to an asymptotic value of $\epsilon {\sim} 0.4$ while the position angle varies between $-63 < \theta < -50$.  There is good agreement between our measured ellipticities and position angles and  the \citet{ja10} results.  

The $BVR$ color profiles (see Table~\ref{tab:chap4_n4406_sfc_phot}) are relatively flat in all three colors, which is consistent with previous surface photometry \citep{pe90,mi99}.  The profiles show systematic blueward or redward deviations at larger radii ($r \gtrsim 300\arcsec$).  These are likely artifacts of the choice of sky background in one or more filters, as small changes in the sky level at faint surface brightness levels {\it in either filter} can significantly change the measured color.  We used a weighted least-squares algorithm to fit a linear function to the $B-R$ profile (over the full radial range) and find a color gradient of $\Delta(B-R)/\Delta \log{(r)} = -0.03 \pm 0.01~{\rm mag~dex}^{-1}$.  This is consistent with the results of \citet{mi99} and \citet{pe90} who found  gradients of $-0.02$ and $-0.03\pm0.02~{\rm mag~dex}^{-1}$, respectively.  \citet{id02}, however, found a gradient of ${\sim} -0.10~{\rm mag~dex}^{-1}$ in $B-R$.  We compare the $B-R$ color profile of the galaxy to the GC system in $\S\ref{sec:chap4_n4406_gc_system}$ below.

\subsubsection{Globular Cluster System}
\label{sec:chap4_n4406_gc_system}

\begin{figure}
\epsscale{1.2}
\plotone{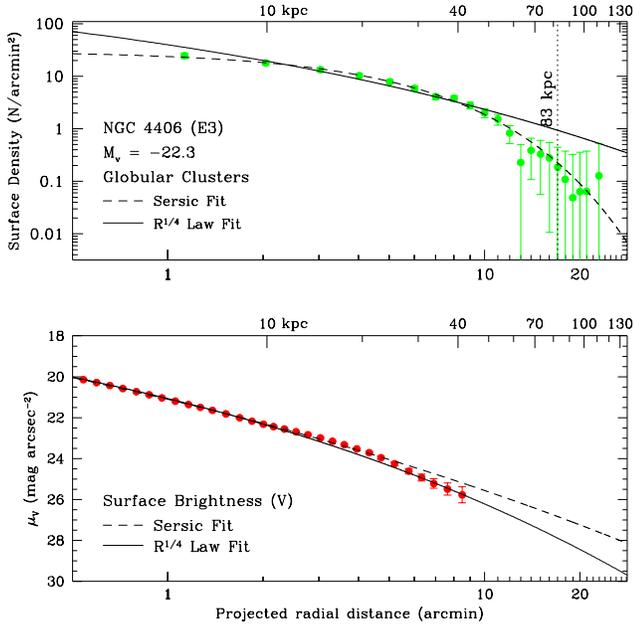}
\caption[Radial GC Surface Density and Galaxy Surface Brightness Profiles for NGC~4406]{Comparison of the GC radial surface density and galaxy $V$ band surface brightness profile for NGC~4406.  The GC profile (top panel; data from RZ04) shows the corrected surface density of GC candidates as a function of projected radial distance from the center of the host galaxy.  The surface density of GC candidates becomes consistent with the background within the errors at $17\arcmin$ (vertical dotted line). The $V$ band surface brightness profile from this work (bottom panel; see also Figure~\ref{fig:chap4_n4406_sfc_phot}) is shown as a function of geometric mean radius $r \equiv \sqrt{ab}$.  The de Vaucouleurs law and S\'ersic fits to both profiles are shown as the solid lines and dashed lines, respectively.  
  \label{fig:chap4_n4406_gc_sfc}}
\end{figure}

RZ04 measured a GC system radial extent of $\sim17\arcmin$ and a total number of GCs of $2900\pm400$ for NGC~4406.  The final, corrected radial surface density profile of the GC system is shown with the $V$-band surface brightness profile  (see $\S\ref{sec:chap4_n4406_sfcphot}$)  in Figure~\ref{fig:chap4_n4406_gc_sfc}.   The GC population has a measured radial extent of $\sim 83$ kpc, a radial distance which would require observations at $\mu_V\gtrsim 28 ~{\rm mag~arcsec}^{-2}$ to probe the halo starlight at the same radius.  The $r^{1/4}$ fit to the profile gives a theoretical effective radius for the GC system of $r_{e{\rm (GC)}} = 20\pm 3 \arcmin ~(96 \pm 15$ kpc) \txred{with a reduced chi-square of $\chi^2/\nu = 3.1$}.  As noted previously, the effective radius -- as determined from model fits -- is the radius that includes half the total number of GCs assuming the profile extends to infinity.  However, the surface density profile shows that the system has a measurable extent of only ${\sim} 17\arcmin$, which is less than the derived $r_{e{\rm (GC)}}$.  Although some GCs will be located at $r>17\arcmin$, the low surface density at large radii implies that a {\it significant} population is not present.  This highlights the sensitivity of the theoretical effective radius to the relatively shallow slope ($-1.58$) of the de Vaucouleurs law fit.  

\txred{A S\'ersic fit to the GC radial profile (Figure~\ref{fig:chap4_n4406_gc_sfc}) gives $r_{e{\rm (GC)}} = 5.8 \pm 0.1\arcmin$ ($28 \pm 1$ kpc) and $n=0.9 \pm 0.1$ with a reduced chi-square of $\chi^2/\nu = 0.5$.  The 3-parameter S\'ersic model fits both the flattening surface density of GCs at small projected radii ($r<3\arcmin$) and the decreasing surface density at large radii ($r>10\arcmin$).  The model value of  $r_{e{\rm (GC)}}$ also agrees much better (than the $r^{1/4}$ model $r_e$) with the empirical effective radius for the GC system of $r_{e{\rm (GC)}} = 6.4\arcmin$ ($32$ kpc).}

\begin{figure}
\epsscale{1.2}
\plotone{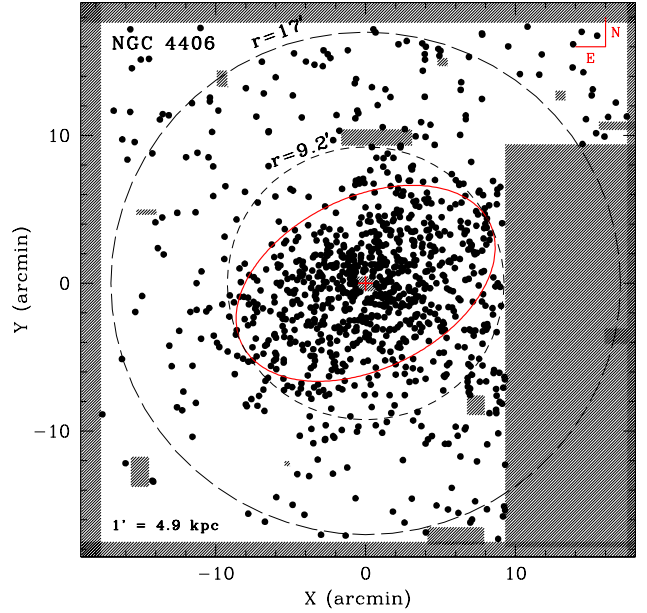}
\caption[Spatial positions of the GC candidates in NGC 4406]{Results of the azimuthal distribution analysis for NGC 4406. The positions of the $90\%$ sample of 1031 GC candidates are shown with respect to the galaxy center (red cross).  The outer dashed circle denotes the radial extent of the GC system ($r\sim 17\arcmin$) and the inner dashed circle denotes the radius over which we are able to study the azimuthal distribution of the GC system ($r= 9.2 \arcmin$).   The red ellipse shows the GC system ellipticity ($\epsilon = 0.38 \pm 0.05$) and position angle ($\theta = -63 \pm 6$)  as determined by the method of moments.  The masked regions of the image are shaded, including a large portion due to the proximity (in projection) of NGC~4374. 
  \label{fig:chap4_n4406_90_sample}}
\end{figure}

\begin{figure}
\epsscale{1.2}
\plotone{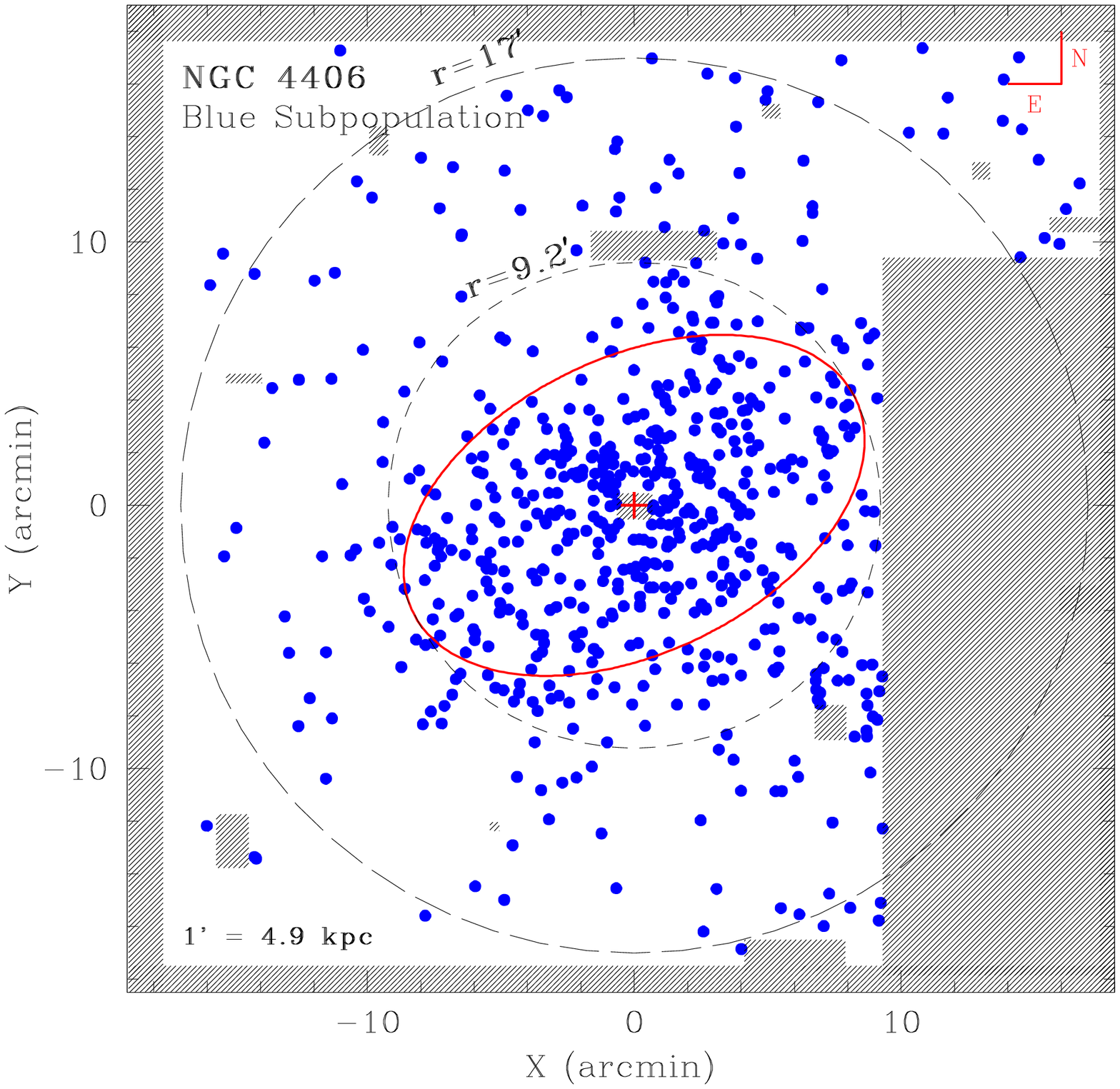}
\caption[Spatial positions of the metal-poor GC candidates in NGC 4406]{Results of the azimuthal distribution analysis of the blue (metal-poor) GC subpopulation for NGC 4406. The positions of the 643 blue GC candidates are shown with respect to the galaxy center (red cross).  The inner and outer dashed circles are the same as in Figure~\ref{fig:chap4_n4406_90_sample}.   The red ellipse shows the GC system ellipticity ($\epsilon = 0.39 \pm 0.06$) and position angle ($\theta = -64 \pm 9$)  as determined by the method of moments.
  \label{fig:chap4_n4406_blue_spatial}}
\end{figure}

\begin{figure}
\epsscale{1.2}
\plotone{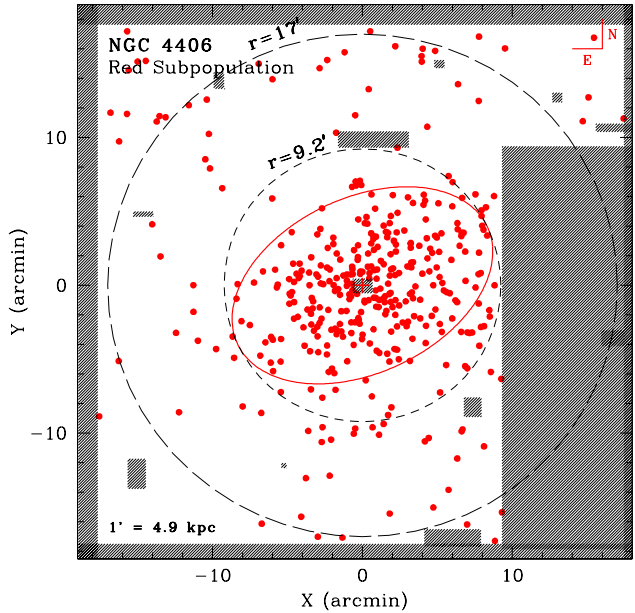}
\caption[Spatial positions of the metal-rich GC candidates in NGC 4406]{Results of the azimuthal distribution analysis of the red (metal-rich) GC subpopulation for NGC 4406. The positions of the 388 red GC candidates are shown with respect to the galaxy center (red cross).  The inner and outer dashed circles are the same as in Figure~\ref{fig:chap4_n4406_90_sample}.   The red ellipse shows the GC system ellipticity ($\epsilon = 0.36 \pm 0.07$) and position angle ($\theta = -64 \pm 7$)  as determined by the method of moments.
  \label{fig:chap4_n4406_red_spatial}}
\end{figure}

The spatial positions of the $90\%$ color-complete sample of 1031 GC candidates around NGC~4406 are shown in Figure~\ref{fig:chap4_n4406_90_sample}.  The elliptical galaxy NGC~4374 appears in close proximity (in projection) to NGC~4406.  Therefore in the RZ04 analysis, a $10\arcmin \times 20\arcmin$ area around NGC~4374 was masked to avoid contamination  from that galaxy's GC population (see Figure~\ref{fig:chap4_n4406_90_sample}).  The radial extent of the masked region limits our study the NGC~4406 GC system azimuthal distribution to $9.2\arcmin$ ($N = 772$ GCs), since the method of moments requires measurement annuli to be located entirely on the frame.  \txred{Note that the exact shape of the masked region does not affect the ellipticity results we derive; the method of moments is sensitive only to the radial extent of the mask.}

Applying the method of moments algorithm to the $r < 9.2\arcmin$ $90\%$ sample yields an ellipticity of $\epsilon = 0.38 \pm 0.05$ and a position angle of $\theta = -63 \pm 6$.  We show the ellipse solution with the GC positions in Figure~\ref{fig:chap4_n4406_90_sample}.  Based on our azimuthally isotropic spatial distribution simulations, we find that the probability of obtaining an ellipticity as big or bigger than the measured value by chance is only $p < 0.01\%$.  A previous analysis of the GC azimuthal distribution in NGC~4406 was done by \citet{fo96} from {\it HST/WFPC2} imaging but no significant ellipticity was found, likely due to the smaller radial extent of the observations.   

To investigate the azimuthal distributions by GC subpopulations, we split the $90\%$ sample at the empirical color cut of $B-R=1.23$.  This yields a fraction of blue GCs of $f_{\rm blue} = 0.62$, consistent with the mixture modeling results from RZ04.  Over the $9.2\arcmin$ radial extent for the blue subpopulation, we find $\epsilon = 0.39 \pm 0.06$ and $\theta = -64 \pm 9$ from $N=379$ GC candidates (see Figure~\ref{fig:chap4_n4406_blue_spatial}).  The probability of obtaining an ellipticity as big or bigger than the measured value by chance is  $p = 0.22\%$. For the red subpopulation, we find $\epsilon = 0.36 \pm 0.07$ and $\theta = -64 \pm 7$ from $N = 266$ GC candidates (see Figure~\ref{fig:chap4_n4406_red_spatial}). Our simulations show a probability of obtaining an ellipticity as big or bigger than the measured value by chance is $p = 0.07\%$.  Both subpopulations show nearly identical azimuthal distributions and good agreement with the host galaxy ($\bar{\epsilon} \approx 0.4, \bar{\theta} \approx -55$ for $r>300\arcsec$; see Figure~\ref{fig:chap4_n4406_sfc_phot} and Table~\ref{tab:chap4_n4406_sfc_phot}).  

\begin{figure}
\epsscale{1.2}
\plotone{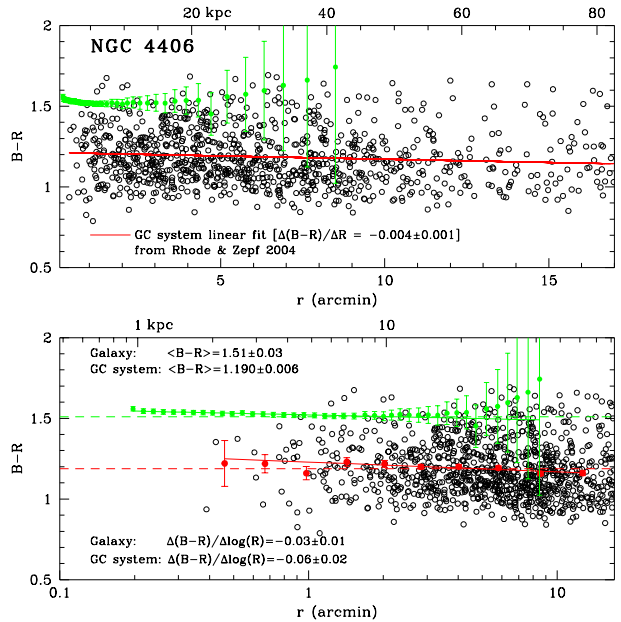}
\caption[GC and Galaxy B-R color profiles for NGC~4406]{Comparison of the GC and galaxy $B-R$ color profile for NGC~4406 out to the measured radial extent of the GC system. The open circles mark the $B-R$ and radial positions (from the galaxy center) for the $90\%$ sample of GC candidates in NGC~4406.  The filled green circles show $B-R$ color profile for the galaxy light.  {\bf Top panel}:  The red line shows the linear fit to the GC $B-R$ colors from RZ04. 
{\bf Bottom panel}: The filled red points show the mean $B-R$ colors of GC candidates within elliptical bins at the same geometric mean radius $R$ of the galaxy isophotes.  Linear fits to the galaxy and GC system profiles are shown as the solid green and red lines, respectively.  The measured slopes (color gradients) are listed.  The weighted mean $B-R$ colors for the galaxy light and GC population are listed and shown as the dashed green and red lines, respectively.  
  \label{fig:chap4_n4406_gc_gal_grad}}
\end{figure}

Figure ~\ref{fig:chap4_n4406_gc_gal_grad} compares the mean $B-R$ colors and color gradients of the galaxy light and GC population. RZ04 found evidence for a small $B-R$ color gradient of $\Delta (B-R)/ \Delta(r) = -0.004 \pm 0.001~{\rm mag~arcmin}^{-1}$ over the full $17\arcmin$ radial extent of the GC system.   Using elliptical bins, we find a color gradient of $\Delta(B-R)/ \Delta \log{(r)} = -0.06 \pm 0.02~{\rm mag~dex}^{-1}$ ($\Delta {\rm [Fe/H]}/\Delta \log{(r)} = -0.18 \pm 0.05$).  Although the GC population shows a slightly larger gradient than the galaxy stellar populations ($\Delta(B-R)/ \Delta \log{(r)} = -0.03 \pm 0.01~{\rm mag~dex}^{-1}$), the difference is not statistically significant. 

\txred{We explored the possibility of GC subpopulation gradients by (1) dividing the $90\%$ color-complete sample at $B-R=1.23$ (see earlier discussion; RZ04), (2) binning the data in circular annuli of various spacings, and (3) performing linear fits (see Equation~\ref{eq_color_grad}) to the binned color profile over various radial ranges.  We find no evidence (over any radial range) for a statistically significant color gradient in either the metal-rich or metal-poor subpopulations.}

The mean $B-R$ color of the host galaxy ($B-R = 1.51 \pm 0.03$) is, however, significantly redder than the GC system.   We find a mean color for the GC system of $B-R = 1.190 \pm 0.006$ (${\rm [Fe/H]} = -1.31\pm 0.02$) with subpopulation peaks at $1.11\pm 0.01$ (blue) and  $1.41\pm 0.01$ (red).  The difference in colors of ${\sim} 0.3$ is consistent with previous results which perform similar comparisons \citep{pe06}.  The red GC subpopulation is $0.1$ mag bluer than the host galaxy, a result which is statistically significant given the low uncertainties in the integrated colors.  The mean colors and color gradients are discussed in greater detail in $\S\ref{sec:chap4_comparisons}$.


\subsection{NGC~4472}
\label{sec:chap4_n4472_results}


\subsubsection{Galaxy Surface Photometry}
\label{sec:chap4_n4472_sfcphot}

\begin{figure}
\epsscale{1.2}
\plotone{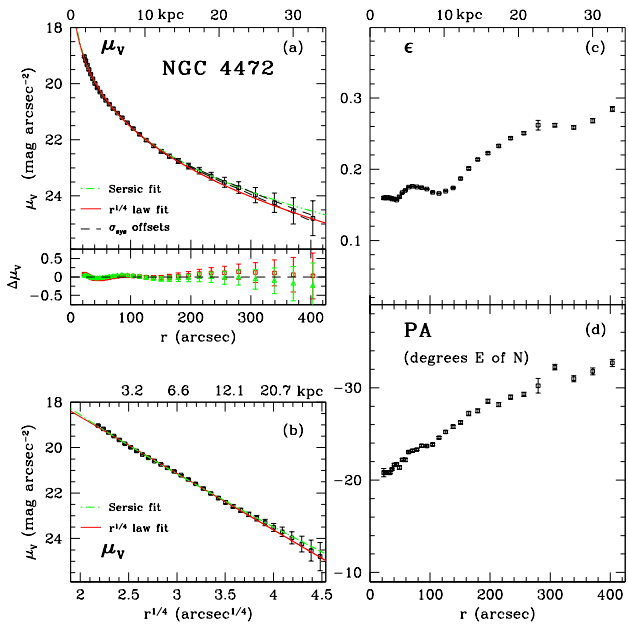}
\caption[Surface Photometry of NGC 4472]{Results of the surface photometry for NGC 4472.  Panels (a) and (b) show the surface brightness as a function of projected radius in (a) linear space and (b) $r^{1/4}$ space. The best-fit de Vaucouleurs law and S\'ersic profiles are shown as the red solid line and dash-dot green line, respectively.  The fit residuals ($\Delta \mu_{V}$) are shown below panel (a) as red squares for the de Vaucouleurs law fit and green triangles for the S\'ersic profile fit. The effects of the systematic uncertainty in the sky background ($\sigma_{\rm sys}$; see $\S \ref{sec:chap4_sfc_phot_errors}$) are shown as dashed lines.  The ellipticity ($\epsilon$) and position angle (PA) are shown in panels (c) and (d), respectively.  The surface photometry data are listed in Table~\ref{tab:chap4_n4472_sfc_phot}.
  \label{fig:chap4_n4472_sfc_phot}}
\end{figure}

We fitted ellipses to the galaxy light from a semi-major axis of ${\sim} 25\arcsec$ 
to $477\arcsec$
and the results are shown in Figure~\ref{fig:chap4_n4472_sfc_phot}. 
The data are listed in Table~\ref{tab:chap4_n4472_sfc_phot}.  The $V$-band surface brightness profile shows differences of less than $0.1$ mag (over the full radial range of our observations) compared to the \citet{ja10} results.  In the outer regions ($r > 300\arcsec$) the profile is slightly fainter by $0.2-0.4$ mag but consistent with their results within the statistical uncertainties.

A de Vaucouleurs model provides a good fit to the data over the full radial range of the observations.  The fit yields an effective radius of $r_e = 118 \pm 3\arcsec$ with a reduced chi-squared value of $\chi^2/\nu = 1.4$.  This is in excellent agreement with the results of \citet{ki00} who found $r_e = 120 \pm 2\arcsec$ from an $r^{1/4}$ law fit to their wide-field surface photometry data.  \txred{A S\'ersic fit to the profile gives a larger effective radius of $180 \pm 25\arcsec$ with $n=5.7 \pm 0.7$ and a reduced chi-squared value of $\chi^2/\nu = 0.70$.  \citet{ko09} find a similar effective radius and S\'ersic index from a fit to their NGC~4472 data ($r_e = 194 \pm 17$ and $n=6.0\pm 0.3$), but \citet{ja10} find larger values of $r_e=311 \pm 20$ and $n=6.9 \pm 0.5$.}

The ellipticity and position angle of the best-fit isophotes (Figure~\ref{fig:chap4_n4472_sfc_phot}; Table~\ref{tab:chap4_n4472_sfc_phot}) vary from the inner regions to the outskirts, with more circular isophotes in the center that become gradually more flattened at larger radii.  The position angle varies ${\sim} 10^{\circ}$ from the center to our outermost isophotes.  We find systematically larger ellipticities (by ${\sim} 0.03-0.05$) than both \citet{ja10} and \citet{ko09}  at $r > 250\arcsec$.  We also find slightly more variation in the position angle ($\sim 5^{\circ}$) over the radial range of our data compared to \citet{ja10} and \citet{ko09}.

As with NGC~4406, the $BVR$ color profiles of NGC~4472 (see Table~\ref{tab:chap4_n4472_sfc_phot}) are relatively flat in each filter in the inner ${\sim} 100\arcsec$ but become systematically redder at larger radii.  This is likely a consequence of the extreme sensitivity of the color at faint surface brightness levels to the sky background determination.  The uncertainties in the color grow rapidly in the same regions and any variations are clearly smaller than our errors.  The weighted linear fit to the $B-R$ versus $\log{(r)}$ data (over the full radial range) yields a best-fit slope of $\Delta(B-R)/\Delta\log{(r)} = -0.02 \pm 0.01~{\rm mag~dex}^{-1}$.  Steeper $B-R$ gradients are found by \citet{pe90}, \citet{mi99}, and \citet{id02}: values of $\Delta(B-R)/\Delta\log{(r)} =-0.05 \pm 0.02$, $-0.06$, and $-0.045 ~{\rm mag~dex}^{-1}$, respectively.  We find that the shallower gradient is caused by the systematic reddening of the color at larger radii ($\gtrsim 100\arcsec$).
The color gradient within $r< 100\arcsec$  is $\Delta(B-R)/\Delta\log{(r)} = -0.04 \pm 0.01~{\rm mag~dex}^{-1}$, which is consistent with the literature results.  In addition, the $B-V$ gradient in this region ($\Delta(B-V)/\Delta\log{(r)} = -0.04 \pm 0.01~{\rm mag~dex}^{-1}$) shows excellent agreement with the deep imaging of NGC~4472 by \citet{mi13}, who find $\Delta(B-V)/\Delta\log{(r)} =-0.03\pm 0.01$. 

\subsubsection{Globular Cluster System}
\label{sec:chap4_n4472_gc_system}

\begin{figure}
\epsscale{1.2}
\plotone{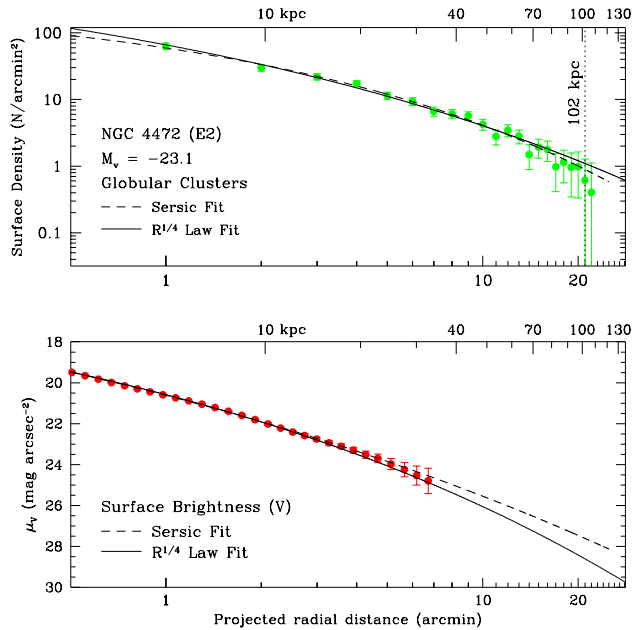}
\caption[Radial GC Surface Density and Galaxy Surface Brightness Profiles for NGC~4472]{Comparison of the GC radial surface density and galaxy $V$ band surface brightness profile for NGC~4472.  The GC profile (top panel; from RZ01) shows the corrected surface density of GC candidates as a function of projected radial distance from the center of the host galaxy.  The surface density of GC candidates becomes consistent with the background within the errors at $\sim 21\arcmin$ (vertical dotted line). The $V$ band surface brightness profile from this work (bottom panel; see also Figure~\ref{fig:chap4_n4472_sfc_phot}) is shown as a function of geometric mean radius $r \equiv \sqrt{ab}$.  The de Vaucouleurs law and S\'ersic fits to both profiles are shown as the solid lines and dashed lines, respectively.  
  \label{fig:chap4_n4472_gc_sfc}}
\end{figure}

RZ01 estimated a radial extent of ${\sim} 21\arcmin$ for the GC system of NGC~4472 and a total number of GCs of $5900\pm 700$.  Figure~\ref{fig:chap4_n4472_gc_sfc} shows the final, corrected radial surface density profile with the $V$-band surface brightness profile ($\S\ref{sec:chap4_n4472_sfcphot}$).  The GC system extends to ${\sim} 102$ kpc, approximately eight times the effective radius of the host galaxy light.  We derive a theoretical effective radius for the GC system of $r_{e{\rm (GC)}} = 21 \pm 3 \arcmin$ ($101 \pm 15$ kpc; \txred{$\chi^2/\nu = 0.64$}) from an $r^{1/4}$ law fit to the radial profile.  As with NGC~4406, the theoretical effective radius is inconsistent with the RZ01 results, as the system has a {\it total} radial extent of ${\sim} 21\arcmin$.  We find an empirical effective radius for the GC system of $r_{e{\rm (GC)}} =  8.1\arcmin$ (39 kpc) from the integration of the $r^{1/4}$ profile.  \txred{A S\'ersic fit to the GC radial profile (Figure~\ref{fig:chap4_n4472_gc_sfc}) does not provide better agreement with this empirical measure.  This fit yields an effective radius of $r_{e{\rm (GC)}} = 12 \pm 2 \arcmin$ and $n=2.5 \pm 0.5$ ($\chi^2/\nu = 0.48$).}

\begin{figure}
\epsscale{1.2}
\plotone{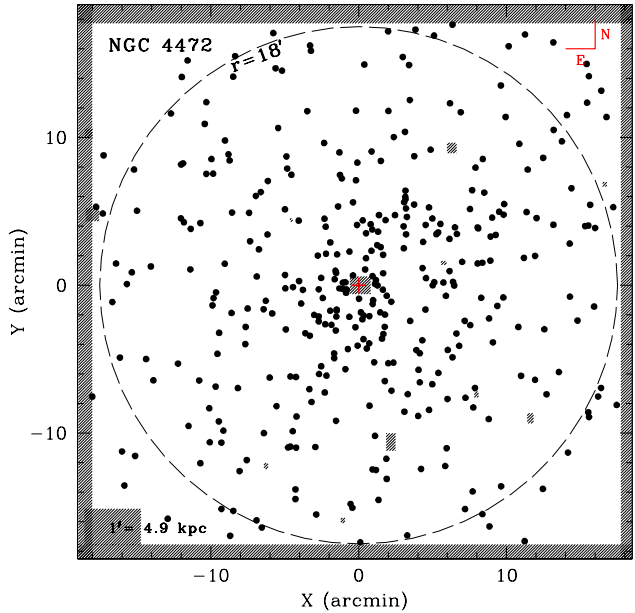}
\caption[Spatial positions of the GC candidates in NGC 4472]{The positions of the $90\%$ sample of 366 GC candidates in NGC~4472 are shown with respect to the galaxy center (red cross).  The outer dashed circle denotes the radial extent of the GC system over which we are able to study the azimuthal distribution ($r\sim 18\arcmin$).  For the $90\%$ sample, the low numbers of GCs hinder the ability to detect an ellipticity $\epsilon < 0.3$ (see the discussion in $\S\ref{sec:chap4_n4472_gc_system}$). 
The masked regions of the image are shaded. 
  \label{fig:chap4_n4472_90_sample}}
\end{figure}

\begin{figure}
\epsscale{1.2}
\plotone{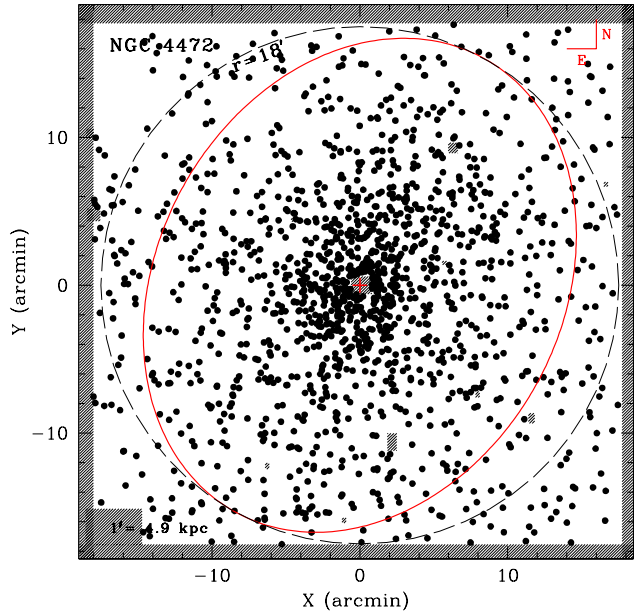}
\caption[Spatial positions of the GC candidates in NGC 4472]{The positions of the full sample of 1465 GC candidates in NGC~4472 are shown with respect to the galaxy center (red cross).  The outer dashed circle is the same as in Figure~\ref{fig:chap4_n4472_90_sample}. 
The red ellipse shows the GC system ellipticity ($\epsilon = 0.22 \pm 0.07$) and position angle ($\theta = -28 \pm 14^\circ$) as determined by the method of moments.
  \label{fig:chap4_n4472_full_sample_r18}}
\end{figure}

\begin{figure}
\epsscale{1.2}
\plotone{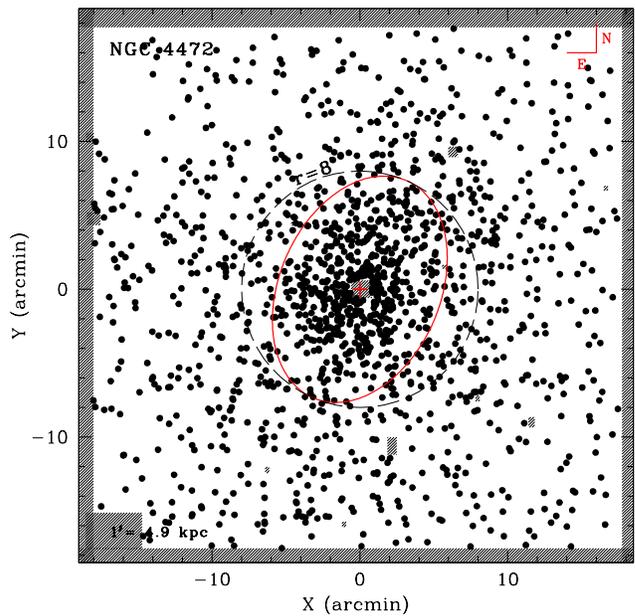}
\caption[Spatial positions of the GC candidates in NGC 4472]{The positions of the full sample of 1465 GC candidates in NGC~4472 are shown with respect to the galaxy center (red cross).  The dashed circle denotes the radial extent of the GC system over which we have reliable surface photometry and have measured the azimuthal distribution ($r=8\arcmin$).  The red ellipse shows the GC system ellipticity ($\epsilon = 0.32 \pm 0.07$) and position angle ($\theta = -23 \pm 40^\circ$) as determined by the method of moments.
  \label{fig:chap4_n4472_full_sample_r8}}
\end{figure}

The spatial positions of the 366 GC candidates in the $90\%$ color-complete sample for NGC~4472 are shown in Figure~\ref{fig:chap4_n4472_90_sample}.  The shallow imaging of NGC~4472 in the $B$ band (see RZ01 for details) resulted in a relatively small number of GC candidates in the $90\%$ sample.  The Mosaic image contained relatively little masked area, so the limiting radius of our azimuthal distribution analysis is the frame edge at $r=18\arcmin$ from the galaxy center.  We find $\epsilon = 0.16 \pm 0.10$ and $\theta = -61 \pm 34^{\circ}$ over this radial region.  The likelihood of obtaining an ellipticity as big or bigger than the measured value by chance is $p=45\%$.
It is important to note that this high probability does {\it not} imply that the system is intrinsically circular.  Rather, this result shows that the distribution is consistent with the expectations from an intrinsically circular distribution.  However, such results could also arise from non-circular distributions.

Could the low ellipticity result for the GC system occur simply due to small numbers from what is otherwise an intrinsically non-zero ellipticity GC system?  That is, we wish to know how frequently one would measure a value of $\epsilon = 0.16 \pm 0.10$ {\it even if the intrinsic spatial distribution is more elliptical}.  
To test this, we simulated 10,000 GC systems with $\epsilon = 0.25$ and $\theta = -30$ (the approximate ellipticity and position angle of the host galaxy outskirts) and the identical number and surface density of GC candidates.  We found that spatial distributions with $\epsilon = 0.16$ or smaller would be found in these simulated distributions with a frequency of $32\%$.  This relatively high frequency of obtaining too small of an ellipticity means that the probability distribution of ellipticities is too broad (due to the relatively low numbers of GCs) to confidently measure a low-$\epsilon$ distribution.  

Lastly, as a simple test, we ran the azimuthal distribution analysis on the full GC candidate sample (with two different radial cuts, $r<18\arcmin$ and $r<8\arcmin$) in order to compare the results to the host galaxy light.  The full GC candidate list contains four times more objects than the $90\%$ sample and should therefore provide a more robust statistical result.  We note that the radial color gradient in the NGC~4472 GC system is relatively modest ($\Delta (B-R) /\Delta(r) = -0.010 \pm 0.007$ over the inner $8\arcmin$; RZ01), so unless there are significant differences in the azimuthal distribution with GC subpopulation, the analysis of the full sample will provide a reasonably unbiased estimate.   The spatial distribution of the full sample is shown in Figures~\ref{fig:chap4_n4472_full_sample_r18} and~\ref{fig:chap4_n4472_full_sample_r8} with the larger and smaller radial cuts, respectively.  The larger radial cut provides a direct comparison to the $90\%$ sample analysis while the smaller radial cut restricts the analysis over the region where we have reliable surface photometry ($r {\sim} 480\arcsec$).  For the $r<18\arcmin$ radial cut, we find $\epsilon = 0.22 \pm 0.07$ and $\theta = -28 \pm 14^\circ$ with an as-large-or-larger probability of only $p=0.26\%$.  The $r<8\arcmin$ radial cut analysis yields a similar solution but with larger uncertainties in the position angle: $\epsilon = 0.32\pm 0.07$ and $\theta = -23 \pm 40^\circ$.  

The results for the GC system ($\epsilon {\sim} 0.2-0.3$ and $\theta {\sim} -25^\circ$) are in good agreement with the host galaxy light (see~\ref{sec:chap4_n4472_sfcphot}).  Previous studies of NGC~4472's GC system have found differing results for the azimuthal distribution shape.  \citet{le98} and \citet{le00} used  $16\arcmin \times 16\arcmin$ ground-based and WFPC2 {\it HST} imaging, respectively, to study the spatial structure of NGC~4472's GC system.  Both studies found that the overall azimuthal distribution of GCs is slightly elliptical (similar to the galaxy light) and driven largely by a non-uniformly distributed metal-rich GC subpopulation; they found the metal-poor GC population to be spherically (i.e., circularly) distributed.  In contrast, \citet{pa13} analyzed the ACSVCS data on NGC~4472 but found no significant ellipticity in either the metal-rich or metal-poor GC subpopulations.  The varying radial coverage of the observations, in combination with the changing numbers of metal-rich and metal-poor GC subpopulations in the central regions, likely contributes to the different conclusions.

\begin{figure}
\epsscale{1.2}
\plotone{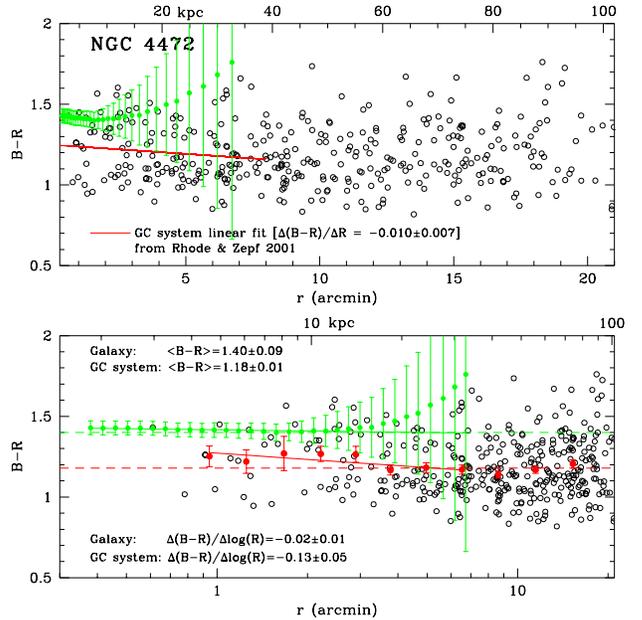}
\caption[GC and Galaxy B-R color profiles for NGC~4472]{Comparison of the GC and galaxy $B-R$ color profile for NGC~4472 out to the measured radial extent of the GC system.  The open circles mark the $B-R$ and radial positions (from the galaxy center) for the $90\%$ sample of GC candidates in NGC~4472.  The filled green circles show $B-R$ color profile for the galaxy light.  {\bf Top panel}:  The red line shows the linear fit to the GC $B-R$ colors from \citet{rh01}.  
{\bf Bottom panel}: The filled red points show the mean $B-R$ colors of GC candidates within elliptical bins at the same geometric mean radius $R$ of the galaxy isophotes.  Linear fits to the galaxy $B-R$ profile and GC system elliptical bins are shown as the solid green and red lines, respectively.  The measured slopes (color gradients) are listed.  The weighted mean $B-R$ colors for the galaxy light and GC population are listed and shown as the dashed green and red lines, respectively.  
  \label{fig:chap4_n4472_gc_gal_grad}}
\end{figure}

 We show the mean $B-R$ color and color gradients for the galaxy and GC populations for NGC~4472 in Figure~\ref{fig:chap4_n4472_gc_gal_grad}.  A color gradient of $\Delta (B-R) /\Delta(r) = -0.010 \pm 0.007$ over the inner $8\arcmin$ of the GC system was measured (RZ01).  Using elliptical bins which match the host galaxy shape, we find $\Delta (B-R) /\Delta\log{(r)} = -0.13\pm 0.05~{\rm mag~dex}^{-1}$  ($\Delta [{\rm Fe/H}]/\Delta\log{(r)} = -0.38 \pm 0.14$) in the inner $8\arcmin$.  \txred{We explored the possibility of individual GC subpopulation color gradients (see $\S\ref{sec:chap4_n4406_gc_system}$) but found no statistically significant gradients when dividing the GC candidates by subpopulation.}

For the galaxy light we found $B-R$ gradients of $\Delta (B-R) /\Delta\log{(r)} = -0.02\pm 0.01~{\rm mag~dex}^{-1}$ over the full radial range of the surface photometry and $\Delta (B-R) /\Delta\log{(r)} = -0.04\pm 0.01~{\rm mag~dex}^{-1}$ over the inner $r< 100\arcsec$. The difference in the galaxy light and global GC system gradients is statistically significant at ${\sim} 9\sigma$ level.

The mean colors are only different at the ${\sim}2.5\sigma$ level: on average the galaxy is more than $0.2$ mag redder than the GC system (integrated color of $<B-R>=1.40\pm 0.09$ for the galaxy; mean color of $<B-R> = 1.18\pm 0.01$ (${\rm [Fe/H]} = -1.33\pm0.03$) for the GC system).  The mean colors of the blue and red GC subpopulations are $B-R=1.09\pm 0.02$ and $1.39\pm 0.02$, respectively.  The red GC system peak shows good agreement with mean color of the galaxy light within the uncertainties.


\subsection{NGC~5813}
\label{sec:chap4_n5813_results}

\subsubsection{Galaxy Surface Photometry}
\label{sec:chap4_n5813_sfcphot}

\begin{figure}
\epsscale{1.2}
\plotone{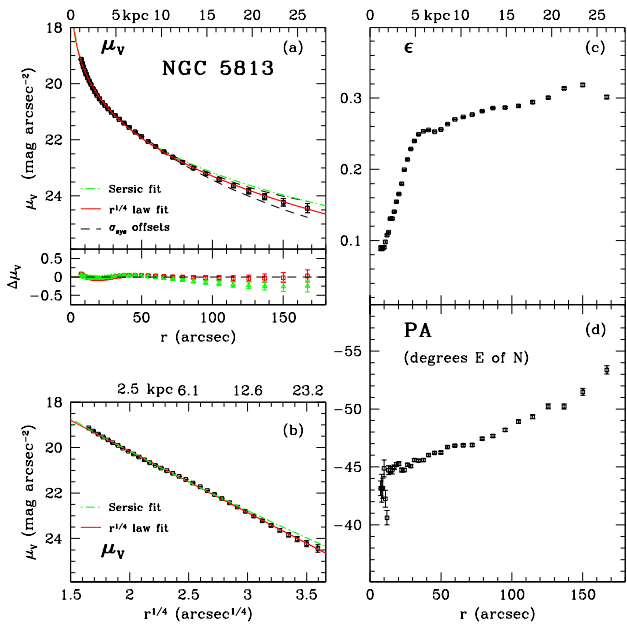}
\caption[Surface Photometry of NGC~5813]{Results of the surface photometry for NGC 5813.  Panels (a) and (b) show the surface brightness as a function of projected radius in (a) linear space and (b) $r^{1/4}$ space. The best-fit de Vaucouleurs law and S\'ersic profiles are shown as the red solid line and dash-dot green line, respectively.  The fit residuals ($\Delta \mu_{V}$) are shown below panel (a) as red squares for the de Vaucouleurs law fit and green triangles for the S\'ersic profile fit. The effects of the systematic uncertainty in the sky background ($\sigma_{\rm sys}$; see $\S \ref{sec:chap4_sfc_phot_errors}$) are shown as dashed lines.  The ellipticity ($\epsilon$) and position angle (PA) are shown in panels (c) and (d), respectively.  The surface photometry data are listed in Table~\ref{tab:chap4_n5813_sfc_phot}.
  \label{fig:chap4_n5813_sfc_phot}}
\end{figure}

For NGC~5813,  the proximity of a $V=8$ mag star (at $9.5\arcmin$ in projection from the galaxy) limits our surface photometry analysis to only $200\arcsec$ from the galaxy center.  Others have encountered similar problems, so no reliable surface photometry outside of ${\sim} 200\arcsec$ exists in the literature \citep{ma05,je09,fa11}.  
The $V$-band surface brightness, ellipticity, and position angle are shown as a function of geometric mean radius in Figure~\ref{fig:chap4_n5813_sfc_phot}.  The data are listed in Table~\ref{tab:chap4_n5813_sfc_phot}.  
Our $V$-band surface photometry shows good agreement with the $V$-band imaging from \citet{je09} and $g$ band Sloan imaging analyzed by \citet{ma05}.

The fit residuals (Figure~\ref{fig:chap4_n5813_sfc_phot}) indicate that the de Vaucouleurs profile is in excellent agreement over the full radial extent of the data.  The reduced chi-squared value, however, is large ($\chi^2/\nu = 9.0$), suggesting that we have underestimated our uncertainties in the surface brightness.  The effective radius derived from the fit is $r_e = 90 \pm 1\arcsec$.  \txred{A S\'ersic fit to the data give a much larger effective radius of $r_e = 175 \pm 15\arcsec$ with $n=5.7\pm 0.2$ and a smaller reduced chi-squared value of  $\chi^2/\nu = 5.8$.  Although the S\'ersic profile gives a formally better statistical fit, the $r^{1/4}$ profile is in better agreement with the outermost isophotes.  We note also that the larger $r_e$ from the S\'ersic fit likely results from the high value of $n$ and the degeneracy of these two parameters in the model.}

Our derived values of the effective radius are larger than most estimates in the literature, which range from $39\arcsec$ \citep{mi94} to ${\sim} 57\arcsec$ \citep{RC3,fa11}.  \citet{ef82} noted this discrepancy as well in their photographic study of NGC~5813, finding that a significantly larger effective radius of $72\arcsec$ (compared to the RC2 value of $44\arcsec$) was necessary to fit the inner $100\arcsec$ of the data with an $r^{1/4}$ profile.  We also find that a smaller effective radius of $40-60\arcsec$ provides a poor fit to the profile {\it at all radii}.  We investigated the possibility that an underestimate of our sky background could produce a surface brightness profile that is consistent with $r_e {\sim} 60\arcsec$.  Even with large adjustments to the sky background -- well outside our estimated systematic uncertainties -- we are unable to obtain a profile consistent with a significantly smaller effective radius.  

We find that the isophotal ellipses become more flattened at larger radii and that the position angle changes smoothly by ${\sim} 10^{\circ}$ over the radial range of our data.  Similar ellipticity and position angle profiles are seen in the surface photometry of \citet{pe90} and \citet{je09}.  
 
The $BVR$ color profiles (see Table~\ref{tab:chap4_n5813_sfc_phot}) are relatively flat and show less of a systematic deviation in the color at large radii compared to NGC~4406 and NGC~4472, although the radial extent of the data are significantly smaller for NGC~5813. Using a weighted least-squares linear fit, we find a color gradient of  $\Delta(B-R)/\Delta \log{(r)} = -0.04 \pm 0.01~{\rm mag~dex}^{-1}$.  This is in good agreement with the results from \citet{pe90}, who find a gradient of $-0.05 \pm 0.02~{\rm mag~dex}^{-1}$, but is shallower than the results from \citet{mi99} and \citet{id02} who find $-0.10$ and $-0.07~{\rm mag~dex}^{-1}$, respectively.

\subsubsection{Globular Cluster System}
\label{sec:chap4_n5813_gc_system}

\begin{figure}
\epsscale{1.2}
\plotone{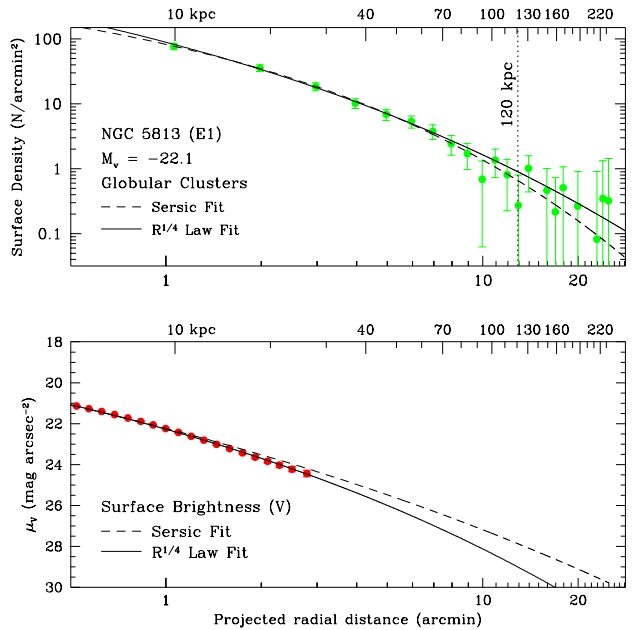}
\caption[Radial GC Surface Density and Galaxy Surface Brightness Profiles for NGC~5813]{Comparison of the GC radial surface density and galaxy $V$ band surface brightness profile for NGC~5813.  The GC profile (top panel; from HR12) shows the corrected surface density of GC candidates as a function of projected radial distance from the center of the host galaxy.  The surface density of GC candidates becomes consistent with the background within the errors at $\sim 13\arcmin$ as denoted by the vertical dotted line. The $V$ band surface brightness profile from this work (bottom panel; see also Figure~\ref{fig:chap4_n5813_sfc_phot}) is shown as a function of geometric mean radius $r \equiv \sqrt{ab}$.   The de Vaucouleurs law and S\'ersic fits to both profiles are shown as the solid lines and dashed lines, respectively.  
  \label{fig:chap4_n5813_gc_sfc}}
\end{figure}

The corrected radial surface density profile for the NGC~5813 GC system from HR12 is shown in Figure~\ref{fig:chap4_n5813_gc_sfc}.  We measured a radial extent of $\sim13\arcmin$ ($\sim 120$ kpc) for the GC system and a total number of GCs of $2900\pm400$.  The de Vaucouleurs law fit to the GC system profile yields a theoretical effective radius of $r_{e{\rm (GC)}} = 5.0 \pm 1.0 \arcmin$ ($47 \pm 9$ kpc) \txred{with a reduced chi-squared of $\chi^2/\nu = 0.33$}.  This galaxy is similar to NGC~4406 in terms of its properties and the number of GCs it hosts.  Even so, the $r^{1/4}$ slope of the NGC~5813 GC system is significantly steeper ($-2.23\pm0.09$) compared to that of NGC~4406's GC system ($-1.58\pm0.06$).   This results in a smaller theoretical effective radius compared to NGC~4406, which is still likely an overestimate given the assumptions of the de Vaucouleurs law fits.  We find an empirical effective radius for the GC system of $r_{e{\rm (GC)}} = 3.6\arcmin$ (34 kpc).  \txred{A S\'ersic fit to the GC radial profile give an effective radius in good agreement with this measurement.  We find $r_{e{\rm (GC)}} =3.9 \pm 0.3\arcmin$ and $n=2.5 \pm 0.8$ with a reduced chi-squared value of $\chi^2/\nu = 0.31$.}

\begin{figure}
\epsscale{1.2}
\plotone{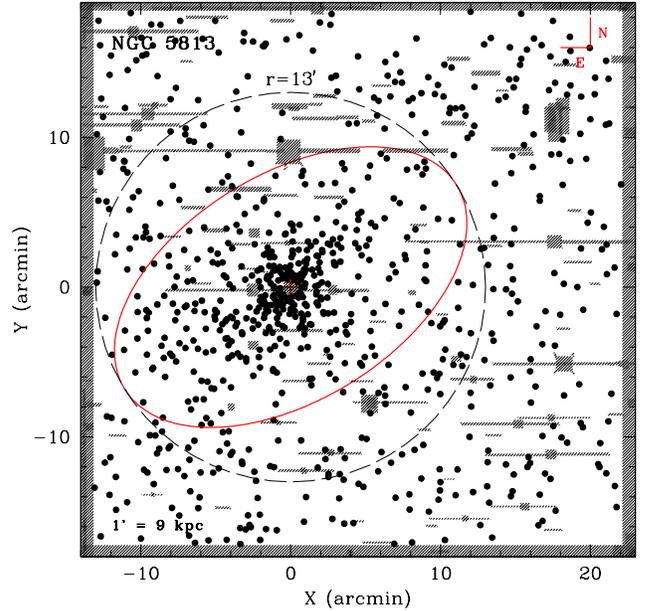}
\caption[Spatial positions of the GC candidates in NGC 5813]{Results of the azimuthal distribution analysis for NGC~5813. The positions of the $90\%$ sample of 809 GC candidates are shown with respect to the galaxy center (red cross).  The outer dashed circle denotes the radial extent of the GC system ($r\sim 13\arcmin$) and is the full extent over which we probe the GC system shape.   The red ellipse shows the GC system ellipticity ($\epsilon = 0.42 \pm 0.08$) and position angle ($\theta = -59 \pm 13$)  as determined by the method of moments.  The masked regions of the image are shaded.  
  \label{fig:chap4_n5813_90_sample}}
\end{figure}

The spatial positions of the 809 GC candidates in the $90\%$ sample of NGC~5813 are shown in Figure~\ref{fig:chap4_n5813_90_sample}.   Although there are numerous masked regions on the image, they constitute a relatively small total area and therefore we are able to study the azimuthal distribution over the full radial extent of the GC system.  We used a series of Monte Carlo simulations (with and without the masked regions in place) and found no significant bias in ellipticity or position angle caused by the presence of the masks.  

The method of moments yields an elliptical spatial distribution with $\epsilon = 0.42 \pm 0.08$ and $\theta = -59 \pm 13$ derived from $N=410$ objects within $r<13\arcmin$ (the approximate radial extent of the GC system).  The likelihood of obtaining an ellipticity as large or larger than the measured value is only $p = 0.67\%$. If we split the GC population into blue and red subpopulations, 
the method of moments does not give a statistically significant result for either subpopulation over the $r<13$ region.  The lower number of GC candidates in the subpopulations spread over a large area likely contributes to this result, since the total GC population clearly shows a non-circular projected shape.  

\begin{figure}
\epsscale{1.2}
\plotone{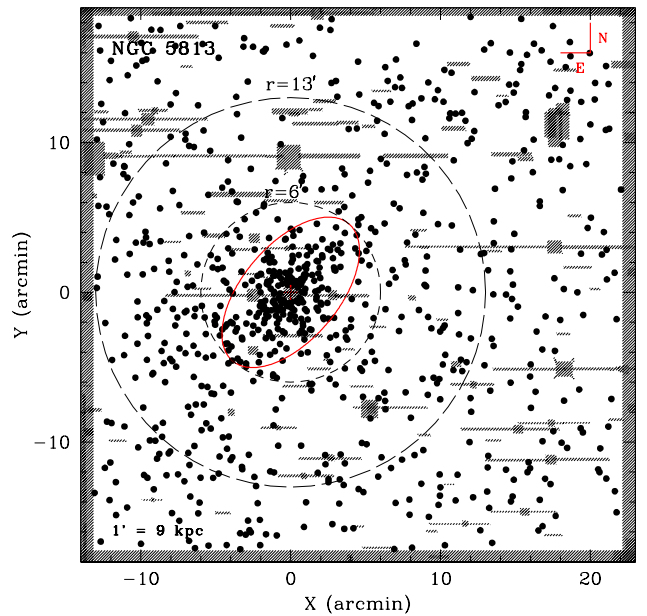}
\caption[Spatial positions of the GC candidates in NGC 5813 and the $r<6\arcmin$ azimuthal distribution solution]{Results of the azimuthal distribution analysis for NGC~5813 for $r<6\arcmin$. The positions of the $90\%$ sample of 809 GC candidates are shown with respect to the galaxy center (red cross).  The outer dashed circle is the same as Figure~\ref{fig:chap4_n5813_90_sample}. The inner dashed circle ($r = 6\arcmin = 4 r_e$) shows the radial extent of the color gradient in the GC system (HR12).   The red ellipse shows the GC system ellipticity ($\epsilon = 0.47 \pm 0.08$) and position angle ($\theta = -40 \pm 9$)  as determined by the method of moments over the $r<6\arcmin$ region.  The masked regions of the image are shaded.  
  \label{fig:chap4_n5813_r6}}
\end{figure}

\begin{figure}
\epsscale{1.2}
\plotone{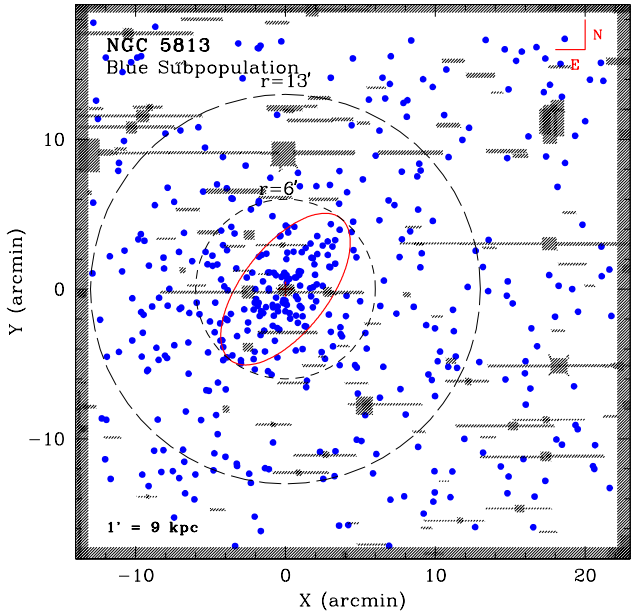}
\caption[Spatial positions of the blue GC candidates in NGC 5813 and the $r<6\arcmin$ azimuthal distribution solution]{Results of the azimuthal distribution analysis for $r<6\arcmin $ for the blue subpopulation of GC candidates in NGC~5813. The positions of the $90\%$ sample of 436 blue GC candidates are shown with respect to the galaxy center (red cross).  The inner and outer dashed circles are the same as Figure~\ref{fig:chap4_n5813_r6}.   The red ellipse shows the GC system ellipticity ($\epsilon = 0.52 \pm 0.15$) and position angle ($\theta = -38 \pm 19$)  as determined by the method of moments.  
  \label{fig:chap4_n5813_spatial_blue}}
\end{figure}

\begin{figure}
\epsscale{1.2}
\plotone{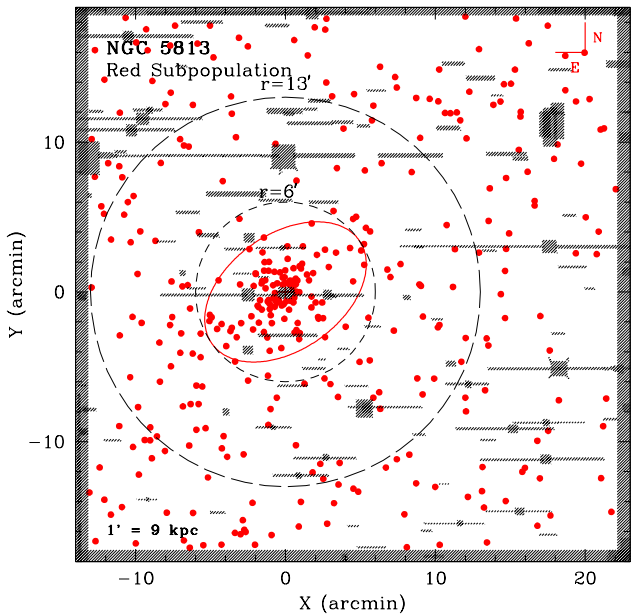}
\caption[Spatial positions of the reed GC candidates in NGC 5813 and the $r<6\arcmin$ azimuthal distribution solution]{Results of the azimuthal distribution analysis for $r<6\arcmin$ for the red GC subpopulation of NGC~5813. The positions of the $90\%$ sample of 373 red GC candidates are shown with respect to the galaxy center (black cross).  The inner and outer dashed circles are the same as Figure~\ref{fig:chap4_n5813_r6}.    The red ellipse shows the GC system ellipticity ($\epsilon = 0.36 \pm 0.11$) and position angle ($\theta = -55 \pm 11$)  as determined by the method of moments.  
  \label{fig:chap4_n5813_spatial_red}}
\end{figure}

We also wish to investigate the azimuthal distribution over a similar radial extent to that of NGC~4406 where we found a significant 2D shape in both GC subpopulations. 
For NGC~5813, this correponds to $\sim 6\arcmin$, which is also the region over which we found a GC system color gradient (HR12).  We find $\epsilon = 0.47 \pm 0.08$ and $\theta = -40 \pm 9$ for the $N=275$ GCs within this radius (see Figure~\ref{fig:chap4_n5813_r6}), which is consistent with the azimuthal distribution results over the full radial extent. The likelihood that an ellipticity as larger or larger than this result could arise by chance is only $p=0.05\%$.  Splitting the GC sample into red and blue subpopulations, we find that both spatial distributions show a non-circular shape consistent with the full $r<6\arcmin$ sample, although at a lower significance level.  For the blue subpopulation, we find $\epsilon = 0.52 \pm 0.15$ and $\theta = -38 \pm 19$ ($N=159$ objects within $r<6\arcmin$) with the probability that an ellipticity this large or larger could arise by chance from a circular distribution of $p=1\%$. For the red subpopulation, we find $\epsilon = 0.36 \pm 0.11$ and $\theta = -55 \pm 11$ ($N=116$ objects within $r<6\arcmin$) with the probability that an ellipticity this large or larger could arise by chance of $p=10\%$. Figures~\ref{fig:chap4_n5813_spatial_blue} and ~\ref{fig:chap4_n5813_spatial_red} show the blue and red spatial distributions with the azimuthal distribution solutions, respectively.  We do not consider the shape differences between the subpopulations to be significant given the resulting uncertainties.

We also explored the 2D shape of the GC system over the radius for which we measured the galaxy light.  Our surface photometry for NGC~5813 extends to only ${\sim} 2\arcmin$ (see $\S\ref{sec:chap4_n5813_sfcphot}$), so we applied the method of moments algorithm to the $N=99$ objects in the $90\%$ sample within $2\arcmin$.  We found a non-uniform shape of $\epsilon = 0.57 \pm 0.22$ and $\theta = -29 \pm 29^{\circ}$, with a probability of obtaining an ellipticity this larger or larger by chance of $p=4\%$.  
Although this result suggests that the inner regions of the GC system have a similar ellipticity and position angle to the $r<6\arcmin$ sample, the smaller number of objects in this region yields large uncertainties.  

Because the azimuthal distribution analysis suggests a slightly higher ellipticity for NGC~5813's GC compared to the host galaxy light, we explored the likelihood that a spatial distribution of GCs that matches the host galaxy light would give the measured values found for our data.  We generated 10,000 simulated GC system spatial distributions with $\epsilon = 0.3$ and $\theta = -50$ as per the results of the NGC~5813 surface photometry.  For the region $r<6\arcmin$, we find that the method of moments will return a solution with $\epsilon \ge 0.47 $ approximately $10\%$ of the time.  So, although the GC system ellipticity results for the $r<6\arcmin$ sample are larger than the galaxy (albeit measured at different radii), we cannot strongly exclude the possibility that the GC system and host galaxy share a comparable shape.

In summary, our analysis shows that the GC system of NGC~5813 has a projected 2D shape with an ellipticity and position angle consistent with the host galaxy.  The GC population shows a high probability of being non-circular over both the full radial extent ($r< 13\arcmin$) and the spatial region of the color gradient ($r< 6\arcmin$).  The red and blue GC subpopulations also have non-circular projected shapes, but at a smaller level of statistical significance.

\begin{figure}
\epsscale{1.2}
\plotone{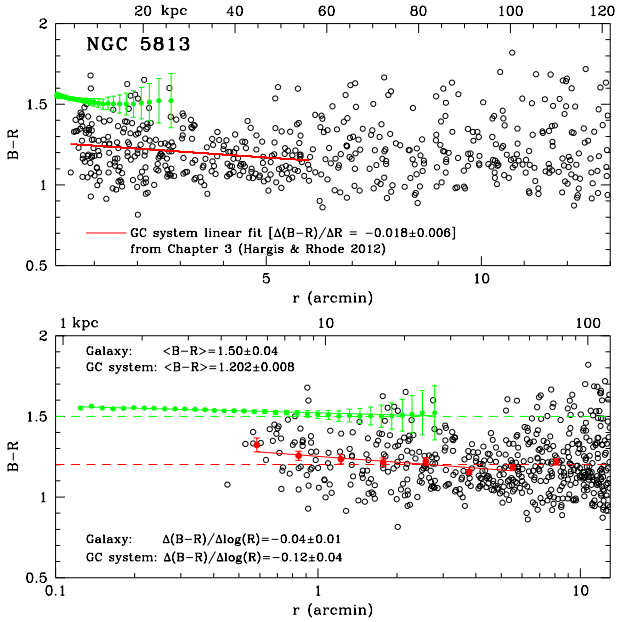}
\caption[GC and Galaxy B-R color profiles for NGC~5813]{Comparison of the GC and galaxy $B-R$ color profile for NGC~5813 out to the measured radial extent of the GC system. The open circles mark the $B-R$ and radial positions (from the galaxy center) for the $90\%$ sample of GC candidates in NGC~5813 used in this study.  The filled green circles show $B-R$ color profile for the galaxy light.  {\bf Top panel}:  The red line shows the linear fit to the GC $B-R$ colors from HR12.   {\bf Bottom panel}: The filled red points show the mean $B-R$ colors of GC candidates within elliptical bins at the same geometric mean radius $R$ of the galaxy isophotes.  Linear fits to the galaxy $B-R$ profile and GC system elliptical bins are shown as the solid green and red lines, respectively.  The measured slopes (color gradients) are listed.  The weighted mean $B-R$ colors for the galaxy light and GC population are listed and shown as the dashed green and red lines, respectively.  
  \label{fig:chap4_n5813_gc_gal_grad}}
\end{figure}

The $B-R$ versus radius profiles for the GC system and galaxy light for NGC~5813 are shown in Figure~\ref{fig:chap4_n5813_gc_gal_grad}.  We previously found a color gradient in the inner $6\arcmin$ of the GC system of $\Delta (B-R)/ \Delta(r) = -0.018 \pm 0.006~{\rm mag ~arcmin}^{-1}$ (HR12).  The weighted linear least-squares fit to the elliptically binned $B-R$ GC profile gives a gradient of $\Delta(B-R)/ \Delta \log{(r)} = -0.12 \pm 0.03~{\rm mag~dex}^{-1}$ ($\Delta {\rm [Fe/H]}/ \Delta \log{(r)} = - 0.36 \pm 0.10$) over the inner $r<6\arcmin$.   Comparing the global GC system gradient to the galaxy light ($\Delta(B-R)/ \Delta \log{(r)} = -0.04\pm 0.01~{\rm mag~dex}^{-1}$; see $\S\ref{sec:chap4_n5813_sfcphot}$), we find a larger GC system radial gradient but only at the ${\sim}2\sigma$ significance level.  \txred{Our analysis of the GC subpopulation color gradients show no statistically significant gradients (over any radial ranges) in the individual blue or red GC populations.}

As with the other three galaxies in our study, we find that the mean $B-R$ color of the galaxy is ${\sim}0.3$ redder than the GC system ($B-R=1.50 \pm 0.04$ for the galaxy; $B-R=1.202\pm 0.008$; ${\rm [Fe/H]} = -1.27 \pm 0.02$ for the GC system).  For the GC system subpopulations, we find color distribution peaks at $B-R = 1.16\pm0.01$ (blue) and $B-R=1.47 \pm 0.03$ (red).  The mean color of the red GCs and the host galaxy are offset by only ${\sim}0.03$, a negligible difference within the errors.


\subsection{NGC~4594}
\label{sec:chap4_n4594_results}

The Sombrero galaxy (NGC~4594, M104) has been classified as an Sa galaxy due to its prominent disk and dust lane \citep{RC3}.  However, the luminous bulge (bulge-to-total ratio of $0.86$; Kent 1988) is more characteristic of lenticular galaxies (that is, intermediate to ellipticals and spirals).  Also, the galaxy has a $B-V$ color of $0.84$ (RC3; \citealt{RC3}), consistent with observations of other S0 galaxies \citep{ro94}.   In addition, the galaxy hosts a large number of GCs ($1900\pm 200$; RZ04) and has a $V$-band normalized specific frequency of $S_N = 2.1 \pm 0.3$, a value which is more characteristic of giant elliptical galaxies than spirals (weighted means of  $S_N = 1.8 \pm 0.1$ for ellipticals, $S_N = 0.6 \pm 0.1$ for spirals; see HR12).  RZ04 therefore classify NGC~4594 as an S0 galaxy for the purposes of our GC system survey.  

\subsubsection{Galaxy Surface Photometry}
\label{sec:chap4_n4594_sfcphot}


\begin{figure}
\epsscale{1.2}
\plotone{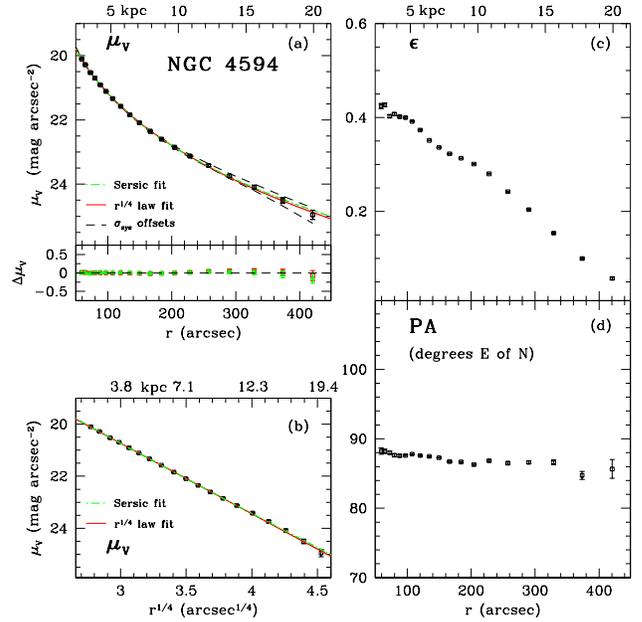}
\caption[Surface Photometry of NGC~4594]{Results of the surface photometry for NGC 4594.  Panels (a) and (b) show the surface brightness as a function of projected radius in (a) linear space and (b) $r^{1/4}$ space. The best-fit de Vaucouleurs law and S\'ersic profiles are shown as the red solid line and dash-dot green line, respectively.  The fit residuals ($\Delta \mu_{V}$) are shown below panel (a) as red squares for the de Vaucouleurs law fit and green triangles for the S\'ersic profile fit. The effects of the systematic uncertainty in the sky background ($\sigma_{\rm sys}$; see $\S \ref{sec:chap4_sfc_phot_errors}$) are shown as dashed lines.  The ellipticity ($\epsilon$) and position angle (PA) are shown in panels (c) and (d), respectively.  The surface photometry data are listed in Table~\ref{tab:chap4_n4594_sfc_phot}.
  \label{fig:chap4_n4594_sfc_phot}}
\end{figure}

We fitted isophotal ellipses to the masked image from a semi-major axis of $78\arcsec$ ($r=59\arcsec$) to $433\arcsec (R=420\arcsec)$ and the results are shown in Figure~\ref{fig:chap4_n4594_sfc_phot}.  
The data are listed in Table~\ref{tab:chap4_n4594_sfc_phot}.  
The bleed trail from the saturated galaxy center limits our ability to probe the regions inward of ${\sim} 70\arcsec$.  The dust lane shows a projected radius of at least $r{\sim} 184\arcsec$, so inside this radius our isophotes overlap the galaxy disk.  Comparing our results to the $V$-band surface photometry of NGC~4594 from the SINGS survey \citep{ke03,mu09}, we find excellent agreement between the data sets, with overall differences of $<0.1~{\rm mag/arcsec}^{2}$.   The galaxy light is well fit by an $r^{1/4}$ law over the full radial extent of our measurements.  In addition, the low reduced chi-square value ($\chi^2/\nu = 0.41$) indicates that we may have overestimated our uncertainties.  The profile fits give an effective radius of $r_e = 89 \pm 2\arcsec$, larger than the RC3 value of $72\arcsec$ \citep{RC3}.   In the region containing the galaxy disk ($r< 184\arcsec$), we find no significant deviations from the de Vaucouleurs profile.  This suggests that the bulge light dominates the profile in the inner regions, as expected given the large bulge-to-disk ratio.  \txred{A S\'ersic fit to the surface brightness profile is in excellent agreement with $r^{1/4}$ law fit: we find $r_e = 85\pm5\arcsec$ and $n=4.7\pm0.5$ with a reduced chi-squared value of $\chi^2/\nu = 0.24$.}

Our surface photometry shows a smooth decrease in the ellipticity with increasing radius and a position angle that is nearly constant (aligned with the galaxy disk) at all radii.   The increasing circularity of the isophotes is also seen in the optical and near-infrared ($3.6\mu$) surface photometry of \citet{bu86} and \citet{ga12}, respectively.  The position angle measurements show some evidence of a slight variation with increasing radius (a change of a few degrees over ${\sim} 350\arcmin$), consistent with the \citet{ga12} observations.

As with the three giant ellipticals in our study, the $BVR$ color profiles for NGC~4594 (see Table~\ref{tab:chap4_n4594_sfc_phot}) are relatively flat in all colors but show systematic deviations in the outer regions ($r \gtrsim 250-300\arcsec$) presumably due to the uncertainties in the sky background subtraction.  
A weighted linear least-squares fit yields a $B-R$ color gradient of $\Delta(B-R)/\Delta \log{(r)} = -0.04 \pm 0.02~{\rm mag~dex}^{-1}$.  We constructed the $B-R$ profile from the SINGS surface photometry \citep{mu09} and found a similar result ($\Delta(B-R)/\Delta \log{(r)} = -0.06 \pm 0.01~{\rm mag~dex}^{-1}$).
We also compared our color gradient to the Local Volume Legacy Survey (LVL; Lee et al. 2008) imaging of the galaxy (L. van Zee, private communication).  The LVL data show a significantly steeper gradient of $\Delta(B-R)/\Delta \log{(r)} = -0.30 \pm 0.01~{\rm mag~dex}^{-1}$ over the same radial region as our data due to redder colors of their isophotes at smaller radii.  Similarly, the {\it HST/ACS} imaging of the Sombrero galaxy analyzed by \citet{sp06} ($BVR$ equivalent filters) shows a strong color gradient of $\Delta(B-R)/\Delta \log{(r)} {\sim} -0.35~{\rm mag~dex}^{-1}$  in the inner $1\arcmin-3\arcmin$ (see their Figure 17).

\subsubsection{Globular Cluster System}
\label{sec:chap4_n4594_gc_system}

\begin{figure}
\epsscale{1.2}
\plotone{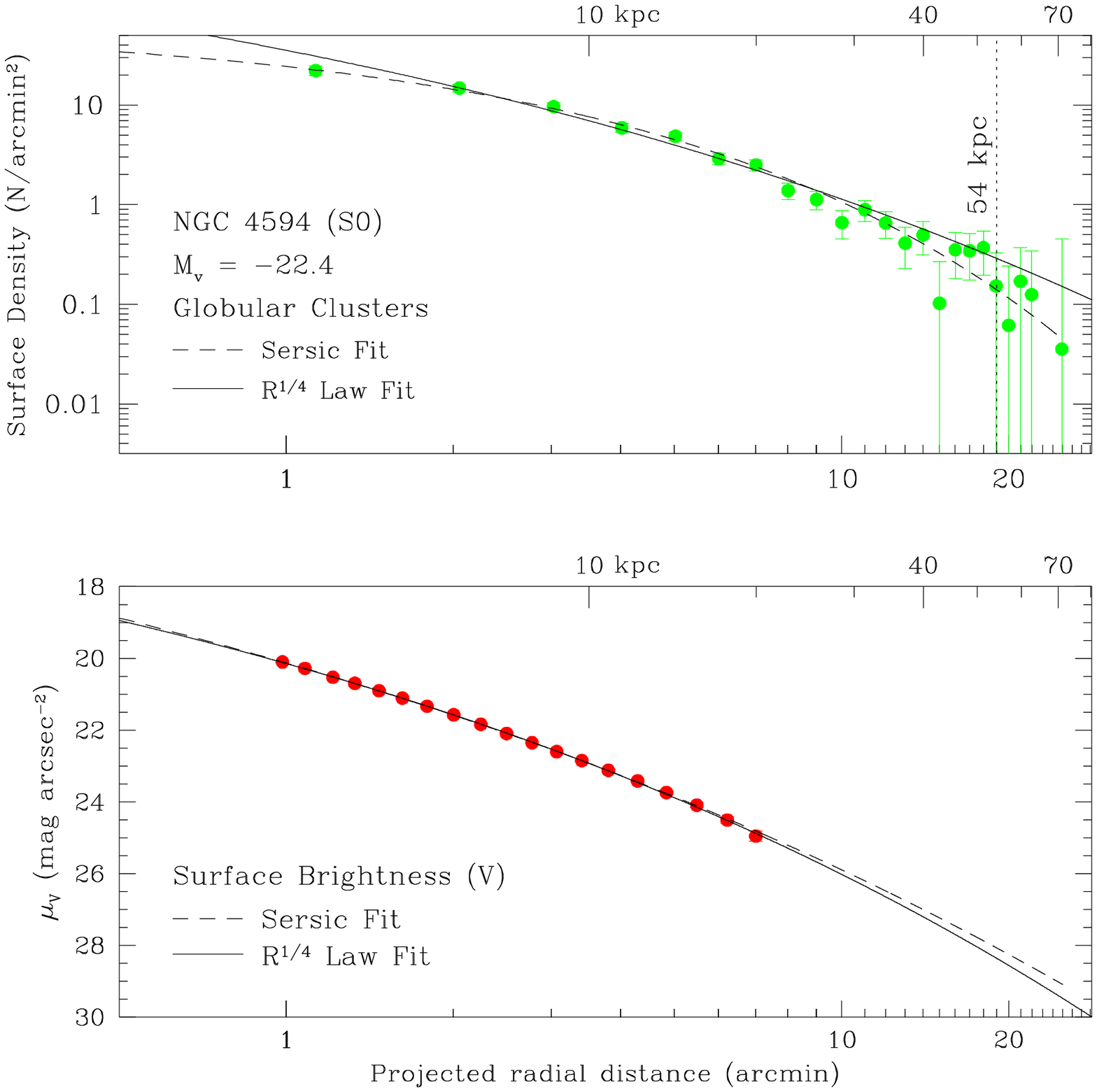}
\caption[Radial GC Surface Density and Galaxy Surface Brightness Profiles for NGC~4594]{Comparison of the GC radial surface density and galaxy $V$ band surface brightness profile for NGC~4594.  The GC profile (top panel; from RZ04) shows the corrected surface density of GC candidates as a function of projected radial distance from the center of the host galaxy.  The surface density of GC candidates becomes consistent with the background within the errors at $\sim 19\arcmin =\sim 12.8 r_e$ as denoted by the vertical dotted line. The $V$ band surface brightness profile from this work (bottom panel; see also Figure~\ref{fig:chap4_n4594_sfc_phot}) is shown as a function of geometric mean radius $r \equiv \sqrt{ab}$. The de Vaucouleurs law and S\'ersic fits to both profiles are shown as the solid lines and dashed lines, respectively.  
  \label{fig:chap4_n4594_gc_sfc}}
\end{figure}

RZ04 measured a GC system radial extent of ${\sim} 19\arcmin$ and total number of GCs of $1900\pm 200$ for NGC~4594.  The final, corrected radial surface density profile for the GC system (derived using our modifed GC candidate list) and the $V$-band surface brightness profile (see $\S\ref{sec:chap4_n4594_sfcphot}$)  are shown in Figure~\ref{fig:chap4_n4594_gc_sfc}.  The GC population extends to almost $13~r_e$ (${\sim} 54$) kpc.  The de Vaucouleurs law fit to the GC radial profile gives a theoretical effective radius of $r_{e{\rm (GC)}} = 9.1 \pm 1.0 \arcmin$ ($25.8 \pm 2.8$ kpc) \txred{with a reduced chi-square value of $\chi^2/\nu = 0.74$}.  We find an empirical effective radius of $r_{e{\rm (GC)}} = 4.3 \arcmin$ ($12$ kpc).  \txred{A S\'ersic fit to the GC radial profile gives an effective radius of $r_{e{\rm (GC)}} =5.9 \pm 0.4\arcmin$ with $n = 1.9 \pm 0.3$ and a reduced chi-square value of $\chi^2/\nu = 1.0$.  Although the effective radius derived from the S\'ersic profile is in better agreement with our empirical measurement, the $r^{1/4}$ law is statistically a slightly better model.}

\begin{figure}
\epsscale{1.2}
\plotone{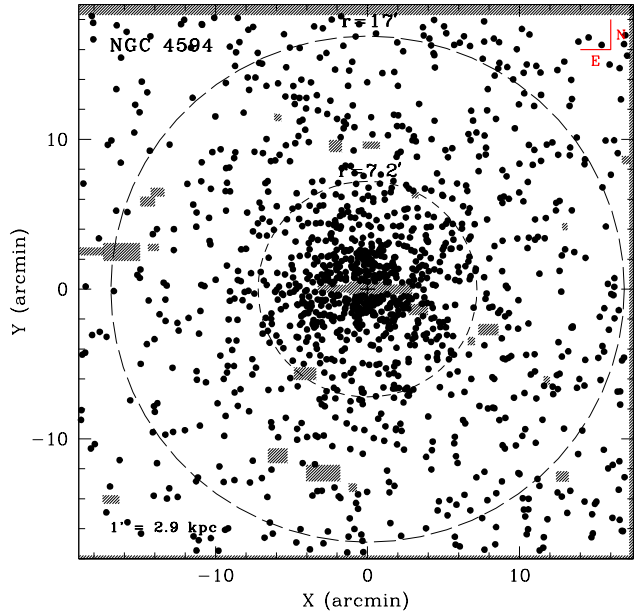}
\caption[Spatial positions of the GC candidates in NGC~4594]{The positions of the $90\%$ sample of 1285 GC candidates in NGC~4594 are shown with respect to the galaxy center (red cross).  The outer dashed circle denotes the radial extent of the GC system over which we are able to study the azimuthal distribution ($r\sim 17\arcmin$). The inner dashed circle shows the $r=7.2\arcmin$ radius over which we explored the azimuthal distribution of the red and blue subpopulations (see Figures~\ref{fig:chap4_n4594_blue_spatial} and~\ref{fig:chap4_n4594_red_spatial}, respectively).  We find no statistically significant shape to the 2D distribution compared to a uniform circular distribution over either radius.  The masked regions of the image are shaded. In the lower left we list the host galaxy effective radius $r_e$ and physical distance (in kpc) in terms of arcminutes on the image.
  \label{fig:chap4_n4594_90_sample}}
\end{figure}

The spatial positions of the 1286 GC candidates in the $90\%$ color-complete sample for NGC~4594 are shown in Figure~\ref{fig:chap4_n4594_90_sample}. The outermost circular annulus which does not lie off the frame is at $17\arcmin$.  We find a 2D distribution that is consistent with our azimuthally uniform simulations over the $r<17\arcmin$ radial extent: $\epsilon = 0.14 \pm 0.06$ and position angle $\theta = 20 \pm 51^{\circ}$ with an as-large-or-larger probability of $p=20\%$.  We applied the method of moments over a series of radial cuts ranging from ${{\sim}}3\arcmin-7\arcmin$.  The results were similar in all cases: the resulting likelihoods (of obtaining distributions with ellipticities as larger or larger than measured by-chance from azimuthally isotropic distributions) are greater than  $25\%$.
In their photographic study of NGC~4594, \citet{ha84} also did not detect a significant ellipticity in the GC system of NGC~4594.

\begin{figure}
\epsscale{1.2}
\plotone{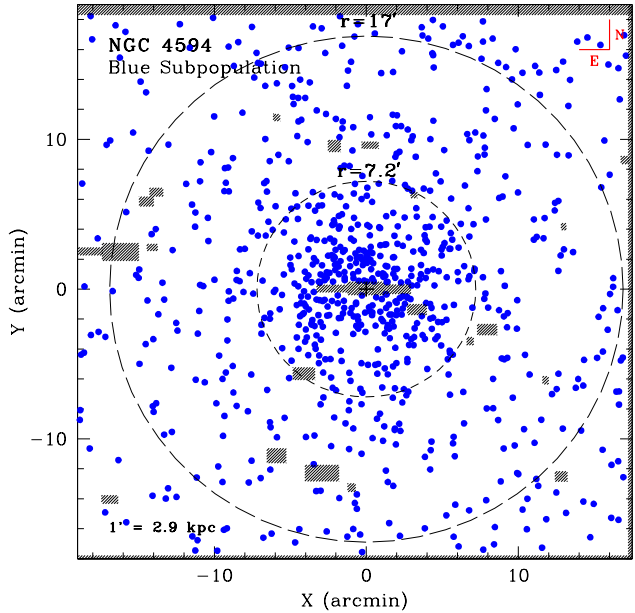}
\caption[Spatial positions of the metal-poor GC candidates in NGC~4594]{The positions of the 795 blue GC candidates in NGC~4594 are shown with respect to the galaxy center (red cross).  The outer dashed circle denotes the radial extent of the GC system over which we are able to study the azimuthal distribution ($r\sim 17\arcmin$). The inner dashed circle shows the $r=7.2\arcmin$ radius over which we explored the azimuthal distribution of the GC subpopulations. We find no statistically significant ellipticity (compared to an azimuthally uniform distribution) in the 2D spatial distribution of the blue subpopulation. The masked regions of the image are shaded. In the lower left we list the host galaxy effective radius $r_e$ and physical distance (in kpc) in terms of arcminutes on the image.
  \label{fig:chap4_n4594_blue_spatial}}
\end{figure}

\begin{figure}
\epsscale{1.2}
\plotone{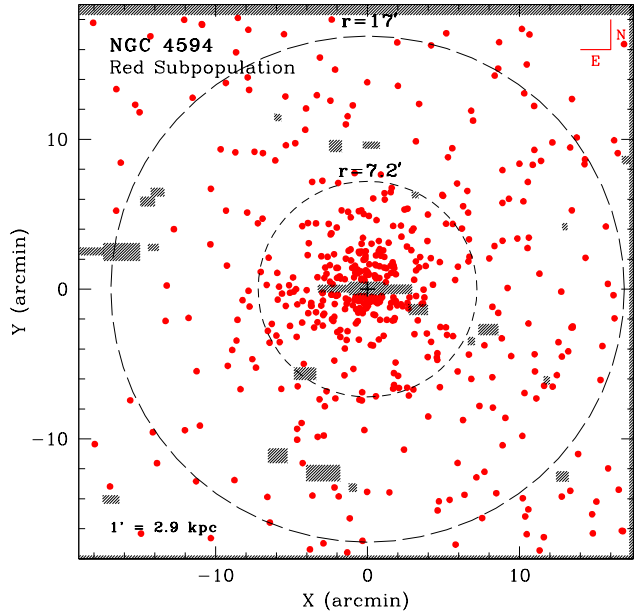}
\caption[Spatial positions of the metal-rich GC candidates in NGC~4594]{The positions of the 490 red GC candidates in NGC~4594 are shown with respect to the galaxy center (black cross).  The outer dashed circle denotes the radial extent of the GC system over which we are able to study the azimuthal distribution ($r\sim 17\arcmin$). The inner dashed circle shows the $r=7.2\arcmin$ radius over which we explored the azimuthal distribution of the GC subpopulations. We find no statistically significant ellipticity (compared to an azimuthally uniform distribution) in the 2D spatial distribution of the red subpopulation. The masked regions of the image are shaded. In the lower left we list the host galaxy effective radius $r_e$ and physical distance (in kpc) in terms of arcminutes on the image.
  \label{fig:chap4_n4594_red_spatial}}
\end{figure}

Although no significant non-zero ellipticity was detected, we analyzed the GC subpopulations to explore the possiblity of projected 2D substructure. RZ04 found a modest color gradient of $\Delta (B-R)/ \Delta(r) = -0.003 \pm 0.001~{\rm mag~arcmin}^{-1}$ in the N4594 GC population that extends over the full radial extent of the system ($r{\sim}19\arcmin$).  \txred{The color gradient is caused by the changing fraction of of red and blue clusters with radius due to the stronger central concentration of red GCs (compare Figures~\ref{fig:chap4_n4594_blue_spatial} and~\ref{fig:chap4_n4594_red_spatial})}.  We chose the $r=7.2\arcmin$ radius to explore the GC subpopulations given that the radially-dependent contamination fraction is $<20\%$ inwards of $7.2\arcmin$.  For the full 90\% sample, we find that the measured ellipticity is not statistically significant ($\epsilon = 0.1\pm0.08,\theta=-72\pm44$), since the probability of obtaining an ellipticity as larger or larger than measured by chance is $p=54\%$.  We find similar results for the red and blue subpopulations over this radius.  

Lastly, we used a series of Monte Carlo simulations to estimate an upper limit on the GC system ellipticity.  The number of GC candidates in NGC~4594 is relatively large, so we expect that if the GC system of NGC~4594 was significantly more elliptical (intrinsically), it would be unlikely that we would measure values of $\epsilon$ near zero.  That is, the large number of GC candidates should allow us to ``resolve" even a modestly elliptical GC system.  For ellipticities of $\epsilon = 0.2, 0.3$, we generated 10,000 simulated GC systems (each) and determined the frequency with which we would observe the measured value of $\epsilon = 0.14, 0.10$ or smaller.  We considered both the $r<17\arcmin$ and $r<7.2\arcmin$ radial cuts in our analysis.  For the $\epsilon = 0.2$ simulations, the measured values of $\epsilon$ occur relatively frequently (of order ${\sim} 10\%$).  For the $\epsilon = 0.3$ simulations, the measured values (or smaller) occur less than $1.4\%$ of the time.  This sets a reasonable upper limit: if the NGC~4594 GC system was as elliptical as $\epsilon{\sim}0.3$, we would have a ${\sim} 98.6\%$ likelihood of measuring an ellipticity {\it larger} than $\epsilon{\sim}0.1$.  

\begin{figure}
\includegraphics[angle=0,width=3.6in]{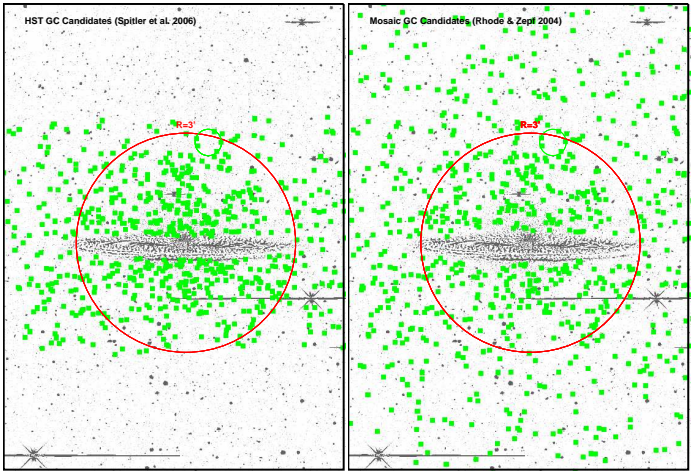}
\caption[HST and Mosaic GC candidates in NGC~4594]{ Spatial positions
  of the {\it HST} (left) and
  Mosaic GC candidates (right) in NGC~4594 from Spitler et al. (2006) and
  Rhode \& Zepf (2004), respectively, shown as green points on the galaxy light
  subtracted $V$ band Mosaic image.  A radius of $3\arcmin$ is shown
  as the red circle. The Mosaic imaging detects GCs
  nearly as
  close to the galaxy disk as the {\it HST} imaging.  Although the
  {\it HST} imaging is deeper, the radial surface density profiles for
  both data sets agree (Spitler et al. 2006).  Note that neither data
  set shows significant flattening of the GC system within $3\arcmin$ from a simple ``by-eye'' investigation.
  \label{fig:chap4_n4594_two_panel}}
\end{figure}

To summarize, over the explored radial extent of the GC system of NGC~4594 we see no strong evidence for a spatial distribution that is inconsistent with a uniform circular distribution.  The galaxy light becomes significantly more circular at larger radii, so in broad terms the GC system and galaxy show similar results.  However, near the galaxy center the bulk starlight of the galaxy becomes more flattened, reaching an ellipticity of ${\sim} 0.4$ in the inner $r \lesssim100\arcsec$ and so we wish to investigate any possible shape to the GC system in this inner region.  Our Mosaic observations probe quite close to the central region of the galaxy (innermost GC candidates at a radial distance of only ${\sim} 25\arcsec$), so we are not limited in studying the shape of the inner regions of the GC system by the inward radial extent of our observations.  In fact, comparing our GC candidate positions to those of Spitler et al. (2006; from {\it HST} ACS imaging) shows that space-based, high resolution imaging does not provide significantly better inward radial coverage (i.e., close to the galaxy disk) than the ground based Mosaic imaging (see Figure~\ref{fig:chap4_n4594_two_panel}).   Thus the limiting factor in analyzing the shape of the central regions is the presence of the galaxy disk.  The positions of the innermost GC candidates ($r< 3\arcmin$) in both the {\it HST} and Mosaic imaging show no obvious flattening of the GC system from a simple ``by-eye'' examination in either the total GC population or the metal-rich and metal-poor subpopulations.

\begin{figure}
\epsscale{1.2}
\plotone{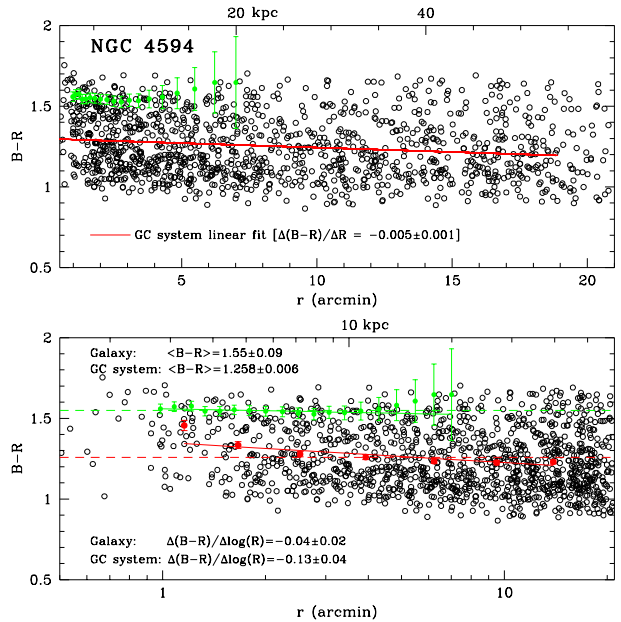}
\caption[GC and Galaxy B-R color profiles for NGC~4594]{Comparison of the GC and galaxy $B-R$ color profile for NGC~4594 in linear space (top panel) and logarithmic space (bottom panel).  We show the profiles out to the measured radial extent of the GC system.  In both panels the open circles mark the $B-R$ and radial positions (from the galaxy center) for the $90\%$ sample of GC candidates in NGC~4594 used in this study.  The filled green circles show the $B-R$ color profile for the galaxy light.  {\bf Top panel}:  The red line shows the linear fit to the GC system $B-R$ colors from \citet{rh04}. The GC population shows a color gradient over the full radial extent of the system.  {\bf Bottom panel}: The filled red points show the mean $B-R$ colors of GC candidates within elliptical bins at the same geometric mean radius $R$ of the galaxy isophotes.  Linear fits to the galaxy $B-R$ profile and GC system elliptical bins are shown as the solid green and red lines, respectively.  The measured slopes (color gradients) are listed.  The weighted mean $B-R$ colors for the galaxy light and GC population are listed and shown as the dashed green and red lines, respectively.  
  \label{fig:chap4_n4594_gc_gal_grad}}
\end{figure}

Figure~\ref{fig:chap4_n4594_gc_gal_grad} show the $B-R$ color profiles for the GC system and galaxy light for NGC~4594.  As noted above, RZ04 found a small $B-R$ color gradient over the full $19\arcmin$ radial extent of the GC system.   Using our modified $90\%$ color-complete sample, we f
ind a slightly steeper gradient of $\Delta (B-R)/ \Delta(r) = -0.005 \pm 0.001~{\rm mag~arcmin}^{-1}$ due to the addition of red GC candidates close to the galaxy center.  A weighted linear least-squares fit to the elliptically binned data gives a slope of $\Delta(B-R)/ \Delta \log{(r)} = -0.13 \pm 0.04~{\rm mag~dex}^{-1}$ ($\Delta {\rm [Fe/H]} / \Delta \log{(r)} = -0.37 \pm 0.12$) over the $19\arcmin$ radial extent. Although the GC system shows a steeper gradient relative to the galaxy ($\Delta(B-R)/ \Delta \log{(r)} = -0.04 \pm 0.02~{\rm mag~dex}^{-1}$), the difference is only significant at the ${\sim} 2\sigma$ level.  

\txred{
In addition to the overall GC color gradient (resulting from the changing mix of red and blue GCs), we find evidence of statistically significant color gradients in the {\it individual} GC subpopulations of NGC~4594.  We divided the $90\%$ color-complete sample of GCs into red or blue subpopulations using the KMM results from RZ04, who found a subpopulation split at $B-R=1.3$.  We then computed the mean GC color and standard error on the mean in linearly spaced circular radial bins for each subpopulation.  The radially binned data were then fit as a function of $\log{(r)}$ using Equation~\ref{eq_color_grad}.}

\begin{figure}
\epsscale{1.2}
\plotone{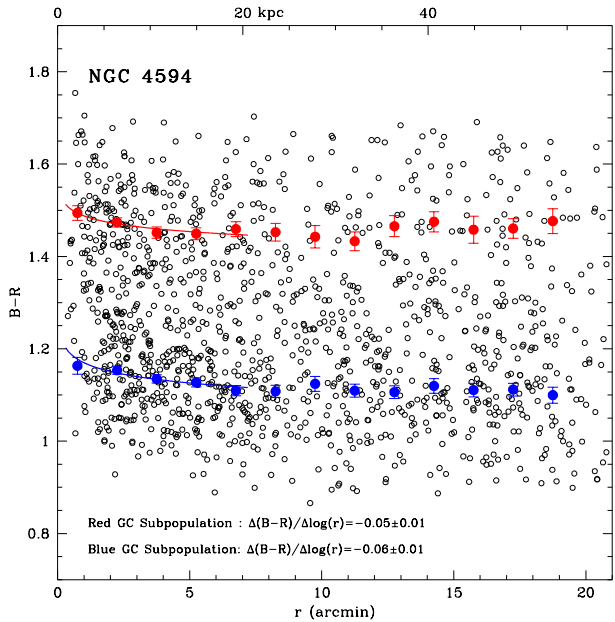}
\caption[GC subpopulation B-R color profiles for NGC~4594]{\txred{GC $B-R$ color profile for NGC~4594 with $\log{(r)}$ fits to the blue and red GC subpopulations.  We show the profile out to the measured radial extent of the GC system.  The open circles mark the $B-R$ and radial positions (from the galaxy center) for the $90\%$ sample of GC candidates in NGC~4594 used in this study. The filled red and blue points show the mean $B-R$ colors the metal-rich and metal poor GC subpopulations, respectively, within linearly spaced circular bins.  Linear fits to the blue and red GC subpopulation $B-R$ profiles are shown as the solid blue and red lines, respectively.  The measured slopes (color gradients) are listed.}
\label{fig:chap4_n4594_gc_gal_grad_subpop}}
\end{figure}

\txred{The $B-R$ GC color profile and $\log(r)$ fits are shown in Figure~\ref{fig:chap4_n4594_gc_gal_grad_subpop}.  Within the inner $7\arcmin$ of the blue GC subpopulation, we find a slope of $\Delta(B-R)/ \Delta \log{(r)} = -0.06 \pm 0.01~{\rm mag~dex}^{-1}$ ($\Delta {\rm [Fe/H]} / \Delta \log{(r)} = -0.18 \pm 0.04$) from the weighted linear least-squares fit.  For the red GC subpopulation, we find a slope of $\Delta(B-R)/ \Delta \log{(r)} = -0.05 \pm 0.01~{\rm mag~dex}^{-1}$ ($\Delta {\rm [Fe/H]} / \Delta \log{(r)} = -0.16 \pm 0.04$) over the same inner radial range.  We explored the sensitivity of these results to the choice of bin shape (elliptical vs. circular) and bin spacings (linear vs. logarithmic). In all case the gradients were statistically significant at the $3\sigma$ level or greater.  For both subpopulations, we find no statistically significant gradient outside of $r\sim 7\arcmin$.  We note that contamination is unlikely to have conspired to create the subpopulation gradients since contamination fraction is less than $20\%$ within $7\arcmin$ (RZ04).  We discuss the implications of GC subpopulation gradients for galaxy formation scenarios in $\S\ref{sec:chap5_compare_colors}$.}

Comparing the mean colors of the galaxy and GC system show a statistically significant difference.  We find an integrated color of NGC~4594 of $B-R = 1.55 \pm 0.09$ compared to a mean GC system color of $B-R = 1.258 \pm 0.006$ (${\rm [Fe/H]} = -1.10 \pm 0.02$), a difference of almost $0.3$ mag.  The mean colors of the blue and red GC subpopulations are $B-R=1.13\pm 0.01$ and $B-R=1.46\pm0.01$, respectively. The difference between the color of the red GC system peak and the host galaxy mean color (${\sim} 0.09$ mag) is not statistically significant given the large error on the integrated color of the galaxy.


\section{Discussion and Conclusions}
\label{sec:chap4_comparisons}

In this paper we present a comparison of the spatial distributions (radial and azimuthal) and color profiles of the GC systems of four giant galaxies to those of their host galaxies.  The main results are summarized in Table~\ref{tab:chap5_table2}.  We focus most of our discussion below on the azimuthal distribution results.  For completeness we also summarize our color profile and mean color results and include a discussion of key points from the literature, as the cause of the galaxy and GC system mean color offsets is currently being debated.

\begin{deluxetable*}{lcccc}
\tablecolumns{5}
\tablewidth{0pc}
\tabletypesize{\footnotesize}
\tablecaption{Comparison of GC System (GCS) and Host Galaxy Properties\label{tab:chap5_table2}}
\tablehead{
\colhead {} & \colhead{NGC~4472} & \colhead{NGC~4406} & \colhead{NGC~5813} & \colhead{NGC~4594} 
}
\startdata
Type                                  & E2 & 			  E3 & 		       	 E1 & 			  	  S0 \\  
$M_V^T$                          & -23.1 & 		          -22.3 & 	        -22.3 & 		  -22.4 \\	     
\\
                                                 &                            & Mean Color and Color Gradient Properties & & \\
\cline{1-5}
\Tstrut\Bstrut
Integrated $B-R$ Galaxy color           & $1.40 \pm 0.09$ & 		$1.51\pm0.03$ & 	$1.50\pm0.04$ &      $1.55\pm0.09$ \\  
Mean $B-R$ GCS color               & $1.18\pm0.01$ &               $1.190\pm0.006$ & 	$1.202\pm0.008$ &      $1.258\pm0.006$ \\ 	     
$B-R$ color of blue color distribution peak\tablenotemark{1}            & $1.09\pm 0.02$  &  $1.11 \pm 0.01$  & $1.16 \pm 0.01$ & $1.13 \pm 0.01$ \\
$B-R$ color of red color distribution peak\tablenotemark{1}             & $1.39\pm 0.02$  &  $1.41 \pm 0.01$  &  $1.47 \pm 0.03$  & $1.46 \pm 0.01$ \\
Galaxy color gradient\tablenotemark{2} & $-0.02\pm0.01$  & 	   $-0.03\pm0.01$  &     $-0.04\pm0.01$  &     $-0.04\pm0.02$  \\ 	     
GCS global color gradient\tablenotemark{2} & $-0.13\pm0.05$  & 	   $-0.06\pm0.02$  &     $-0.12\pm0.03$  &     $-0.13\pm0.04$   \\    
GCS blue subpopulation color gradient\tablenotemark{2} & -- & -- & -- & $-0.05 \pm 0.01$\\
GCS red subpopulation color gradient\tablenotemark{2} & -- & -- & -- & $-0.04 \pm 0.01$ \\
\cline{1-5}

\\
                                                 &                            & Spatial Distribution Properties & & \\
\cline{1-5}
\Tstrut\Bstrut
Theoretical $r_e$ for GCS ($r^{1/4}$ law)         & $21 \pm 3 \arcmin$  & $20\pm 3 \arcmin$  &   	 $5 \pm 1 \arcmin$ & $9.1\pm1.0\arcmin$ \\	     
Theoretical $r_e$ for GCS (S\'ersic) &  $12 \pm 2\arcmin$ & $5.8 \pm 0.1\arcmin$ & $3.9 \pm 0.3\arcmin$ & $5.9 \pm 0.4\arcmin$ \\
Empirical $r_e$ for GCS          & $8.1\arcmin$   & $6.4\arcmin$ 	     &  $3.6\arcmin$    	   &  $4.3\arcmin$   \\ 	     
Theoretical $r_e$ for galaxy ($r^{1/4}$ law)       & $1.97 \pm 0.05\arcmin$  & $2.52\pm0.02\arcmin$  &	 $1.50\pm0.01\arcmin$ &	 $1.48\pm0.03\arcmin$ \\            
Theoretical $r_e$ for galaxy (S\'ersic)       & $3.0 \pm 0.4\arcmin$ & $5.8 \pm 0.4\arcmin$ & $2.9 \pm 0.3\arcmin$ & $1.42 \pm 0.08\arcmin$ \\            
S\'ersic index for GCS & $2.5 \pm 0.5$ & $0.9 \pm 0.1$ & $2.5 \pm 0.8$ & $1.9 \pm 0.3$ \\
S\'ersic index for galaxy & $5.7 \pm 0.7$ & $6.5\pm0.2$ & $5.7 \pm 0.2$ & $4.7 \pm 0.5$\\
GCS ellipticity                    & --            & $0.38 \pm 0.05$ & $0.42\pm 0.08$ &	${\sim} 0$\tablenotemark{3} \\    
Galaxy ellipticity\tablenotemark{5} & $0.25$   & $0.4$ & $0.3$ 	& $0.05$ \\                      
GCS PA (degrees)\tablenotemark{4}           & --             & $-63 \pm 6$ & $-59\pm13$ &  --  \\		  
Galaxy PA (degrees)\tablenotemark{4,5}        & $-25$ & $-55$ &  $-53$ & 	$86$ 
\enddata
\tablenotetext{1}{Homoscedastic KMM results}
\tablenotetext{2}{Defined as $\Delta(B-R)/\Delta\log{(r)}$ in units of ${\rm mag~dex}^{-1}$}
\tablenotetext{3}{Method of moments results for NGC~4954 are consistent with an azimuthally uniform distribution (see $\S\ref{sec:chap4_n4594_gc_system}$).}
\tablenotetext{4}{Position angle (PA) is defined as degrees east of north.}
\tablenotetext{5}{Adopted mean values are for the the outermost galaxy isophotes \txred{(see discussion in $\S\ref{sec:chap4_ellip_analysis}$)}.}
\end{deluxetable*}

\subsection{Colors of GC Systems and their Host Galaxies}
\label{sec:chap5_compare_colors}

Comparing the global mean colors and color gradients of the GC systems to their host galaxies, we find that (i) the mean colors of the GC systems are ${\sim} 0.3$ mag bluer than their host galaxies and (ii) that the global color gradients in GC systems are slightly steeper (e.g., more negative, i.e., bluer colors with increasing radius) than the color gradients in their host galaxies.  \txred{Both results are consistent with a number of previous galaxy/GC system comparisons in the literature (for mean color offsets see Harris et al. 1991, Ostrov et al. 1993, Peng et al. 2006; for color gradient comparisons see Liu et al. 2011)}.  While the first result is found at a high level of statistical significance, the larger uncertainties in the gradients mean the latter result is only significant at the ${\sim} 2\sigma$ level.  

\txred{When examining the individual metal-rich and metal-poor GC subpopulation colors, we find that the ${\sim 0.3}$ mag offset between the global GC system and integrated host galaxy colors is largely due to the presence of the blue (metal-poor) subpopulation.  Although this may be a first-order cause for these differences, recent work suggests the GC subpopulation color/host galaxy color offsets may be more complicated.}  \citet{go13} compared the mean colors of the metal-rich GC subpopulation in seven giant ellipticals to the host galaxy color profile.  They found that the host galaxies are redder by a (statistically significant) ${\sim} 0.1-0.2$ mag (in $g-z$ or $B-I$) but that the luminosity-weighted ages and metallicities (as measured from spectroscopic Lick indices) of the galaxy and GC systems are consistent.  \citet{pe06} also noted the color difference between the host galaxies and their GC systems in their study of the ACSVCS galaxies, \txred{as did \citet{sp10}, who studied the offset in the context of the galaxy bulge/metal-rich GC connection}.  We find offsets of ${\sim} 0.1$ mag in two of our four galaxies (NGC~4406 and NGC~4594), but the result is only statistically significant for NGC~4406.  \citet{go13} argue that the observed color offset can be explained by a bottom-heavy (steep) initial mass function (IMF) and the dynamical evolution of the metal-rich clusters.  \citet{ch13} present an alternative explanation of the observations in terms of the multiple generations of stars in GCs.  \citet{ch13} use stellar population synthesis models to show that the enhanced helium in the second generation stars (which are assumed to form within a few hundred Myr of the first generation; Gratton et al. 2012) can cause the color offsets via contributions from blue horizontal branch stars without differences in the IMF.  Additional work, both observationally and theoretically, will be necessary to understand the origin of the mean color offsets between galaxies and their GC systems.  

\txred{
We also explored the presence of color gradients in the individual GC subpopulations.  Only NGC~4594 shows evidence of a statistically significant gradient, which is found in both GC subpopulations (with similar magnitudes of $\Delta({\rm Fe/H})/ \Delta \log{(r)} \sim -0.17 \pm 0.04$) over the inner $\sim 7\arcmin$ ($\sim 5 r_e; \sim 20$ kpc).  At larger radii the GC system colors remain relatively constant and we find no evidence of a radial gradient.  \citet{fo11} found a similar result from their wide-field imaging of the massive elliptical NGC~1407.  Here, both the metal-rich and metal-poor subpopulations show similar gradients of $\Delta({\rm Fe/H})/ \Delta \log{(r)} \sim -0.40 \pm 0.06$ over the inner $\sim5-9 r_e$ ($\sim40-70$ kpc), but no gradient is detected at larger radii.  Radial color gradients in both GC subpopulations have also been observed in the S0 galaxy NGC~3115 \citep{ar11}, the giant cD elliptical M87 \citep[][but see Strader et al. 2011]{ha09b}, and the Milky Way \citep{ha01}.

\citet{fo11} argue that the break in the color gradients are qualitative evidence of the ``two phase'' formation scenario for early-type galaxy formation (see $\S\ref{sec:chap4_intro}$).  The results for NGC~4594 may be consistent with this picture.  On the other hand, few wide-field GC system imaging studies have had subpopulation gradient analyses.  Furthermore, other studies examining GC system radial trends have found evidence for subpopulation gradients only in {\it stacks} of GC population color profiles \citep{ha09a,li11}.  Additional wide-field imaging studies of GC populations, for a large number of galaxies, will be important in assessing the prevalence of GC radial subpopulation gradients in early-type galaxies \citep[e.g.][]{br14}.  
}

\subsection{GC and Galaxy Azimuthal Distributions}
\label{sec:chap5_compare_azimuth}

Our analysis of the azimuthal distribution of the GC systems of the giant galaxies shows some commonalities between systems.  For two giant ellipticals (NGC~4406 and NGC~5813), we find that the GC systems show an elliptical projected spatial distribution that is consistent with that of the host galaxy light.   For NGC~4472, the low numbers of GC candidates in the $90\%$ color-complete subsample gave inconclusive azimuthal distribution analysis results. In addition to the analysis of the total GC population, we also examine the azimuthal distributions of the metal-rich (red) and metal-poor (blue) GC subpopulations in each galaxy.  For NGC~4406, we find that both the blue and red GC subpopulations show a statistically significant elliptical azimuthal distribution with measured ellipticities and position angles that are consistent with the host galaxy light.  For NGC~5813, we find that the blue and red GC subpopulations have an elliptical spatial distribution, but the smaller number of GC candidates in the subpopulations results in larger uncertainties in the shape parameters and a smaller overall statistical significance (${\sim}2\sigma$ versus $3\sigma$ for NGC~4406) of non-circularity.

For NGC~4594, we do not detect an elliptical distribution that is significantly different from circular over the full radial extent of the GC system nor in the inner regions of the galaxy surrounding the disk.  This result holds for our analysis of the red and blue subpopulations as well.  The circular shape of the GC system is in contrast to the galaxy light which becomes as flattened as $\epsilon = 0.4$ in the inner regions while becoming circular at large radii.  In this sense there is broad agreement between the projected shapes of the GC system and the galaxy halo light {\it at large radii} in NGC~4594.  However, if the shapes of the GC system and the galaxy were strongly coupled in NGC~4594 (as they appear to be in the giant ellipticals), one might expect to observe a more significant flattening of the GC system (or at least one of the GC subpopulations) in the inner regions.  

\subsubsection{Comparisons with Galaxy Formation Scenarios}

The similarities of the projected shapes of the GC systems and their host galaxies in the two ellipticals in our sample may give insights into the formation history of the systems.  To first order, these similarities may suggest a common dynamical origin for the ellipticities of the GC systems and their host galaxies. The agreement between the projected shapes may imply that the formation and evolutionary processes that produce the non-spherical shape of giant ellipticals may also cause the elliptical shape of the GC system.  \txred{\citet{ki96} also noted this possible connection in their study of the ellipticity of the GC system of NGC~720.}  

It is well established that the shapes of luminous ellipticals are caused not by flattening due to rotation (like lower-luminosity ellipticals; Emsellem et al. 2007, Cappellari et al. 2007) but by velocity anisotropy (see \citealt{bi08} and references therein).  Velocity anisotropy arises from the rapidly changing gravitational potential and resulting violent relaxation of the system, as is demonstrated in both collapse simulations of clumpy, spherical distributions of stars \citep{va82,bi08} and simulations of galaxy mergers \citep{na03,hi12}.  The effect of large-scale dynamical processes during E/S0 galaxy formation on the observed spatial properties of galaxy GC systems was explored using numerical simulations by Bekki et al. (2005) and may provide insights into the kinematic origin of GC system azimuthal distributions.

Bekki et al. ran a series of $N$-body simulations that are qualitatively similar to the major merger GC scenario model proposed by \citet{as92}, although as we discuss below, Bekki et al. did not include dissipative processes.  The GC systems of the progenitor spirals were chosen to match the observed spatial distribution of the Milky Way GC system and included both the metal-rich and metal-poor subpopulations.  In addition, the total number of GCs in the simulations was held constant, since star formation and GC destruction models are not included in the simulations.  Bekki et al.  find that the both the stars and GC systems in the merger remnants show nearly identical, modestly flattened spatial distributions.  In their fiducial model (two equal-mass galaxies) they find $\epsilon =0.18$ for the galaxy and $\epsilon = 0.22$ for the GC system, although they note that the difference in $\epsilon$ is insignificant outside of $1~r_e$.  The projected shapes of both GC subpopulations are also elliptical, and show similar flattening to each other.  The resulting position angles of the GC system and host galaxy remnant are nearly aligned (differences of $< 10^\circ$).  The position angles of the individual GC subpopulations are also coincident with the host galaxy light.  

In a qualitative sense, these simulations are in good agreement with the observations of the NGC~4406 GC system and host galaxy light.  However, the Bekki et al.  simulations do not include new star or GC formation (i.e. a dissipational gaseous component) nor a model for GC destruction.  If we consider the influence of star formation via gas dissipation in particular, simulations have shown that the inclusion or exclusion of dissipative processes significantly changes the observed properties of merger remnants.  For example, \citet{co06} showed that disk-disk merger remnants, in which gas dissipation and star formation are included, have more isotropic velocity distributions than remnants formed in mergers without dissipation.  The degree to which dissipative processes influence the properties of a galaxy's GC system will require more detailed simulations. Ideally such modeling would be done in a cosmological context which considers GC formation, destruction, and dissipative gas physics.

Another interesting result from the azimuthal distribution analysis of NGC~4406 and NGC~5813 is that both the metal-rich (red) and metal-poor (blue) subpopulations have similar ellipticities and position angles.  This implies that the physical processes that gave rise to the non-spherical spatial distributions were not (somehow) limited to one subpopulation.  Observations of the {\it radial} distributions of GC subpopulations, however, have shown differences in their spatial distributions; the metal-rich GCs are sometimes more centrally concentrated than the metal-poor GCs in giant ellipticals (see Figure 21 in HR12 for NGC~5813; Brodie \& Strader 2006 and references therein).  Comparisons also sometimes show that the radial surface density profile of the metal-rich population follows the host galaxy surface brightness in the inner regions (e.g., \citealt{st11}).  The broad interpretation of these latter observations is that the metal-rich GC population formed in the dissipational processes that built up the bulk starlight of the host galaxy \citep{br06}.  As \citet{br06} note, if the connection between red GCs and the host galaxy light is strong, one might also expect the {\it azimuthal} distribution to match the host galaxy as well, while the metal-poor population will have a spatial distribution which reflects the earliest stages of galaxy formation.  While we do see this coincidence between the red GC/host galaxy azimuthal distributions in NGC~4406 (and at a smaller significance level in NGC~5813), {\it the blue subpopulation shape is also consistent with the shape of the host galaxy light}.  So, despite the fact that the metal-poor nature of the blue subpopulation implies that they formed from gas with a different chemical enrichment from the red GC population, the similar azimuthal distribution to both the host galaxy and the red GCs \txred{suggests they share a common history.}

\txred{In addition to the two galaxies studied here, we note that many other giant galaxies show evidence of a metal-poor GC population with an elliptical spatial distribution consistent with (or more elliptical than) the host galaxy light, including M87 \citep{st11}, NGC~1316 \citep{ri12}, NGC~2768 \citep{ka13}, NGC~4635 \citep{bl12}, NGC~4636 \citep{di05}, and a number of early-type Virgo galaxies studied by \citet{pa13}.  Although many galaxies do not show this trend (e.g., NGC~720, NGC~1023, NGC~4649; \citealt{ka13}), it is clear that elliptical spatial distributions of metal-poor GC populations are not uncommon among giant early-type galaxies.}

\txred{How do the observations of non-spherical GC subpopulation spatial distributions fit in the context of proposed galaxy/GC formation scenarios? In models where blue GCs form in more chaotic/accretion processes, the implication is that the resulting spatial distribution should be spherical.  This connection is consistent with observational studies of the MW metal-poor GC population, where \citet{zi85} noted their spherical spatial distribution and where subsequent work has shown that the \textit{outer halo} metal-poor GCs likely have an accretion origin (e.g., \citealt{zi93,ma04,fo10,do11,ke12}).  This accretion+spherical spatial distribution picture for the blue GC population was also proposed by \citet{fo97} in their multi-phase dissipational collapse model for GC bimodality in giant ellipticals.  We note that the observations of the blue GC populations in NGC~4406, NGC~5813 (although at a smaller level of significance), and other giant galaxies (see above) disagree with this model prediction, although the metal-rich GC populations do indeed follow the azimuthal distribution of the host galaxy light in these galaxies.  The collapse plus accretion model of \citet{co98}, however, makes no specific predictions for the GC subpopulation azimuthal distributions.

The expectations for the projected spatial distribution of GC populations in a ``two-phase'' galaxy formation model (see $\S\ref{sec:chap4_intro}$) are less well understood, as current $N$-body $+$ hydrodynamical simulations of this specific scenario do not include GC formation and destruction.  Further complicating the picture, at a qualitative level, both \textit{in-situ} and accretion scenarios might be able to explain the azimuthal distribution of metal-poor GCs in some giant ellipticals.  \citet{wa13} suggest that, because merging/accretion may occur along preferential directions (i.e., along local filamentary structures), this may explain the alignment of the blue GCs with the host galaxy. This picture might be consistent with the results of cosmologically-motivated dark matter plus baryon simulations -- both of massive central + satellite galaxies \citep{do14} and late-type galaxies + satellite dwarf galaxies \citep{de11}.  On both simulation scales, the satellites were found to be spatially anisotropic and aligned with either the massive central galaxy \citep{do14} or the outer dark matter halo \citep{de11}.  Simulations which can probe lower mass satellites around massive galaxies will be necessary to explore whether such alignments are expected in GC populations.

On the other hand, \citet{br14} suggest that a significant population of metal-poor GCs may have formed \textit{in-situ} with the host galaxy rather than ``cosmologically'' at $z\sim 9-13$.  This picture synthesizes a number of observations related to GC ages, the orbits of blue GCs from kinematic studies, and the evidence for both inner and outer metal-poor halo components in the MW and other giant galaxies \citep{br14}.   In this scenario, both metal-poor and metal-rich GCs form early and \textit{in situ} in major bursts of star formation.  This ``dual \textit{in situ}'' model may also explain the similar azimuthal distributions of blue and red GCs in giant ellipticals like NGC~4406, if indeed the major dynamical processes of galaxy formation (e.g., violent relaxation, velocity anisotropy) influence the resulting projected shape of a galaxy's GC system. 
}

\subsubsection{Comparisons with GC System Kinematics}

\txred{
The kinematic analysis of a galaxy's GC system, as derived from follow-up spectroscopy and modeling, should provide additional insights into the possible connection between the projected spatial distribution of GCs and their formation history.  In recent years the number of giant galaxies for which follow-up, large sample ($N_{\rm GC}\gtrsim 100$) GC spectroscopy exists has rapidly increased (see \citealt{br14} for a recent review), although for many of these galaxies no study of the GC azimuthal distributions has been done.  Two galaxies in our survey, NGC~4594 and NGC~7457, have published GC kinematic studies, so we can make direct comparisons of the azimuthal distributions to the kinematic data. 
}

Our analysis of the GC system of NGC 4594 does not show statistically-significant evidence for a flattened spatial distribution (see $\S\ref{sec:chap4_n4594_gc_system}$).  Analyses of the red and blue GC subpopulations individually also do not show a statistically significant difference from a projected spherical distribution.  The kinematics of NGC~4594's GC system have recently been studied by Dowell et al. (2014) who use a sample of ${\sim}360$ spectroscopically-confirmed GCs to study the galaxy mass profile, mass-to-light ratio profile, and rotation of the GC system out to ${\sim} 41$ kpc (${\sim} 15\arcmin = 9~ r_e$).  These authors find no evidence of significant rotation in the total GC system out to large radius; the same holds true for the red and blue subpopulations.  This is consistent with the previous kinematic study of 108 GC velocities in NGC~4594 by \citet{br07}, who argue that the lack of observed rotation may be evidence that angular momentum has been transported outward from the system via mergers.  \citet{br07} also note that the lack of rotation in the GC system is in contrast to the stars and gas, which show significant amounts of rotation in the inner ${\sim} 3\arcmin$ (9 kpc).  These kinematic differences between the GC system and stellar light may also be linked to the differences observed in the projected (2D) spatial distributions.  The inner $3\arcmin$ of the galaxy light shows evidence of a large ellipticity and significant rotation while the GC population shows little rotation or spatial flattening.  

For the lower luminosity S0 galaxy NGC~7457, we found evidence of a highly flattened GC system ($\epsilon = 0.66 \pm 0.14$; H11) which is broadly consistent with the ellipticity profile of the host galaxy starlight (mean ellipticity of $\bar{\epsilon} {\sim} 0.46$ between $0.3\arcmin$ and $2.3\arcmin$).  Follow-up  spectroscopy of GC candidates was presented by \citet{po13}, who analyzed the kinematics of $27$ confirmed GC candidates (the WIYN GC candidates from H11 and an {\it HST} study by \citealt{ch08}) to $2~ r_e$ and found evidence of a strongly rotating GC system.  At more than one effective radius in NGC~7457, the rotation curve of the GC population remains flat at ${\sim} 80~{\rm km/s}$, while the velocity dispersion is low at $\lesssim 50~{\rm km/s}$.  In addition, these GC kinematics are consistent with the galaxy dynamics from analyses of both long-slit spectra and integral field unit data \citep{po13}.  Comparing these kinematic results with the azimuthal distribution and galaxy ellipticity profiles, we find good agreement between the galaxy and GC systems: both components show significant spatial flattening and rotation.  Due to the small numbers in the GC kinematic sample and the lack of observed color bimodality (see H11 and references therein), no GC subpopulation analysis was done by \citet{po13} for NGC~7457. 

\txred{
Lastly, we note that although these lenticular galaxies show similarities between their GC system kinematics and azimuthal spatial distributions, for well-studied ellipticals the picture is more complex.  The giant ellipticals M87 \citep{st11} and NGC~4365 \citep{bl12} have had similar GC kinematic+azimuthal distribution studies, and in both cases the red and blue GC subpopulations are spatially flattened.  However, the GC dynamical studies of these systems show evidence of kinematic \textit{differences} between the subpopulations despite the similar projected spatial distributions. Larger numbers of studies which combine imaging and spectroscopy will be helpful in exploring the similarities or differences between GC spatial distributions and population kinematics.  In general, our results suggest that detailed studies of GC azimuthal distributions out to several host galaxy effective radii can provide important constraints to galaxy formation and evolution scenarios.   As emphasized earlier, cosmological dark matter $N$-body $+$ hydrodynamic baryonic simulations which include GC formation will be critical in understanding the connection between the host galaxy, GC system dynamics, and the projected shapes of galaxy GC systems.

}

\acknowledgments 

This research was supported by a Dissertation Year fellowship to JRH from the College of Arts and Sciences of Indiana University and by an NSF Faculty Early Career Development (CAREER) award (AST-0847109) to KLR.  JRH would like to thank Steven Janowiecki, Liese van Zee, and Enrico Vesperini for useful conversations during the course of this work and Beth Willman for financial support during the preparation and revision of this manuscript (NSF AST-1151462).  JRH would also like to thank Liese van Zee for sharing the LVL on NGC~4594 prior to publication.  We would like to thank the anonymous referee for comments which significantly improved this manuscript.  This research has made use of the NASA/IPAC Extragalactic Database (NED), which is operated by the Jet Propulsion Laboratory, California Institute of Technology, under contract with the National Aeronautics and Space Administration.

{}




\section*{Appendix A}
\label{sec:chap4_appendix}

Appendix A presents the galaxy surface photometry results for NGC~4406, NGC~4472, NGC~5813, and NGC~4594 in tabular form (Tables~\ref{tab:chap4_n4406_sfc_phot}-\ref{tab:chap4_n4594_sfc_phot}, respectively).  The data are shown graphically in Figures~\ref{fig:chap4_n4406_sfc_phot} (NGC~4406),~\ref{fig:chap4_n4472_sfc_phot} (NGC~4472),~\ref{fig:chap4_n5813_sfc_phot} (NGC~5813), and~\ref{fig:chap4_n4594_sfc_phot} (NGC~4594).

\input{n4406_sfc_table.tex}

\input{n4472_sfc_table.tex}

\input{n5813_sfc_table.tex}

\input{n4594_sfc_table.tex}

\end{document}

%% file: n4406_sfc_table.tex
{\footnotesize
\begin{longtable*}{cccccccc}
\label{tab:chap4_n4406_sfc_phot}
\\
\caption[Surface photometry for NGC 4406]{Surface photometry for NGC 4406}\\
\hline
\hline
$R$ & $\mu_{V} \pm \sigma$  & $\epsilon \pm \sigma$ & ${\rm PA} \pm \sigma$  & SMA  & $(B-V) \pm \sigma$ & $(V-R) \pm \sigma$ & $(B-R) \pm \sigma$ \\
(arcsec) & $({\rm mag ~arcsec}^{-2})$ & & (degrees E of N) & (arcsec) & (mag) & (mag) & (mag) \\
\hline
\endfirsthead
\caption[]{Surface photometry for NGC 4406, continued.}\\
\hline
\hline
$R$ & $\mu_{V} \pm \sigma_{\mu_{V}}$  & $\epsilon \pm \sigma_{\epsilon}$ & ${\rm PA} \pm \sigma_{\rm PA}$  & SMA  & $(B-V) \pm \sigma_{B-V}$ & $(V-R) \pm \sigma_{V-R}$ & $(B-R) \pm \sigma_{B-R}$ \\
(arcsec) & $({\rm mag ~arcsec}^{-2})$ & & (degrees) & (arcsec) & (mag) & (mag) & (mag) \\
\hline
\endhead 
 11.82 & $ 18.76 \pm  0.01 $ & $ 0.1943 \pm 0.0016 $ & $ -59.84 \pm 0.20 $ &   13.17 & $ 0.964 \pm 0.010 $ &  $ 0.595 \pm 0.010 $ & $ 1.559 \pm 0.014 $ \\          
 13.01 & $ 18.90 \pm  0.01 $ & $ 0.1943 \pm 0.0009 $ & $ -59.84 \pm 0.17 $ &   14.49 & $ 0.949 \pm 0.010 $ &  $ 0.596 \pm 0.010 $ & $ 1.545 \pm 0.014 $ \\          
 14.27 & $ 19.02 \pm  0.01 $ & $ 0.1984 \pm 0.0009 $ & $ -59.69 \pm 0.16 $ &   15.94 & $ 0.946 \pm 0.010 $ &  $ 0.597 \pm 0.010 $ & $ 1.543 \pm 0.014 $ \\          
 15.66 & $ 19.15 \pm  0.01 $ & $ 0.2021 \pm 0.0008 $ & $ -59.72 \pm 0.15 $ &   17.53 & $ 0.943 \pm 0.010 $ &  $ 0.597 \pm 0.010 $ & $ 1.540 \pm 0.014 $ \\          
 17.20 & $ 19.28 \pm  0.01 $ & $ 0.2051 \pm 0.0008 $ & $ -59.40 \pm 0.14 $ &   19.29 & $ 0.942 \pm 0.010 $ &  $ 0.596 \pm 0.010 $ & $ 1.539 \pm 0.014 $ \\          
 18.86 & $ 19.41 \pm  0.01 $ & $ 0.2095 \pm 0.0008 $ & $ -59.31 \pm 0.14 $ &   21.21 & $ 0.942 \pm 0.010 $ &  $ 0.596 \pm 0.010 $ & $ 1.538 \pm 0.014 $ \\          
 20.68 & $ 19.54 \pm  0.01 $ & $ 0.2148 \pm 0.0009 $ & $ -58.95 \pm 0.15 $ &   23.33 & $ 0.940 \pm 0.010 $ &  $ 0.596 \pm 0.010 $ & $ 1.536 \pm 0.014 $ \\          
 22.62 & $ 19.66 \pm  0.01 $ & $ 0.2232 \pm 0.0008 $ & $ -58.85 \pm 0.13 $ &   25.67 & $ 0.941 \pm 0.010 $ &  $ 0.595 \pm 0.010 $ & $ 1.536 \pm 0.014 $ \\          
 24.71 & $ 19.78 \pm  0.01 $ & $ 0.2342 \pm 0.0008 $ & $ -58.51 \pm 0.12 $ &   28.24 & $ 0.939 \pm 0.010 $ &  $ 0.595 \pm 0.010 $ & $ 1.534 \pm 0.014 $ \\          
 27.04 & $ 19.90 \pm  0.01 $ & $ 0.2420 \pm 0.0008 $ & $ -57.62 \pm 0.12 $ &   31.06 & $ 0.936 \pm 0.010 $ &  $ 0.595 \pm 0.010 $ & $ 1.530 \pm 0.014 $ \\          
 29.61 & $ 20.02 \pm  0.01 $ & $ 0.2489 \pm 0.0009 $ & $ -57.26 \pm 0.12 $ &   34.17 & $ 0.936 \pm 0.010 $ &  $ 0.593 \pm 0.010 $ & $ 1.529 \pm 0.014 $ \\          
 32.51 & $ 20.14 \pm  0.01 $ & $ 0.2516 \pm 0.0010 $ & $ -57.64 \pm 0.13 $ &   37.58 & $ 0.941 \pm 0.010 $ &  $ 0.593 \pm 0.010 $ & $ 1.534 \pm 0.014 $ \\          
 35.77 & $ 20.29 \pm  0.01 $ & $ 0.2512 \pm 0.0009 $ & $ -57.58 \pm 0.11 $ &   41.34 & $ 0.935 \pm 0.010 $ &  $ 0.592 \pm 0.010 $ & $ 1.527 \pm 0.014 $ \\          
 39.40 & $ 20.43 \pm  0.01 $ & $ 0.2493 \pm 0.0007 $ & $ -57.83 \pm 0.10 $ &   45.47 & $ 0.935 \pm 0.010 $ &  $ 0.592 \pm 0.010 $ & $ 1.526 \pm 0.014 $ \\          
 43.31 & $ 20.58 \pm  0.01 $ & $ 0.2503 \pm 0.0007 $ & $ -57.09 \pm 0.10 $ &   50.02 & $ 0.932 \pm 0.010 $ &  $ 0.591 \pm 0.010 $ & $ 1.522 \pm 0.015 $ \\          
 47.76 & $ 20.73 \pm  0.01 $ & $ 0.2467 \pm 0.0008 $ & $ -56.19 \pm 0.11 $ &   55.02 & $ 0.932 \pm 0.010 $ &  $ 0.590 \pm 0.011 $ & $ 1.521 \pm 0.015 $ \\          
 52.43 & $ 20.88 \pm  0.01 $ & $ 0.2498 \pm 0.0007 $ & $ -55.43 \pm 0.10 $ &   60.53 & $ 0.930 \pm 0.010 $ &  $ 0.588 \pm 0.011 $ & $ 1.519 \pm 0.015 $ \\          
 57.51 & $ 21.03 \pm  0.01 $ & $ 0.2539 \pm 0.0008 $ & $ -55.04 \pm 0.09 $ &   66.58 & $ 0.931 \pm 0.010 $ &  $ 0.589 \pm 0.011 $ & $ 1.520 \pm 0.015 $ \\          
 63.42 & $ 21.20 \pm  0.01 $ & $ 0.2502 \pm 0.0008 $ & $ -55.80 \pm 0.10 $ &   73.24 & $ 0.929 \pm 0.010 $ &  $ 0.589 \pm 0.011 $ & $ 1.518 \pm 0.015 $ \\          
 69.81 & $ 21.36 \pm  0.01 $ & $ 0.2491 \pm 0.0008 $ & $ -57.10 \pm 0.10 $ &   80.56 & $ 0.927 \pm 0.011 $ &  $ 0.588 \pm 0.012 $ & $ 1.514 \pm 0.016 $ \\          
 76.04 & $ 21.50 \pm  0.01 $ & $ 0.2637 \pm 0.0009 $ & $ -58.80 \pm 0.10 $ &   88.62 & $ 0.928 \pm 0.011 $ &  $ 0.586 \pm 0.012 $ & $ 1.514 \pm 0.016 $ \\          
 83.06 & $ 21.64 \pm  0.01 $ & $ 0.2740 \pm 0.0010 $ & $ -60.10 \pm 0.11 $ &   97.48 & $ 0.929 \pm 0.011 $ &  $ 0.586 \pm 0.013 $ & $ 1.515 \pm 0.017 $ \\          
 91.60 & $ 21.82 \pm  0.01 $ & $ 0.2702 \pm 0.0012 $ & $ -61.36 \pm 0.13 $ &  107.23 & $ 0.926 \pm 0.011 $ &  $ 0.587 \pm 0.014 $ & $ 1.514 \pm 0.018 $ \\          
101.23 & $ 22.01 \pm  0.01 $ & $ 0.2635 \pm 0.0012 $ & $ -62.51 \pm 0.14 $ &  117.95 & $ 0.935 \pm 0.013 $ &  $ 0.587 \pm 0.016 $ & $ 1.521 \pm 0.021 $ \\          
110.69 & $ 22.17 \pm  0.01 $ & $ 0.2722 \pm 0.0013 $ & $ -62.59 \pm 0.15 $ &  129.75 & $ 0.932 \pm 0.015 $ &  $ 0.586 \pm 0.019 $ & $ 1.518 \pm 0.024 $ \\          
120.28 & $ 22.32 \pm  0.02 $ & $ 0.2898 \pm 0.0014 $ & $ -62.65 \pm 0.14 $ &  142.72 & $ 0.928 \pm 0.018 $ &  $ 0.585 \pm 0.022 $ & $ 1.513 \pm 0.028 $ \\          
129.44 & $ 22.44 \pm  0.02 $ & $ 0.3202 \pm 0.0013 $ & $ -61.96 \pm 0.12 $ &  156.99 & $ 0.933 \pm 0.020 $ &  $ 0.587 \pm 0.024 $ & $ 1.520 \pm 0.031 $ \\          
140.05 & $ 22.55 \pm  0.02 $ & $ 0.3423 \pm 0.0012 $ & $ -61.57 \pm 0.12 $ &  172.69 & $ 0.940 \pm 0.022 $ &  $ 0.584 \pm 0.027 $ & $ 1.524 \pm 0.035 $ \\          
152.49 & $ 22.69 \pm  0.02 $ & $ 0.3556 \pm 0.0012 $ & $ -60.85 \pm 0.12 $ &  189.96 & $ 0.932 \pm 0.025 $ &  $ 0.588 \pm 0.030 $ & $ 1.520 \pm 0.039 $ \\          
166.25 & $ 22.83 \pm  0.03 $ & $ 0.3670 \pm 0.0012 $ & $ -60.23 \pm 0.11 $ &  208.96 & $ 0.932 \pm 0.028 $ &  $ 0.588 \pm 0.035 $ & $ 1.520 \pm 0.045 $ \\          
181.89 & $ 22.99 \pm  0.03 $ & $ 0.3738 \pm 0.0012 $ & $ -59.71 \pm 0.11 $ &  229.85 & $ 0.927 \pm 0.033 $ &  $ 0.594 \pm 0.040 $ & $ 1.521 \pm 0.052 $ \\          
198.54 & $ 23.16 \pm  0.03 $ & $ 0.3834 \pm 0.0013 $ & $ -58.67 \pm 0.12 $ &  252.84 & $ 0.922 \pm 0.038 $ &  $ 0.597 \pm 0.046 $ & $ 1.519 \pm 0.060 $ \\          
216.35 & $ 23.32 \pm  0.04 $ & $ 0.3949 \pm 0.0012 $ & $ -57.69 \pm 0.11 $ &  278.12 & $ 0.932 \pm 0.044 $ &  $ 0.600 \pm 0.054 $ & $ 1.532 \pm 0.070 $ \\          
237.03 & $ 23.53 \pm  0.05 $ & $ 0.3997 \pm 0.0015 $ & $ -56.40 \pm 0.14 $ &  305.93 & $ 0.934 \pm 0.054 $ &  $ 0.601 \pm 0.065 $ & $ 1.535 \pm 0.085 $ \\          
259.30 & $ 23.71 \pm  0.06 $ & $ 0.4063 \pm 0.0013 $ & $ -56.31 \pm 0.12 $ &  336.53 & $ 0.920 \pm 0.067 $ &  $ 0.617 \pm 0.079 $ & $ 1.537 \pm 0.103 $ \\          
282.60 & $ 23.96 \pm  0.07 $ & $ 0.4172 \pm 0.0014 $ & $ -55.87 \pm 0.12 $ &  370.18 & $ 0.836 \pm 0.090 $ &  $ 0.616 \pm 0.097 $ & $ 1.452 \pm 0.132 $ \\          
311.85 & $ 24.26 \pm  0.09 $ & $ 0.4135 \pm 0.0015 $ & $ -55.78 \pm 0.13 $ &  407.20 & $ 0.932 \pm 0.105 $ &  $ 0.627 \pm 0.126 $ & $ 1.559 \pm 0.164 $ \\          
345.77 & $ 24.62 \pm  0.13 $ & $ 0.4041 \pm 0.0019 $ & $ -55.36 \pm 0.17 $ &  447.92 & $ 0.933 \pm 0.147 $ &  $ 0.641 \pm 0.175 $ & $ 1.574 \pm 0.229 $ \\          
379.64 & $ 24.93 \pm  0.17 $ & $ 0.4063 \pm 0.0024 $ & $ -55.90 \pm 0.22 $ &  492.71 & $ 0.964 \pm 0.197 $ &  $ 0.633 \pm 0.234 $ & $ 1.597 \pm 0.306 $ \\          
414.68 & $ 25.22 \pm  0.23 $ & $ 0.4146 \pm 0.0032 $ & $ -55.34 \pm 0.28 $ &  541.98 & $ 0.997 \pm 0.260 $ &  $ 0.632 \pm 0.308 $ & $ 1.629 \pm 0.404 $ \\          
458.32 & $ 25.49 \pm  0.30 $ & $ 0.4090 \pm 0.0039 $ & $ -54.35 \pm 0.35 $ &  596.18 & $ 1.094 \pm 0.344 $ &  $ 0.568 \pm 0.413 $ & $ 1.662 \pm 0.538 $ \\          
509.96 & $ 25.77 \pm  0.40 $ & $ 0.3953 \pm 0.0066 $ & $ -52.37 \pm 0.61 $ &  655.80 & $ 1.195 \pm 0.461 $ &  $ 0.548 \pm 0.553 $ & $ 1.743 \pm 0.720 $\\          
\hline       
\end{longtable*}
}

%% file: n4472_sfc_table.tex
{\footnotesize
\begin{longtable*}{cccccccc}
\\
\caption[Surface photometry for NGC 4472]{Surface photometry for NGC 4472} \label{tab:chap4_n4472_sfc_phot} \\
\hline
\hline
$R$ & $\mu_{V} \pm \sigma$  & $\epsilon \pm \sigma$ & ${\rm PA} \pm \sigma$  & SMA  & $(B-V) \pm \sigma$ & $(V-R) \pm \sigma$ & $(B-R) \pm \sigma$ \\
(arcsec) & $({\rm mag ~arcsec}^{-2})$ & & (degrees E of N) & (arcsec) & (mag) & (mag) & (mag) \\
\hline
\endfirsthead
\caption[]{Surface photometry for NGC 4472, continued.}\\
\hline
\hline
$R$ & $\mu_{V} \pm \sigma_{\mu_{V}}$  & $\epsilon \pm \sigma_{\epsilon}$ & ${\rm PA} \pm \sigma_{\rm PA}$  & SMA  & $(B-V) \pm \sigma_{B-V}$ & $(V-R) \pm \sigma_{V-R}$ & $(B-R) \pm \sigma_{B-R}$ \\
(arcsec) & $({\rm mag ~arcsec}^{-2})$ & & (degrees) & (arcsec) & (mag) & (mag) & (mag) \\
\hline
\endhead
 22.78 & $ 19.03 \pm  0.03 $ & $ 0.1599 \pm 0.0016 $ & $ -20.81 \pm 0.44 $ &   24.85 & $ 0.829 \pm 0.030 $ &  $ 0.599 \pm 0.030 $ & $ 1.429 \pm 0.043$ \\          
 25.06 & $ 19.18 \pm  0.03 $ & $ 0.1599 \pm 0.0007 $ & $ -20.81 \pm 0.18 $ &   27.34 & $ 0.827 \pm 0.030 $ &  $ 0.600 \pm 0.030 $ & $ 1.427 \pm 0.043$   \\         
 27.56 & $ 19.33 \pm  0.03 $ & $ 0.1599 \pm 0.0005 $ & $ -20.81 \pm 0.12 $ &   30.07 & $ 0.828 \pm 0.030 $ &  $ 0.600 \pm 0.030 $ & $ 1.428 \pm 0.043$ \\           
 30.32 & $ 19.49 \pm  0.03 $ & $ 0.1599 \pm 0.0005 $ & $ -20.81 \pm 0.11 $ &   33.08 & $ 0.828 \pm 0.030 $ &  $ 0.600 \pm 0.030 $ & $ 1.427 \pm 0.043 $ \\           
 33.35 & $ 19.66 \pm  0.03 $ & $ 0.1599 \pm 0.0005 $ & $ -20.81 \pm 0.11 $ &   36.39 & $ 0.828 \pm 0.030 $ &  $ 0.599 \pm 0.031 $ & $ 1.427 \pm 0.043 $ \\           
 36.72 & $ 19.83 \pm  0.03 $ & $ 0.1585 \pm 0.0005 $ & $ -21.18 \pm 0.10 $ &   40.02 & $ 0.826 \pm 0.030 $ &  $ 0.600 \pm 0.031 $ & $ 1.426 \pm 0.043 $ \\           
 40.40 & $ 19.99 \pm  0.03 $ & $ 0.1582 \pm 0.0006 $ & $ -21.67 \pm 0.12 $ &   44.03 & $ 0.824 \pm 0.030 $ &  $ 0.599 \pm 0.031 $ & $ 1.423 \pm 0.043 $ \\           
 44.47 & $ 20.15 \pm  0.03 $ & $ 0.1569 \pm 0.0007 $ & $ -21.72 \pm 0.13 $ &   48.43 & $ 0.818 \pm 0.030 $ &  $ 0.598 \pm 0.031 $ & $ 1.416 \pm 0.043 $ \\           
 48.79 & $ 20.30 \pm  0.03 $ & $ 0.1613 \pm 0.0011 $ & $ -21.36 \pm 0.19 $ &   53.27 & $ 0.820 \pm 0.030 $ &  $ 0.597 \pm 0.032 $ & $ 1.417 \pm 0.044 $ \\           
 53.48 & $ 20.45 \pm  0.03 $ & $ 0.1671 \pm 0.0008 $ & $ -22.20 \pm 0.14 $ &   58.60 & $ 0.816 \pm 0.030 $ &  $ 0.598 \pm 0.032 $ & $ 1.413 \pm 0.044 $ \\           
 58.71 & $ 20.59 \pm  0.03 $ & $ 0.1705 \pm 0.0009 $ & $ -22.19 \pm 0.17 $ &   64.46 & $ 0.820 \pm 0.030 $ &  $ 0.595 \pm 0.033 $ & $ 1.415 \pm 0.045 $ \\           
 64.40 & $ 20.74 \pm  0.03 $ & $ 0.1751 \pm 0.0006 $ & $ -23.03 \pm 0.10 $ &   70.91 & $ 0.818 \pm 0.031 $ &  $ 0.596 \pm 0.034 $ & $ 1.414 \pm 0.045 $ \\           
 70.80 & $ 20.89 \pm  0.03 $ & $ 0.1761 \pm 0.0007 $ & $ -23.18 \pm 0.13 $ &   78.00 & $ 0.815 \pm 0.031 $ &  $ 0.596 \pm 0.035 $ & $ 1.410 \pm 0.046 $ \\           
 77.92 & $ 21.05 \pm  0.03 $ & $ 0.1752 \pm 0.0006 $ & $ -23.31 \pm 0.11 $ &   85.80 & $ 0.814 \pm 0.031 $ &  $ 0.596 \pm 0.036 $ & $ 1.410 \pm 0.048 $ \\           
 85.76 & $ 21.21 \pm  0.03 $ & $ 0.1743 \pm 0.0007 $ & $ -23.72 \pm 0.13 $ &   94.38 & $ 0.813 \pm 0.031 $ &  $ 0.595 \pm 0.038 $ & $ 1.407 \pm 0.049 $ \\           
 94.45 & $ 21.39 \pm  0.03 $ & $ 0.1723 \pm 0.0006 $ & $ -23.70 \pm 0.11 $ &  103.82 & $ 0.802 \pm 0.032 $ &  $ 0.598 \pm 0.041 $ & $ 1.399 \pm 0.052 $ \\           
104.18 & $ 21.60 \pm  0.03 $ & $ 0.1678 \pm 0.0007 $ & $ -23.86 \pm 0.13 $ &  114.20 & $ 0.804 \pm 0.033 $ &  $ 0.599 \pm 0.045 $ & $ 1.404 \pm 0.055 $ \\           
114.72 & $ 21.80 \pm  0.04 $ & $ 0.1660 \pm 0.0007 $ & $ -24.60 \pm 0.14 $ &  125.62 & $ 0.805 \pm 0.039 $ &  $ 0.599 \pm 0.054 $ & $ 1.404 \pm 0.066 $ \\           
125.93 & $ 22.01 \pm  0.04 $ & $ 0.1694 \pm 0.0008 $ & $ -25.22 \pm 0.15 $ &  138.18 & $ 0.809 \pm 0.047 $ &  $ 0.602 \pm 0.065 $ & $ 1.411 \pm 0.080 $ \\           
138.16 & $ 22.22 \pm  0.05 $ & $ 0.1738 \pm 0.0010 $ & $ -25.78 \pm 0.17 $ &  152.00 & $ 0.807 \pm 0.057 $ &  $ 0.604 \pm 0.078 $ & $ 1.410 \pm 0.097 $ \\           
150.75 & $ 22.41 \pm  0.06 $ & $ 0.1871 \pm 0.0011 $ & $ -26.25 \pm 0.19 $ &  167.20 & $ 0.805 \pm 0.068 $ &  $ 0.611 \pm 0.093 $ & $ 1.416 \pm 0.115 $ \\           
164.38 & $ 22.58 \pm  0.07 $ & $ 0.2012 \pm 0.0015 $ & $ -27.22 \pm 0.24 $ &  183.92 & $ 0.809 \pm 0.080 $ &  $ 0.614 \pm 0.109 $ & $ 1.424 \pm 0.135 $ \\           
179.38 & $ 22.75 \pm  0.09 $ & $ 0.2138 \pm 0.0014 $ & $ -27.50 \pm 0.21 $ &  202.31 & $ 0.813 \pm 0.093 $ &  $ 0.619 \pm 0.127 $ & $ 1.431 \pm 0.158 $ \\           
196.21 & $ 22.94 \pm  0.10 $ & $ 0.2226 \pm 0.0015 $ & $ -28.53 \pm 0.22 $ &  222.54 & $ 0.803 \pm 0.110 $ &  $ 0.628 \pm 0.150 $ & $ 1.432 \pm 0.186 $ \\           
214.41 & $ 23.11 \pm  0.12 $ & $ 0.2328 \pm 0.0016 $ & $ -28.19 \pm 0.22 $ &  244.79 & $ 0.823 \pm 0.129 $ &  $ 0.633 \pm 0.175 $ & $ 1.455 \pm 0.218 $ \\           
234.17 & $ 23.30 \pm  0.14 $ & $ 0.2437 \pm 0.0017 $ & $ -29.00 \pm 0.22 $ &  269.27 & $ 0.826 \pm 0.155 $ &  $ 0.644 \pm 0.208 $ & $ 1.470 \pm 0.259 $ \\           
256.38 & $ 23.52 \pm  0.17 $ & $ 0.2508 \pm 0.0018 $ & $ -29.29 \pm 0.22 $ &  296.20 & $ 0.827 \pm 0.190 $ &  $ 0.671 \pm 0.252 $ & $ 1.498 \pm 0.315 $ \\           
279.90 & $ 23.70 \pm  0.21 $ & $ 0.2620 \pm 0.0068 $ & $ -30.22 \pm 0.80 $ &  325.82 & $ 0.854 \pm 0.227 $ &  $ 0.665 \pm 0.302 $ & $ 1.519 \pm 0.378 $ \\           
307.89 & $ 23.97 \pm  0.27 $ & $ 0.2620 \pm 0.0025 $ & $ -32.23 \pm 0.29 $ &  358.40 & $ 0.869 \pm 0.294 $ &  $ 0.702 \pm 0.382 $ & $ 1.570 \pm 0.482 $ \\           
339.41 & $ 24.24 \pm  0.35 $ & $ 0.2588 \pm 0.0026 $ & $ -30.98 \pm 0.32 $ &  394.24 & $ 0.885 \pm 0.382 $ &  $ 0.726 \pm 0.490 $ & $ 1.611 \pm 0.621 $ \\           
370.95 & $ 24.53 \pm  0.47 $ & $ 0.2683 \pm 0.0031 $ & $ -31.77 \pm 0.38 $ &  433.66 & $ 0.938 \pm 0.513 $ &  $ 0.745 \pm 0.649 $ & $ 1.682 \pm 0.827 $ \\           
403.48 & $ 24.80 \pm  0.62 $ & $ 0.2846 \pm 0.0035 $ & $ -32.69 \pm 0.39 $ &  477.03 & $ 1.005 \pm 0.685 $ &  $ 0.755 \pm 0.857 $ & $ 1.760 \pm 1.097 $ \\
\hline       
\end{longtable*}
}

%% file: n5813_sfc_table.tex
{\footnotesize
\begin{longtable*}{cccccccc}
\\
\caption[Surface photometry for NGC 5813]{Surface photometry for NGC 5813} \label{tab:chap4_n5813_sfc_phot} \\
\hline
\hline
$R$ & $\mu_{V} \pm \sigma$  & $\epsilon \pm \sigma$ & ${\rm PA} \pm \sigma$  & SMA  & $(B-V) \pm \sigma$ & $(V-R) \pm \sigma$ & $(B-R) \pm \sigma$ \\
(arcsec) & $({\rm mag ~arcsec}^{-2})$ & & (degrees E of N) & (arcsec) & (mag) & (mag) & (mag) \\
\hline
\endfirsthead
\caption[]{Surface photometry for NGC 5813, continued.}\\
\hline
\hline
$R$ & $\mu_{V} \pm \sigma_{\mu_{V}}$  & $\epsilon \pm \sigma_{\epsilon}$ & ${\rm PA} \pm \sigma_{\rm PA}$  & SMA  & $(B-V) \pm \sigma_{B-V}$ & $(V-R) \pm \sigma_{V-R}$ & $(B-R) \pm \sigma_{B-R}$ \\
(arcsec) & $({\rm mag ~arcsec}^{-2})$ & & (degrees) & (arcsec) & (mag) & (mag) & (mag) \\
\hline
\endhead 
  7.47 & $ 19.13 \pm  0.01 $ & $ 0.0893 \pm 0.0029 $ & $ -43.16 \pm 0.63 $ &    7.83 & $ 0.989 \pm 0.011 $ &  $ 0.570 \pm 0.012 $ & $ 1.552 \pm 0.007 $ \\          
  8.22 & $ 19.27 \pm  0.01 $ & $ 0.0893 \pm 0.0037 $ & $ -43.16 \pm 1.21 $ &    8.61 & $ 0.983 \pm 0.011 $ &  $ 0.584 \pm 0.011 $ & $ 1.563 \pm 0.006 $ \\          
  9.04 & $ 19.40 \pm  0.01 $ & $ 0.0893 \pm 0.0022 $ & $ -43.16 \pm 0.86 $ &    9.47 & $ 0.974 \pm 0.010 $ &  $ 0.581 \pm 0.011 $ & $ 1.552 \pm 0.005 $ \\          
  9.93 & $ 19.53 \pm  0.01 $ & $ 0.0906 \pm 0.0015 $ & $ -44.87 \pm 0.73 $ &   10.42 & $ 0.972 \pm 0.010 $ &  $ 0.579 \pm 0.010 $ & $ 1.547 \pm 0.004 $ \\          
 10.88 & $ 19.66 \pm  0.01 $ & $ 0.0980 \pm 0.0014 $ & $ -42.26 \pm 0.73 $ &   11.46 & $ 0.971 \pm 0.010 $ &  $ 0.582 \pm 0.010 $ & $ 1.550 \pm 0.004 $ \\          
 11.91 & $ 19.79 \pm  0.01 $ & $ 0.1075 \pm 0.0013 $ & $ -40.62 \pm 0.62 $ &   12.60 & $ 0.968 \pm 0.010 $ &  $ 0.588 \pm 0.010 $ & $ 1.554 \pm 0.004 $ \\          
 13.07 & $ 19.91 \pm  0.01 $ & $ 0.1116 \pm 0.0008 $ & $ -44.72 \pm 0.38 $ &   13.87 & $ 0.969 \pm 0.010 $ &  $ 0.586 \pm 0.010 $ & $ 1.552 \pm 0.003 $ \\          
 14.22 & $ 20.02 \pm  0.01 $ & $ 0.1307 \pm 0.0008 $ & $ -44.72 \pm 0.28 $ &   15.25 & $ 0.966 \pm 0.010 $ &  $ 0.588 \pm 0.010 $ & $ 1.551 \pm 0.004 $ \\          
 15.64 & $ 20.16 \pm  0.01 $ & $ 0.1307 \pm 0.0013 $ & $ -44.72 \pm 0.33 $ &   16.78 & $ 0.964 \pm 0.010 $ &  $ 0.585 \pm 0.010 $ & $ 1.547 \pm 0.004 $ \\          
 17.11 & $ 20.28 \pm  0.01 $ & $ 0.1405 \pm 0.0009 $ & $ -44.97 \pm 0.20 $ &   18.45 & $ 0.967 \pm 0.011 $ &  $ 0.583 \pm 0.010 $ & $ 1.547 \pm 0.004 $ \\          
 18.66 & $ 20.41 \pm  0.01 $ & $ 0.1545 \pm 0.0009 $ & $ -45.21 \pm 0.17 $ &   20.30 & $ 0.964 \pm 0.011 $ &  $ 0.585 \pm 0.010 $ & $ 1.546 \pm 0.004 $ \\          
 20.40 & $ 20.54 \pm  0.01 $ & $ 0.1654 \pm 0.0010 $ & $ -45.28 \pm 0.17 $ &   22.33 & $ 0.963 \pm 0.011 $ &  $ 0.582 \pm 0.010 $ & $ 1.543 \pm 0.005 $ \\          
 22.25 & $ 20.66 \pm  0.01 $ & $ 0.1797 \pm 0.0012 $ & $ -44.73 \pm 0.18 $ &   24.56 & $ 0.962 \pm 0.011 $ &  $ 0.582 \pm 0.010 $ & $ 1.541 \pm 0.006 $ \\          
 24.18 & $ 20.78 \pm  0.01 $ & $ 0.1993 \pm 0.0008 $ & $ -44.73 \pm 0.14 $ &   27.02 & $ 0.960 \pm 0.011 $ &  $ 0.579 \pm 0.011 $ & $ 1.536 \pm 0.006 $ \\          
 26.35 & $ 20.90 \pm  0.01 $ & $ 0.2138 \pm 0.0009 $ & $ -45.19 \pm 0.13 $ &   29.72 & $ 0.960 \pm 0.011 $ &  $ 0.577 \pm 0.011 $ & $ 1.534 \pm 0.007 $ \\          
 28.72 & $ 21.02 \pm  0.01 $ & $ 0.2285 \pm 0.0009 $ & $ -45.06 \pm 0.11 $ &   32.69 & $ 0.959 \pm 0.012 $ &  $ 0.579 \pm 0.011 $ & $ 1.535 \pm 0.008 $ \\          
 31.35 & $ 21.14 \pm  0.01 $ & $ 0.2398 \pm 0.0009 $ & $ -45.59 \pm 0.11 $ &   35.96 & $ 0.960 \pm 0.012 $ &  $ 0.576 \pm 0.011 $ & $ 1.533 \pm 0.008 $ \\          
 34.28 & $ 21.27 \pm  0.01 $ & $ 0.2490 \pm 0.0008 $ & $ -45.54 \pm 0.11 $ &   39.56 & $ 0.956 \pm 0.013 $ &  $ 0.576 \pm 0.011 $ & $ 1.529 \pm 0.009 $ \\          
 37.60 & $ 21.41 \pm  0.01 $ & $ 0.2534 \pm 0.0007 $ & $ -45.59 \pm 0.11 $ &   43.51 & $ 0.959 \pm 0.013 $ &  $ 0.573 \pm 0.012 $ & $ 1.529 \pm 0.011 $ \\          
 41.31 & $ 21.56 \pm  0.01 $ & $ 0.2552 \pm 0.0007 $ & $ -46.02 \pm 0.11 $ &   47.86 & $ 0.953 \pm 0.015 $ &  $ 0.574 \pm 0.013 $ & $ 1.524 \pm 0.012 $ \\          
 45.51 & $ 21.72 \pm  0.01 $ & $ 0.2528 \pm 0.0006 $ & $ -46.22 \pm 0.09 $ &   52.65 & $ 0.954 \pm 0.018 $ &  $ 0.572 \pm 0.016 $ & $ 1.523 \pm 0.014 $ \\          
 49.96 & $ 21.89 \pm  0.01 $ & $ 0.2559 \pm 0.0007 $ & $ -46.24 \pm 0.12 $ &   57.91 & $ 0.952 \pm 0.021 $ &  $ 0.570 \pm 0.018 $ & $ 1.519 \pm 0.017 $ \\          
 54.68 & $ 22.06 \pm  0.02 $ & $ 0.2634 \pm 0.0007 $ & $ -46.72 \pm 0.10 $ &   63.71 & $ 0.947 \pm 0.024 $ &  $ 0.571 \pm 0.021 $ & $ 1.515 \pm 0.020 $ \\          
 59.87 & $ 22.24 \pm  0.02 $ & $ 0.2701 \pm 0.0007 $ & $ -46.83 \pm 0.09 $ &   70.08 & $ 0.943 \pm 0.028 $ &  $ 0.570 \pm 0.025 $ & $ 1.511 \pm 0.023 $ \\          
 65.71 & $ 22.43 \pm  0.03 $ & $ 0.2734 \pm 0.0008 $ & $ -46.87 \pm 0.10 $ &   77.08 & $ 0.941 \pm 0.034 $ &  $ 0.570 \pm 0.030 $ & $ 1.508 \pm 0.027 $ \\          
 72.11 & $ 22.62 \pm  0.03 $ & $ 0.2768 \pm 0.0008 $ & $ -46.89 \pm 0.11 $ &   84.79 & $ 0.939 \pm 0.040 $ &  $ 0.565 \pm 0.035 $ & $ 1.501 \pm 0.032 $ \\          
 79.05 & $ 22.80 \pm  0.04 $ & $ 0.2817 \pm 0.0010 $ & $ -47.43 \pm 0.12 $ &   93.27 & $ 0.943 \pm 0.048 $ &  $ 0.565 \pm 0.042 $ & $ 1.505 \pm 0.039 $ \\          
 86.70 & $ 23.01 \pm  0.04 $ & $ 0.2860 \pm 0.0009 $ & $ -47.67 \pm 0.12 $ &  102.60 & $ 0.938 \pm 0.058 $ &  $ 0.567 \pm 0.051 $ & $ 1.502 \pm 0.047 $ \\          
 95.31 & $ 23.22 \pm  0.05 $ & $ 0.2868 \pm 0.0011 $ & $ -48.20 \pm 0.13 $ &  112.86 & $ 0.935 \pm 0.070 $ &  $ 0.569 \pm 0.062 $ & $ 1.502 \pm 0.056 $ \\          
104.69 & $ 23.43 \pm  0.06 $ & $ 0.2889 \pm 0.0012 $ & $ -48.91 \pm 0.14 $ &  124.15 & $ 0.929 \pm 0.084 $ &  $ 0.573 \pm 0.075 $ & $ 1.501 \pm 0.068 $ \\          
114.72 & $ 23.64 \pm  0.08 $ & $ 0.2943 \pm 0.0015 $ & $ -49.33 \pm 0.17 $ &  136.56 & $ 0.919 \pm 0.102 $ &  $ 0.582 \pm 0.091 $ & $ 1.502 \pm 0.082 $ \\          
125.63 & $ 23.84 \pm  0.09 $ & $ 0.3006 \pm 0.0016 $ & $ -50.22 \pm 0.21 $ &  150.22 & $ 0.920 \pm 0.123 $ &  $ 0.586 \pm 0.109 $ & $ 1.507 \pm 0.098 $ \\          
136.89 & $ 24.03 \pm  0.11 $ & $ 0.3137 \pm 0.0017 $ & $ -50.22 \pm 0.23 $ &  165.24 & $ 0.909 \pm 0.146 $ &  $ 0.599 \pm 0.129 $ & $ 1.513 \pm 0.116 $ \\          
150.06 & $ 24.23 \pm  0.13 $ & $ 0.3184 \pm 0.0023 $ & $ -51.48 \pm 0.29 $ &  181.76 & $ 0.901 \pm 0.175 $ &  $ 0.615 \pm 0.155 $ & $ 1.522 \pm 0.138 $ \\          
167.09 & $ 24.44 \pm  0.16 $ & $ 0.3016 \pm 0.0025 $ & $ -53.37 \pm 0.35 $ &  199.94 & $ 0.899 \pm 0.213 $ &  $ 0.617 \pm 0.188 $ & $ 1.522 \pm 0.168 $ \\                  
\hline       
\end{longtable*}
}

%% file: n4594_sfc_table.tex
{\footnotesize
\begin{longtable*}{cccccccc}
\label{tab:chap4_n4594_sfc_phot}
\\
\caption[Surface photometry for NGC~4594]{Surface photometry for NGC~4594} \\
\hline
\hline
$R$ & $\mu_{V} \pm \sigma$  & $\epsilon \pm \sigma$ & ${\rm PA} \pm \sigma$  & SMA  & $(B-V) \pm \sigma$ & $(V-R) \pm \sigma$ & $(B-R) \pm \sigma$ \\
(arcsec) & $({\rm mag ~arcsec}^{-2})$ & & (degrees E of N) & (arcsec) & (mag) & (mag) & (mag) \\
\hline
\endfirsthead
\endhead 
 59.09 & $ 20.10 \pm  0.02 $ & $ 0.4243 \pm 0.0056 $ & $  88.23 \pm 0.45 $ &   77.88 & $ 0.914 \pm 0.021 $ &  $ 0.645 \pm 0.021 $ & $ 1.559 \pm 0.030 $ \\          
 64.87 & $ 20.28 \pm  0.02 $ & $ 0.4266 \pm 0.0038 $ & $  88.21 \pm 0.30 $ &   85.66 & $ 0.925 \pm 0.021 $ &  $ 0.647 \pm 0.021 $ & $ 1.572 \pm 0.030 $ \\          
 72.83 & $ 20.53 \pm  0.02 $ & $ 0.4026 \pm 0.0033 $ & $  88.01 \pm 0.25 $ &   94.23 & $ 0.914 \pm 0.021 $ &  $ 0.663 \pm 0.021 $ & $ 1.576 \pm 0.030 $ \\          
 79.79 & $ 20.70 \pm  0.02 $ & $ 0.4075 \pm 0.0026 $ & $  87.65 \pm 0.19 $ &  103.65 & $ 0.919 \pm 0.021 $ &  $ 0.625 \pm 0.021 $ & $ 1.544 \pm 0.029 $ \\          
 88.19 & $ 20.90 \pm  0.02 $ & $ 0.4018 \pm 0.0021 $ & $  87.57 \pm 0.16 $ &  114.02 & $ 0.917 \pm 0.021 $ &  $ 0.629 \pm 0.021 $ & $ 1.545 \pm 0.029 $ \\          
 97.19 & $ 21.11 \pm  0.02 $ & $ 0.3995 \pm 0.0014 $ & $  87.59 \pm 0.11 $ &  125.42 & $ 0.924 \pm 0.020 $ &  $ 0.628 \pm 0.021 $ & $ 1.553 \pm 0.029 $ \\          
107.61 & $ 21.34 \pm  0.02 $ & $ 0.3916 \pm 0.0012 $ & $  87.79 \pm 0.09 $ &  137.96 & $ 0.921 \pm 0.020 $ &  $ 0.628 \pm 0.021 $ & $ 1.549 \pm 0.030 $ \\          
120.13 & $ 21.58 \pm  0.02 $ & $ 0.3734 \pm 0.0010 $ & $  87.59 \pm 0.09 $ &  151.76 & $ 0.918 \pm 0.021 $ &  $ 0.626 \pm 0.022 $ & $ 1.544 \pm 0.030 $ \\          
134.48 & $ 21.84 \pm  0.02 $ & $ 0.3511 \pm 0.0008 $ & $  87.47 \pm 0.08 $ &  166.94 & $ 0.917 \pm 0.021 $ &  $ 0.624 \pm 0.023 $ & $ 1.541 \pm 0.031 $ \\          
149.62 & $ 22.09 \pm  0.02 $ & $ 0.3361 \pm 0.0008 $ & $  87.27 \pm 0.07 $ &  183.63 & $ 0.908 \pm 0.021 $ &  $ 0.623 \pm 0.025 $ & $ 1.531 \pm 0.033 $ \\          
166.27 & $ 22.35 \pm  0.02 $ & $ 0.3224 \pm 0.0008 $ & $  86.71 \pm 0.08 $ &  201.99 & $ 0.903 \pm 0.022 $ &  $ 0.623 \pm 0.027 $ & $ 1.526 \pm 0.035 $ \\          
184.17 & $ 22.60 \pm  0.02 $ & $ 0.3130 \pm 0.0007 $ & $  86.66 \pm 0.08 $ &  222.19 & $ 0.905 \pm 0.023 $ &  $ 0.631 \pm 0.031 $ & $ 1.537 \pm 0.038 $ \\          
204.34 & $ 22.85 \pm  0.02 $ & $ 0.3010 \pm 0.0008 $ & $  86.27 \pm 0.09 $ &  244.41 & $ 0.899 \pm 0.025 $ &  $ 0.634 \pm 0.035 $ & $ 1.534 \pm 0.043 $ \\          
228.13 & $ 23.12 \pm  0.03 $ & $ 0.2800 \pm 0.0009 $ & $  86.83 \pm 0.11 $ &  268.86 & $ 0.901 \pm 0.032 $ &  $ 0.644 \pm 0.045 $ & $ 1.545 \pm 0.055 $ \\          
257.51 & $ 23.42 \pm  0.03 $ & $ 0.2418 \pm 0.0011 $ & $  86.50 \pm 0.14 $ &  295.74 & $ 0.904 \pm 0.042 $ &  $ 0.654 \pm 0.059 $ & $ 1.557 \pm 0.072 $ \\          
290.28 & $ 23.74 \pm  0.05 $ & $ 0.2038 \pm 0.0014 $ & $  86.61 \pm 0.20 $ &  325.31 & $ 0.903 \pm 0.056 $ &  $ 0.678 \pm 0.078 $ & $ 1.581 \pm 0.096 $ \\          
329.14 & $ 24.09 \pm  0.06 $ & $ 0.1540 \pm 0.0015 $ & $  86.62 \pm 0.30 $ &  357.85 & $ 0.910 \pm 0.078 $ &  $ 0.697 \pm 0.107 $ & $ 1.607 \pm 0.132 $ \\          
373.47 & $ 24.51 \pm  0.09 $ & $ 0.0998 \pm 0.0020 $ & $  84.73 \pm 0.58 $ &  393.63 & $ 0.914 \pm 0.114 $ &  $ 0.732 \pm 0.153 $ & $ 1.647 \pm 0.190 $ \\          
420.45 & $ 24.95 \pm  0.14 $ & $ 0.0571 \pm 0.0027 $ & $  85.66 \pm 1.35 $ &  432.99 & $ 0.879 \pm 0.170 $ &  $ 0.768 \pm 0.227 $ & $ 1.647 \pm 0.284 $ \\          
\hline       
\end{longtable*}
}